\documentclass[onecolumn]{aastex631}
\usepackage{amsmath,amssymb}
\def\msol{\hbox{$\hbox{M}_\odot$}}
\def\lsol{\hbox{$\hbox{L}_\odot$}}

\newcommand{\sgra}{Sgr~A*}  
\newcounter{subfig}

\begin{document}
\title{
Non-stop Variability of Sgr A* using JWST at {\bf2.1} and 4.8 micron Wavelengths:\\ 
Evidence for Distinct  Populations of Faint  and Bright Variable Emission}

\author[0000-0001-8403-8548]{F. Yusef-Zadeh} 
\affiliation{Dept Physics and Astronomy, CIERA, Northwestern University, 2145 Sheridan Road, Evanston , IL 60207, USA
(zadeh@northwestern.edu)}
\author[0000-0001-8403-8548]{H. Bushouse} 
\affiliation{Space Telescope Science Institute, 3700 San Martin Drive, Baltimore,  MD 21218  (bushouse@stsci.edu)}
\author[0000-0001-8403-8548]{R. G. Arendt} 
\affiliation{Center for Space Sciences and Technology, University of Maryland, Baltimore County, Baltimore, MD 21250, USA}
\affiliation{Code 665, NASA/GSFC, 8800 Greenbelt Road, Greenbelt, MD 20771, USA}
\affiliation{Center for Research and Exploration in Space Science and Technology, NASA/GSFC, Greenbelt, MD 20771, USA (Richard.G.Arendt@nasa.gov)}
\author[0000-0001-8403-8548]{M. Wardle} 
\affiliation{School of Mathematical and Physical Sciences,  Centre for Astronomy and Space Technology,\\
Macquarie University, Sydney NSW 2109, Australia (mark.wardle@mq.edu.au)}
\author[0000-0003-3503-3446]{J. M. Michail}
\affiliation{Center for Astrophysics, Harvard \& Smithsonian, 60 Garden St., Cambridge, MA 02138}
\affiliation{NSF Astronomy and Astrophysics Postdoctoral Fellow (joseph.michail@cfa.harvard.edu)}
\author[0000-0002-7570-5596]{C. J. Chandler} 
\affiliation{National Radio Astronomy Observatory, P.O. Box O,  Socorro, NM 87801, USA (cchandle@nrao.edu)}

\begin{abstract}
We present first results of JWST Cycle 1 and 2 observations of Sgr A* using NIRCam taken simultaneously at 2.1 and 4.8 $\mu$m for a total of $\sim48$h
over seven different epochs in 2023 and 2024.
We find correlated variability at 2.1 and 4.8 $\mu$m in
all epochs, continual short-time scale (a few seconds) variability and epoch-to-epoch variable emission
implying long-term ($\sim$ days to months) variability of Sgr A*. A highlight of this analysis is the evidence for sub-minute, horizon-scale time
variability of Sgr A*, probing inner accretion disk size scales. The power spectra of the light curves in each observing epoch also indicate long-term
variable emission. With continuous observations,
JWST data suggest that the flux of \sgra\ is fluctuating constantly. The flux density correlation exhibits
 a distinct break in the slope at $\sim$3 mJy at 2.1 $\mu$m. The analysis indicates two different processes contributing to the
variability of Sgr A*. Brighter emission trends towards shallower spectral indices than the fainter emission. Cross correlation of the light curves
indicates for the first time, a time delay of 3 - 40 sec in the 4.8 $\mu$m variability with respect to 2.1 $\mu$m. This phase
shift leads to loops in plots of flux density vs spectral index as the emission  rises and  falls. Modeling suggests that the synchrotron emission from the
evolving, age-stratified electron population reproduces the shape of the observed light curves with a direct estimate of
the magnetic field strengths in the range between
40--90\,G, and upper cutoff energy, $E_c$,  between 420 and 720\,MeV.
\end{abstract}

%\keywords{cosmology: observations --- diffuse radiation --- zodiacal dust}

\section{Introduction} 
\label{sec:intro}

The center of our Galaxy hosts a 4$\times10^6$\msol\ supermassive black hole, \sgra, presenting an unparalleled opportunity 
to study up close how material is irradiated, captured, accreted, and ejected by a black hole. The accretion disk of \sgra\ 
is fueled by winds from young, mass-losing stars orbiting within a few arc-seconds (0.1pc) of the black hole, with yearly to 
decadal orbital time scales. The bolometric luminosity of \sgra\ is only $\sim$100\lsol, suggesting that a small fraction of 
the infalling material reaches the event horizon \citep{quataert04,yuan14,wang15}. The exact nature of the accretion flow is 
still not well understood \citep{yuan03,goldston05,moscib13,ressler20a,ressler20b,dexter20,ressler23}.

%Now that the size  of the \sgra\ event horizon shadow silhouetted against its accretion disk, 26.13 $\mu$as (micro-arcsec) with zero spin, has been measured by the Event 
%mechanisms at work within it \citep{genzel03,ghez05,eckart08,zadeh06b,michail19,witzel21,michail24}. 

Now that the size of the \sgra\ event horizon shadow silhouetted against its accretion disk, 26.13 $\mu$as (micro-arcsec) with zero spin, 
has been measured by the Event 
Horizon Telescope (EHT) (\cite{eht1}), the focus is on the variability of the emission spanning from radio to X-ray wavelengths 
\citep{genzel03,zadeh06a,neilsen13,wielgus22}. The variability of near-IR (NIR) emission is a key tracer of the dynamics of the inner accretion flow and the 
mechanisms at work within it \citep{genzel03,ghez05,eckart08,zadeh06b,michail19,witzel21,michail24}. 
When combined with other wavelengths, NIR probes the ejection, expansion, and orbital 
motion of material (hot spots) occurring near \sgra's event horizon. NIR observations have revealed several bright, hourly-timescale flaring events per day 
that are localized in the accretion flow to within a few gravitational radii (r$_g$)
\citep{Do2009}. Projected orbits of 
hot spots in the accretion flow located within a few $\mu$as of \sgra\ have been determined on hourly time scales \citep{gravity18}. 
GRAVITY measurements have 
also reported hot spot orbital velocity of $\sim0.3$c with an orbital radius  of $\sim$5 r$_g$ corresponding to  2.5 Schwarszchild radii (R$_s$)  
(light travel time of 1R$_s$ is $\sim38$ sec) \citep{gravity18,gravity21}. 

%also reported hot spot orbital velocity of $\sim0.3$c with an orbital radius  of $\sim$5 r$_g$ corresponding to  2.5 Schwarszchild radii (R$_s$)   
%(light travel time of 1R$_s$ is $\sim38$ sec) \citep{gravity18,gravity21}. 
%\bibitem[\protect\citeauthoryear{GRAVITY Collaboration et al.}{2021}]{gravity21} GRAVITY Collaboration, Abuter R., Amorim A., Baub{\"o}ck M., Baganoff F., Berger J.~P., Boyce H., et al., 2021, A\&A, 654, A22. doi:10.1051/0004-6361/202140981

The flux dependence of the spectral index 
$\alpha$, 
where $\alpha \propto \log\, {F_\nu / \log\,{\nu}}$, 
has 
been studied in the past but there is a debate on how  the spectral index behaves at low flux densities. 
Steeper spectral indices at low flux densities have been reported 
\citep{eisen05,gillessen06,krabbe06,fyz06,fyz09,bremer11,ponti17,witzel18,paugnat24}, whereas some studies 
report  that the spectral index is independent of  flux density 
\citep{horn07,trap11,witzel14}. 

Long-term variability  of faint flickering activity has  also been reported, which provides global details of the accretion flow 
\citep{weldon23}.  Thus, time variability analysis of Sgr A* provides details of compact spots with high plasma density and magnetic field strength, 
but also large-scale flow in the  accretion disk. 
Magnetohydrodynamic (MHD) simulations of the accretion flow predict flux variability on a wide range of time scales, from 
minutes to years \citep{dexter20,ressler20a,ressler20b,chatter21,weldon23}.

There is a wealth of observations  using mainly the Keck, VLT, and Spitzer  observatories,  but also often ignored HST measurements,  
studying the physical characteristics of the variable emission at NIR wavelength 
\citep{eisen05,ghez05,gillessen06,krabbe06,horn07,eckart08,dodds-eden09,fyz09,bremer11,trap11,hora14,witzel14,ponti17,witzel18}. 
Because of a large concentration of stellar sources in the Galactic center, there is 
considerable confusion in subtracting the background from the variable emission from Sgr A*. 
The complexity of the region in the immediate vicinity of Sgr A* has created  debates 
 on the flux distribution, power spectrum, spectral index and periodicity  of the emission. 
JWST observations have an advantage over  ground-based adaptive optics instruments, 
in  observing for a long, continuous periods, 
 and monitoring  the spectral evolution of the variable emission through simultaneous observations at multiple wavelengths.  
Here we  focus only on NIRCam 2.1 and 4.8 $\mu$m 
data presenting the results and the analysis 
of the measurements.  We then compare them  with other measurements in the discussion.

In $\S2$  we  describe observational details of seven 
epochs of observations in 2023 and 2024 followed by calibration and data processing. 
In $\S3.1$,   we show  background subtraction, reference stars and extinction corrected  light curves  of  Sgr A* 
in seven epochs  separated by a few days, months and a year. 
Our analysis indicates continuous short-time scale variability on seconds-to-hours 
and day-to-day variable emission implying no steady component of the 
 emission from Sgr A*.  We focus on the intrinsic short and long-time variability
associated with the  hot spot picture  and 
the large-scale flow of the accretion disk, respectively.  
These are followed  by  
modeling of the flux distribution at 2.1 and 4.8 $\mu$m. 
In $\S3.2$,  we fit  the power spectrum on each epoch of observation  and discuss
 two different statistics applied to low and high-frequency variable emission. 
The histogram of intensities of the variable emission suggests  a bimodal distribution 
with  different statistics in short- and long-term variability of Sgr A*. 
In 
$\S3.3$,  
we concentrate on the time evolution of the spectral index of Sgr A*  and show 
that the spectral index evolves differently at fainter and brighter flux densities. 
In $\S3.4$, we present cross-correlation analysis of  2.1  vs 4.8 $\mu$m emission. 
We note time delays of up to 40 sec
 between 2.1  and 4.8 $\mu$m emission. 
We examine the flux density as a function of the spectral index, 
and show loop diagrams indicating the  evidence that 
the  mean  spectral index tends to  steepen with increasing brightness up to 
$\sim 3$ mJy at 2.1 $\mu$m, without any further steepening at higher brightnesses.  
In  $\S4$,  we discuss  models of  the variable emission from Sgr A* and  loop diagrams,  
giving  estimates of the electron density and the magnetic field. 

% 1) Continous variability
% 2) long term varianility on wide range of time scales 
% 3) power spectrum showing 2 different statistiscs as well as submin time scale variabnility

%\vfill\eject 
\clearpage
\section{JWST NIRCam Observations}

Near-IR photometric monitoring data of Sgr A* were obtained with the NIRCam instrument \citep{rieke23}
 of JWST on  7 days
from April 2023 through April 2024. The observations were obtained under JWST Cycle 1 and 2
programs 2235 and 3559, respectively. All observations used the NIRCam instrument to
acquire imaging of the region near Sgr A*, using detectors in module B of the instrument.
Simultaneous imaging was obtained in the NIRCam short- and long-wavelength channels, using
the medium band filters F210M and F480M. These filters have central wavelengths of
2.096 and 4.814 $\mu$m, with bandwidths of 0.205 and 0.299 $\mu$m, respectively.

Two different observing methods were employed. All observations obtained in the 5 days
of 2023 used conventional imaging of the field, employing a 4-point dither pattern in order
to avoid possible effects of detector persistence and to increase the effective sampling
of the instrumental PSF. Based on analysis of the 2023 observations it was determined
that 1) persistence was not an issue, 2) dithering increased the risk of failing to
reacquire the guide star after each move, and 3) that more precise relative photometry could be
obtained by not dithering. Hence all observations obtained in 2024 used the NIRCam
Time Series Imaging method, where the source is left at a fixed location in the field and
many individual images are obtained over the hours-long duration of the observation.

The observations obtained on 2023 April 13, 16, and 18 used the 
$``$SUB640'' subarray detector
readout, along with 10 groups (non-destructive detector read outs) per integration and 18 or 19
integrations per exposure.  
The 2023 September 22 observations
also used the $``$SUB640''  readout, with 5 groups per integration and 33 integrations per
exposure. This results in images with an effective integration time of 41.9 sec (same as in Table 1) and 951 sec
intervals between dither steps (see Table 1). The observations taken in time series mode in March and April
2024 used a somewhat smaller subarray readout ("SUB400P"), with 10 groups per integration,
resulting in images with an effective integration time of 16.6 sec. The total span of
observations on 13, 16 and 18 April  2023 was $\sim$7h, on  22 September 2023 was $\sim$5h, on 
22 March  2024 was $\sim$6 hours, while those on 5 and 7 April 2024 were $\sim$8.5h.

Columns 1 to 3 of Table 1 list the dates and the wavelengths that were used in all seven epochs of observations. 
The photon collection times of individual images used in the analysis are shown in column 5, whereas the cadence between subsequent images (which 
includes overheads and dead time) is shown in column 4.
The last column shows the observing technique that was used in 2023 and 2024. For days 1-3, the total time spent taking multiple integrations within a single exposure 
before moving to a new dither point on the sky is $\sim$795 sec, while the cadence between successive exposures is $\sim$954 sec. For the last 3 epochs, 
where we used the time series mode without any dithering, there is no information for exposure cadence because all the data were taken as part of a 
single exposure.
 
Some problems were encountered with the observatory during the first 3 epochs,  mainly due to the difficulty in 
finding and locking onto guide stars in this severely crowded region of the sky.
In particular, the observations obtained on 2023 April 18 locked onto the wrong guide star,
resulting in Sgr A* being just outside the field-of-view of the NIRCam long-wavelength channel.
We therefore only have 2.1 $\mu$m  data available on that day. In addition, the data obtained
on 2023 April 16 showed evidence of poor guiding during some of the exposures, resulting in
smeared images. Data from those exposures has been excluded from the analysis, resulting in
some temporal gaps on that day. Finally, during the observations on 2023 September 22, the
observatory lost lock on the guide star about half way through the planned observations,
resulting in a truncation of the time coverage for that day.

%Tab1
\begin{deluxetable}{lccccc}   
\tablewidth{0pt}
\tablecaption{NIRCam observation dates}
\tablehead{
\colhead{Epoch} &  
\colhead{Date}  &
\colhead{Wavelength ($\mu$m)} &
\colhead{Sampling time (sec)} &
\colhead{Exposure time (sec)} &
\colhead{Observing technique} 
}
\startdata
 Day 1  &  2023/4/13  &   2.1 \& 4.8 & 46.066  & 41.89 & Dither\\
 Day 2  &  2023/4/16  &   2.1 \& 4.8 & 46.066 & 41.89   & Dither \\
 Day 3  &  2023/4/18  &   2.1        & 46.066 & 41.89   & Dither \\
 Day 4  &  2023/9/22  &   2.1 \& 4.8 & 25.135 &  20.93 & Dither   \\
 Day 5  &  2024/3/21  &   2.1 \& 4.8 & 18.241 &  16.56 & Time series \\
 Day 6  &  2024/4/5   &   2.1 \& 4.8 & 18.241 & 16.56 & Time series \\
 Day 7  &  2024/4/7   &   2.1 \& 4.8 & 18.241  & 16.56 & Time series \\
\enddata
\end{deluxetable}

%Tab2
\begin{deluxetable}{llccc}   
\tablewidth{0pt}
\tablecaption{Coordinates of Sgr A* and reference stars}
\tablehead{
\colhead{Name} &
\colhead{Epoch}  &
\colhead{Right Ascension (J2000)} &
\colhead{Declination (J2000)} 
}
\startdata
  Sgr A*  & August to September 2023                            & $17^h\, 45^m\, 40^s.0323$  & $-29^\circ\, 00'\, 28''.258$ \\
  S0-17  &  April to September 2023      & $17^h\, 45^m\, 40^s.042$  & $-29^\circ\, 00'\, 27''.95$ \\
  S0-17  &  March to April  2024         & $17^h\, 45^m\, 40^s.042$  & $-29^\circ\, 00'\, 27''.92$ \\
  Reference star  & April to September  2023         & $17^h\, 45^m\, 40^s.153$  & $-29^\circ\, 00'\, 25''.02$ \\
  Reference star  & March to April  2024         & $17^h\, 45^m\, 40^s.154$  & $-29^\circ\, 00'\, 25''.02$ \\
\enddata
\end{deluxetable}

\subsection{JWST Data Processing and Analysis}

The standard JWST calibration pipeline software and calibration reference files were still
changing rapidly throughout 2023, so the data from all NIRCam observations of Sgr A* were
uniformly reprocessed using the STScI pipeline from ``jwst" package version 1.14, using the
latest calibration reference files. All datasets were reprocessed from their raw form
through Stages 1 and 2 of the pipeline, yielding fully recalibrated images from each
individual integration in all exposures.

The absolute pointing accuracy of JWST is $\sim$0.1 arcsec rms. In addition to this random uncertainty, some of the NIRCam observations
mis-identified the planned guide star, due to the extremely crowded nature of the Galactic center. This led to an even larger offset in
pointing of several arcsec for some exposures. Because Sgr A* is partially hidden by NIR diffuse background emission, as well as the
overlapping PSF's of neighboring stars, accurate astrometric registration is needed in order to positively identify the exact position of Sgr
A* in the images. We therefore used the highly accurate coordinates of several radio sources visible in an 
a 230 GHz image of the Galactic center using the Atacama Large Millimeter/sub-millimeter Array (ALMA). 
ALMA image is  an average of the observations taken on 2023 August 25, and 2023 September 22, the latter of which  
was taken simultaneously with the Day 4 (September 22, 2023) NIRCam observations to astrometrically register the NIR images. There are a
number of ALMA sources that have IR counterparts in the NIRCam images. The sky coordinates (RA/Dec) of the radio sources were used as an
absolute reference catalog to feed to the ``tweakreg" task in the jwst pipeline package, which matches the locations of those sources in the
NIRCam images and updates the World Coordinate System (WCS) information in each image with the new absolute pointing information. This process
could only be directly applied to the 2.1 $\mu$m NIRCam images, because the counterparts of the radio sources were all saturated in the 4.8 $\mu$m
images. So the registration of the 4.8 $\mu$m images was accomplished by boot-strapping source positions from the already registered 2.1 $\mu$m images,
using the same ``tweakwcs" task. This process was used to astrometrically register all NIRCam data from the 2023-2024 campaigns. After
alignment, the rms residuals in the positions of the IR counterparts to the radio sources are $\sim$0.2 pixel, which corresponds to $\sim$0.006 and
$\sim$0.01 arcsec in the 2.1 and 4.8 $\mu$m images, respectively.

Once the recalibrated images were updated with accurate WCS information, photometry was
performed on Sgr A* and other stellar sources in the field using routines in the Astropy
$``$photutils'' software package. Circular apertures with a radius of 1.64 and 1.61 pixels
with a pixel size of 0.03$''$ were   
used for the 2.1 and 4.8 $\mu$m images, respectively, which is equivalent to aperture diameters
of 0.10 and 0.20 arcsec, respectively. Appropriate aperture corrections were applied to the
measured fluxes, based on the latest aperture correction data from STScI, and converted to
units of mJy.

\subsection{Adjusting timing tags between  2.1 and 4.8 $\mu$m data}

An important part of our analysis of the Sgr A* light curves is the comparison in timing between flares in the 2.1 and
4.8 $\mu$m bands, hence it is necessary to have accurate time stamps for each data point in the light curves. For all data we
use the midpoint of the integration from which the photometry is obtained. There is a discrepancy, however, in effective
midpoints of the 2.1 and 4.8 $\mu$m  integrations. The 2.1 $\mu$m signal level in all pixels near Sgr A* is always low enough that
detector saturation is never reached and hence all groups from the "up the ramp" sampling for each integration are used to
compute the flux. In the 4.8 $\mu$m data, however, signal levels are significantly higher and reach saturation sometime during
each integration. The data reduction software automatically rejects the saturated groups when computing the signal for each
integration, but this means that the effective time at which the data were obtained is shifted relative to the overall
midpoint of the entire integration. We have determined at what point during the 4.8 $\mu$m integrations the data went into
saturation and computed corresponding timestamp corrections.

The data from Days 1-3 are affected most severely by this issue, because longer integrations were used and hence a larger
fraction of each integration was rejected due to saturation. For Days 1-3 each integration had a length of 41.8 secs, with
the midpoint at 20.9 secs. The 4.8 $\mu$m data saturated 12.6 secs into each integration, so the effective midpoint is at 6.3
secs, 14.6 secs earlier than the corresponding 2.1 $\mu$m data that used the full integration. Similarly, the shorter
integrations used in Day 4 resulted in the effective midpoint of the 4.8 $\mu$m data being 4.2 secs earlier, and for Days 5-7
the effective midpoint is 2.5 secs earlier.

All subsequent analysis presented below has taken these timing offsets into account. In particular, for cases where the
fluxes in the two bands are directly compared, the 4.8 $\mu$m  data have been interpolated onto the same time grid as the 2.1 $\mu$m
data.

%f1
\begin{figure}[htbp]
\centering
   \includegraphics[width=5in]{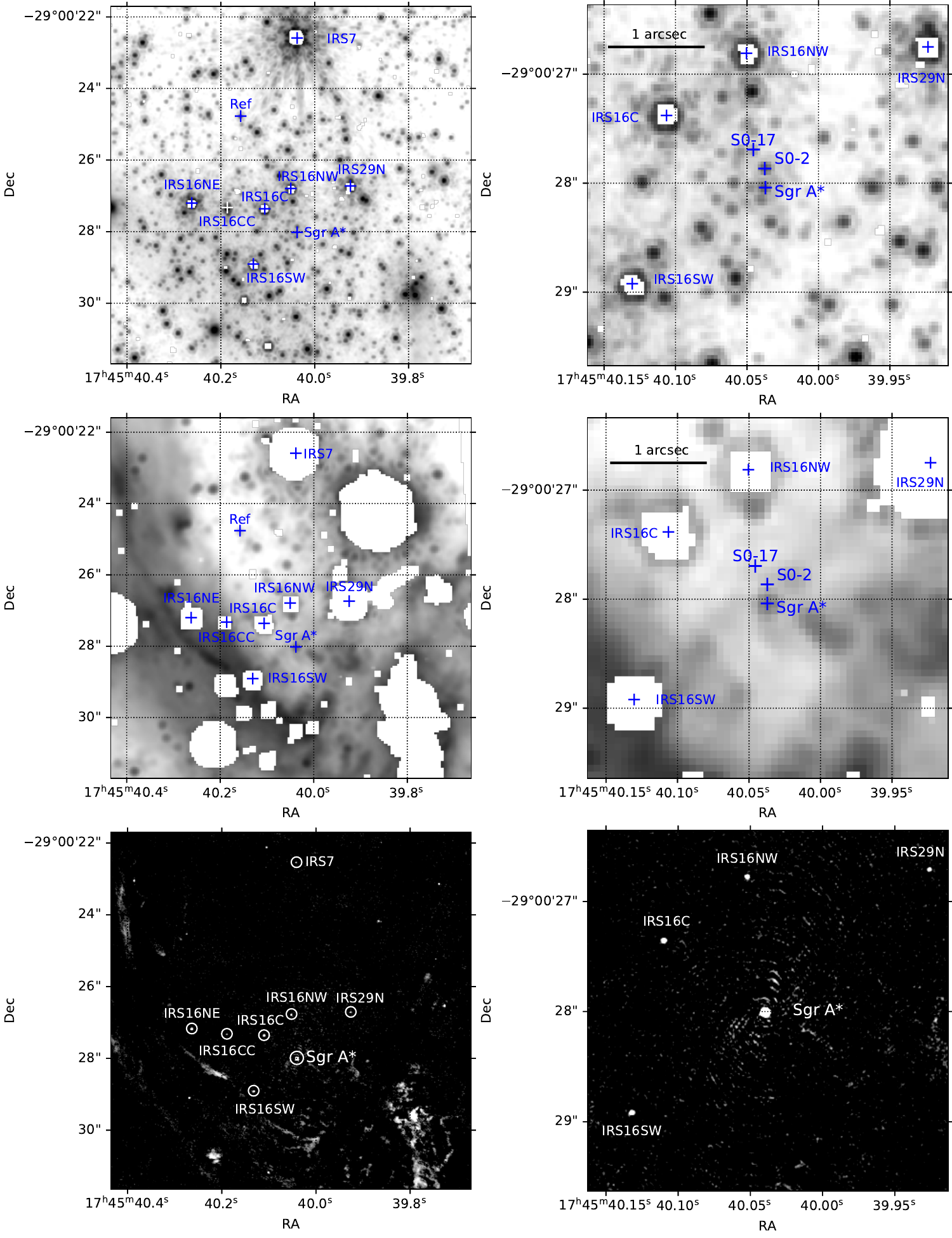}
   \caption{
Astrometrically corrected NIRCam 2.1 $\mu$m (top) and 4.8 $\mu$m (middle), and ALMA 230 GHz (bottom) images of the Galactic center region. 
These images are simultaneously taken together on 
September 22, 2023 and  are astrometrically aligned  with each other. 
Images on the left are zoomed out to show 
the locations of the well-known members of the stellar cluster, such as S0-2,  orbiting  Sgr A*, as well as the unidentified reference star
 (``Ref'') used to compare to Sgr A*. 
Images on the right are zoomed in to 
show the immediate vicinity of Sgr A* along with the comparison star S0-17. White pixels in the NIRCam images are completely 
saturated and have no usable signal values.
} 
\end{figure}

%These  images are constructed by subtracting two  images  when Sgr A* was and was  not flaring. 

\subsection{Reference stars and the position of Sgr A* aided by 230 GHz observations}

The NIR images shown in Figure 1 have used the astrometric registration technique
described above (\S2.1) to align them to the ALMA 230 GHz images. 
The actual coordinates for the aperture locations that we used for S0-17 and the reference ``Ref" (Fig.\ 1) in 2023 and 2024 are given in Table 2. 
The coordinates of these sources are  derived relative to the ALMA image. 
The J2000 position of Sgr  A* at epoch 2023.7 is derived from observations using ALMA obtained on 2023 August 25 and 2023 September 22 
under project code 2022.A.00029.S using baselines up to 15 km. 
This position is within 10 mas (approximately one third of the synthesized CLEAN beam) of the position predicted by the VLBI proper motion study of Sgr A* performed using 
the Very Long Baseline Array (VLBA) \citep{gordon23}.

\subsection{Background subtraction}

The aperture-based flux measurements of Sgr A* contain not only emission from Sgr A* itself,
but also diffuse background emission that is  present throughout the images of the Galactic center,
as well as contaminating flux from at least one nearby stellar source. 
The use of traditional
background measurement techniques, such as the measurement of signal in an annulus around
the Sgr A* aperture, do not result in an accurate estimate of the contaminating signal, due
to the highly non-uniform nature of the background components. We employed a combination of
methods to estimate the total amount of extra signal in the Sgr A* aperture. This included
an estimate of the diffuse background in the pixels immediately adjacent to the Sgr A*
aperture in images taken when Sgr A* was at its lowest signal levels, as well as an estimate
of the amount of contaminating flux from the immediately adjacent stars based on a model of
its PSF and the amount of flux contained in the wings of that PSF that overlapped the
Sgr A* aperture. This stellar contamination had to be estimated individually for each epoch
of observations, due to the fact that the nearby stars are moving relative to Sgr A*. We
used a constant background signal level for each day of data and subtracted it directly from
the flux measured within the Sgr A* aperture.
The final lists of corrected Sgr A* fluxes as a function of time were used to construct
light curves for each day.

The actual measurements of $``$background'' level arrived at doing photometry in background regions around Sgr A* were typically 0.16-0.19 mJy at 2.1 $\mu$m  
(across all 7 days) and 4.0-4.6 mJy at 4.8 $\mu$m. Note that at both wavelengths the actual measurements increased by $\sim$15\% between 
the April 2023 and March-April 
2024 observations, due to movement of the nearby stars, 
thus we picked up more flux from stars in the 2024 
photometry using the same background region locations.

We attempted estimating the amount of additional contamination contained in the Sgr A* aperture due to some members of the S cluster being adjacent to
Sgr A* and spilling some of their fluxes into the Sgr A* aperture. We identified stellar sources in the S-cluster by evolving their orbits to the epoch
of our observations.  The orbital parameters are taken from \citep{gillessen17}. S0-2 is identified to be offset from Sgr A* by $0.1745''\pm0.0013$ and
$0.0006''\pm0.0013$ to the North and East of Sgr A*, respectively, on August 22, 2023.  Figure 1 (top right panel) shows the position of S0-2 with
respect to Sgr A*. S0-2 was too far from Sgr A*, 
thus contributed little to no spilled flux in any epoch of these observations at 2.1 $\mu$m.
However, two other members of the cluster,
S24 and S29, which are relatively faint compared to S0-2 have  approached closer (in projection) and hence spilled more flux in 2004 compared to 2003.

Similarly, at 4.8 $\mu$m, we estimated total extra signal in the Sgr A* aperture of $\sim$6.0 mJy in the 2023 data and $\sim$6.3 mJy in the 2024 data. 
This is an  increase over the actual measurements from the nearby background regions of $\sim$2 mJy, with the extra 2 mJy 
being  a ball park estimate for the extra contamination  due to S0-2.
So given the huge uncertainties in this process of estimating the extra contamination due to stellar contamination, possibly by S24 and S29, 
we have increased   the 4.8 $\mu$m background flux by 1  mJy, thus accounting for possible over-subtraction at 4.8 $\mu$m.

\subsection{Noise and limits on photometry}
 
 It is essentially impossible to empirically measure the random noise in the signal from Sgr A*, because of its nearly constant variability in intrinsic flux level. Thus,
the best proxies for estimating the noise limits of our photometry come from the data for S0-17, the additional reference star (“Ref”), and several randomly selected
regions that have only faint, diffuse background emission. S0-17 and the background regions all have relatively low flux levels, comparable to or less than that of Sgr
A*, and hence represent our ability to detect faint variations in Sgr A* emission. The measured noise levels at 2.1 $\mu$m are consistent across all 7 epochs, while the data
at 4.8 $\mu$m seem to be affected by our use of dithered (Days 1-4) versus non-dithered (Days 5-7) observations. S0-17 and the background regions have rms scatter of
$\sim$0.0015 mJy at 2.1 $\mu$m, suggesting a 3-$\sigma$ detection limit of $\sim$0.005 mJy. At 4.8 $\mu$m the rms scatter is $\sim$0.03-0.04 for both S0-17 and the
background regions, but drops to $\sim$0.01 mJy for the last 3 sets of non-dithered observations, resulting in 3-$\sigma$ detection limits of $\sim$0.1 and $\sim$0.03
mJy for the dithered and non-dithered data, respectively.  When Sgr A* is in a flare state, we expect the total noise to be larger, due to increased Poisson noise. The
reference star (“Ref”) has a 2.1 $\mu$m flux level comparable to the bright flare states of Sgr A* and has 1-$\sigma$ noise of $\sim$0.0035 mJy, which is a little more
than twice the 1-$\sim$ noise level of the fainter S0-17. Thus we expect that the noise in the 2.1 $\mu$m fluxes from Sgr A* when it’s in a flare state to also be higher by
a similar amount, with a 3-$\sigma$ limit of $\sim$0.01 mJy. We have not measured the 4.8 $\mu$m noise level in a reference star that is comparable in flux level to the
flare states of Sgr A*, but we expect a similar increase relative to the noise in faint 4.8 $\mu$m sources, and hence predict 4.8$\mu$m 3-$\sigma$ noise levels of
$\sim$0.2 and $\sim$0.07 mJy for the dithered and non-dithered data, respectively.

%The best proxies for estimating the noise limits of our photometry come from the data for S0-17 and several randomly selected regions that have only faint, diffuse 
%background emission. These all have relatively low flux levels, comparable to or less than that of Sgr A* and hence represent our ability to detect faint variations in 
%Sgr A* emission. The results at 2.1 $\mu$m are consistent across all 7 epochs, while the results at 4.8 $\mu$m seem to be affected by dithered (Days 1-4) and non-dithered 
%(Days 5-7) data. S0-17 and the background regions have rms scatter of 0.001-0.002 mJy at 2.1 $\mu$m. At 4.8 $\mu$m, the rms scatter is $\sim$0.03-0.04 mJy for both S0-17 
%and the background regions, but that drops to $\sim$0.01 mJy for the last 3 sets of non-dithered observations. If we use 0.0015 mJy as 1$\sigma$ at 2.1 $\mu$m, then the 
%3$\sigma$ detection limit is $\sim$0.005 mJy at 2.1 $\mu$m.  Similarly, the 3-$\sigma$ detection limits at 4.8 $\mu$m are $\sim$0.1 mJy and $\sim$0.03 mJy for dithered 
%and non-dithered data, respectively.

\subsection{Extinction}

%\bibitem[\protect\citeauthoryear{Sch{\"o}del et al.}{2011}]{schodel11} Sch{\"o}del R., Morris M.~R., Muzic K., Alberdi A., Meyer L., Eckart A., Gezari D.~Y., 2011, A\&A, 532, A83. doi:10.1051/0004-6361/201116994
%$A_M = 1.0\pm0.3$ (for F480M)  \citep{schodel11}. 

Flux densities have been adjusted for extinction assuming
$A_{K_s} = 2.46\pm0.03$ (for F210M) and
$A_M = 1.0\pm0.3$ (for F480M)  \citep{schodel11}. 
We do not  account for the differences between the $K_s$ and F210M
or  between the M and F480M bands. The mean wavelengths only differ by $\lesssim$ 3\%, which implies a $\lesssim$ 
5\% change in the extinction,
and different choices of  extinction law  can entail larger differences than
 this. Application of the adopted extinction 
correction increases  the reported 2.1 and 4.8 $\mu$m flux densities by factors of 11.16 and 3.94, respectively. 
For the values reported  here,   the extinction correction is applied.

\section{Results}
We first concentrate on the  analysis of Sgr A* variability, followed by the  power spectrum analysis 
and  the evolution of the spectral index variation  as a function of time and flux density.  

\subsection{Light curves}

Short and long-term variable emission from Sgr A* are presented followed by flux distribution of flaring activity. 
We considered a subjective way to classify the morphology of the light curves by:   
flare,  subflare,  weak fluctuations  and  pedestal emission, with the latter referring to 
slow variation of the intrinsic emission 
from Sgr A*. 
The pedestal level is distinct 
from  constant background emission  due to stellar light contamination that is subtracted from the flux of Sgr A*. 
Variations in the pedestal  are fitted by a slope  that varies from 
epoch to epoch.  

We identify flares and subflares having extinction-corrected peak 2.1 $\mu$m flux  densities
$\ge$3  or $\lesssim$3 mJy (corresponding to an observed flux density threshold of 0.27 mJy), respectively.
Subflares generally have  shorter durations $\le$20 minutes, either isolated or  superimposed on rising and falling flares. 
The rationale for choice of $\sim$3  mJy becomes clearer 
when we discuss 
the flux and  spectral index distributions indicating 
a transition from steep to shallow spectral index at a flux density of $\sim$3 mJy.
Finally, weak fluctuations in  emission are identified at a level of $\lesssim$1 mJy  with short durations 
 detected all the time.  

\subsubsection{Reference stars light curves}
   
Figure 2 shows a comparison of linear- and logarithm-scale light curves of the reference star 
S0-17 and reference stars  at 2.1 and 4.8 $\mu$m and those of Sgr A* on all seven days. 
The gaps between different epochs are not displayed in order to show the nature of long-term variability of Sgr A* on daily, 
monthly and yearly time scales. S0-17 acts as a common reference star in 
all epochs of data. 
The light curves of S0-17 indicate that the 
fluxes are very 
stable, and flat with a flux density (uncorrected for  extinction) 
of $0.409\pm0.001$ mJy and $4.44\pm0.02$ mJy at 2.1 and 4.8 $\mu$m, respectively.

The flatness and stability of the reference 
star light curves provide strong support on the continuous low-level weak and strong variable  activity of Sgr A* at 
both wavelengths on all seven days. 
Due to the proximity of S0-17 and Sgr A*, the measured light curve of S0-17 is slightly contaminated by Sgr A* flare emission, especially at 4.8 $\mu$m where 
the PSF of Sgr A* is more extended.

\begin{figure}[p]
\centering \includegraphics[width=4.25in,angle=90]{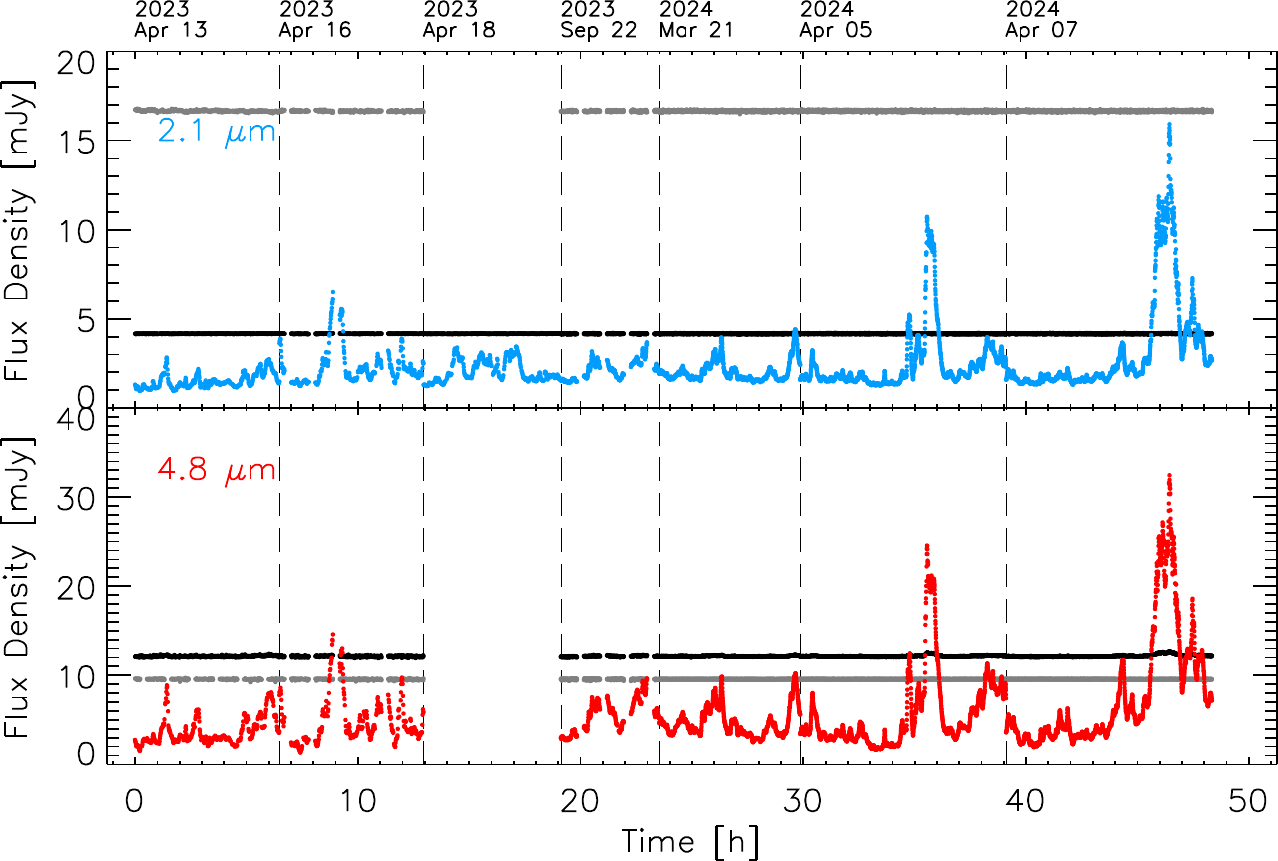}
\centering \includegraphics[width=4.25in,angle=90]{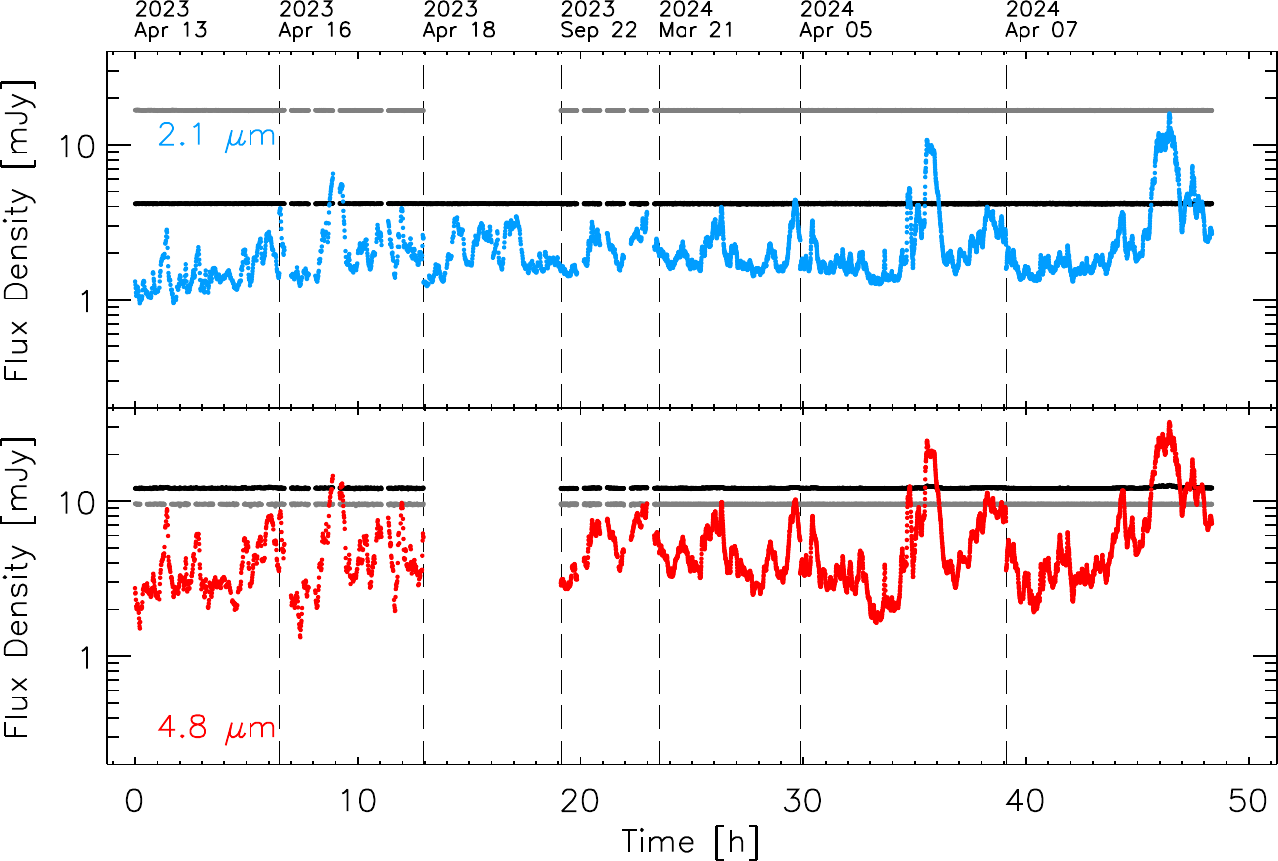}
    \caption{
Extinction corrected light curves of Sgr A* and reference stars for 
all 7 days concatenated together at 2.1 and 4.8 $\mu$m, are presented in linear and logarithmic  scales in the top and 
bottom panels, respectively. 
The light curves of the reference stars S0-17 and ``Ref'',  a random  nearby star in the field  with similar signal level as Sgr A*,  
are displayed on the top panel 
as black and gray points, respectively.  The  reference stars are stable whereas Sgr A* is always fluctuating.
Vertical dashed lines separate seven different non-contiguous days. 
These plots  indicate  that the observed variations in Sgr A* at both wavelengths are real, and not an artifact or noise.
In addition, the amplitude of the quiescent flux of  Sgr A*, and  pedestal emission 
vary on daily, monthly and yearly time scales. 
}
\end{figure}

\subsubsection{Short-term flux variability}

Figure 3   shows extinction-corrected light curves of the seven epochs of observations in 2023 and 2024, respectively. 
There is clearly 
correlated variability at 2.1 and 4.8 $\mu$m in all epochs. 
Day 2 showed the strongest 
flare emission in 2023 epochs at $\ge 6$ mJy at 2.1 $\mu$m. However, the peak of this flare was missed due to 
poor guiding of the telescope.
Only 2.1 $\mu$m data are available on Day 3, due to Sgr A* being outside the 4.8 $\mu$m field-of-view, 
which was caused by an error in guide star acquisition.
 {\it Three} flares each lasting  for a total duration of about one  hour are 
detected.   A number of subflares and weak fluctuations are superimposed on the three flares. 
Similarly, Day 4 observations suffered from 
intermittent poor guiding and was terminated early due to complete loss of lock on the guide star.
 We note {\it two} flares,  each lasting for more than an hour, with a number of superimposed subflares. 

%Day 1 showed continuous variability displaying at least {\it five} flares as well as a number of subflares that are either isolated or superimposed on 
%strong flares. Low-level fluctuating events are noted throughout the observation.

%\setcounter{subfig}{1}
%\begin{figure}[htbp]

%f2b  
%\addtocounter{figure}{-1}
%\stepcounter{subfig}
\begin{figure}[htbp] 
   \centering
   \includegraphics[width=3.4in]{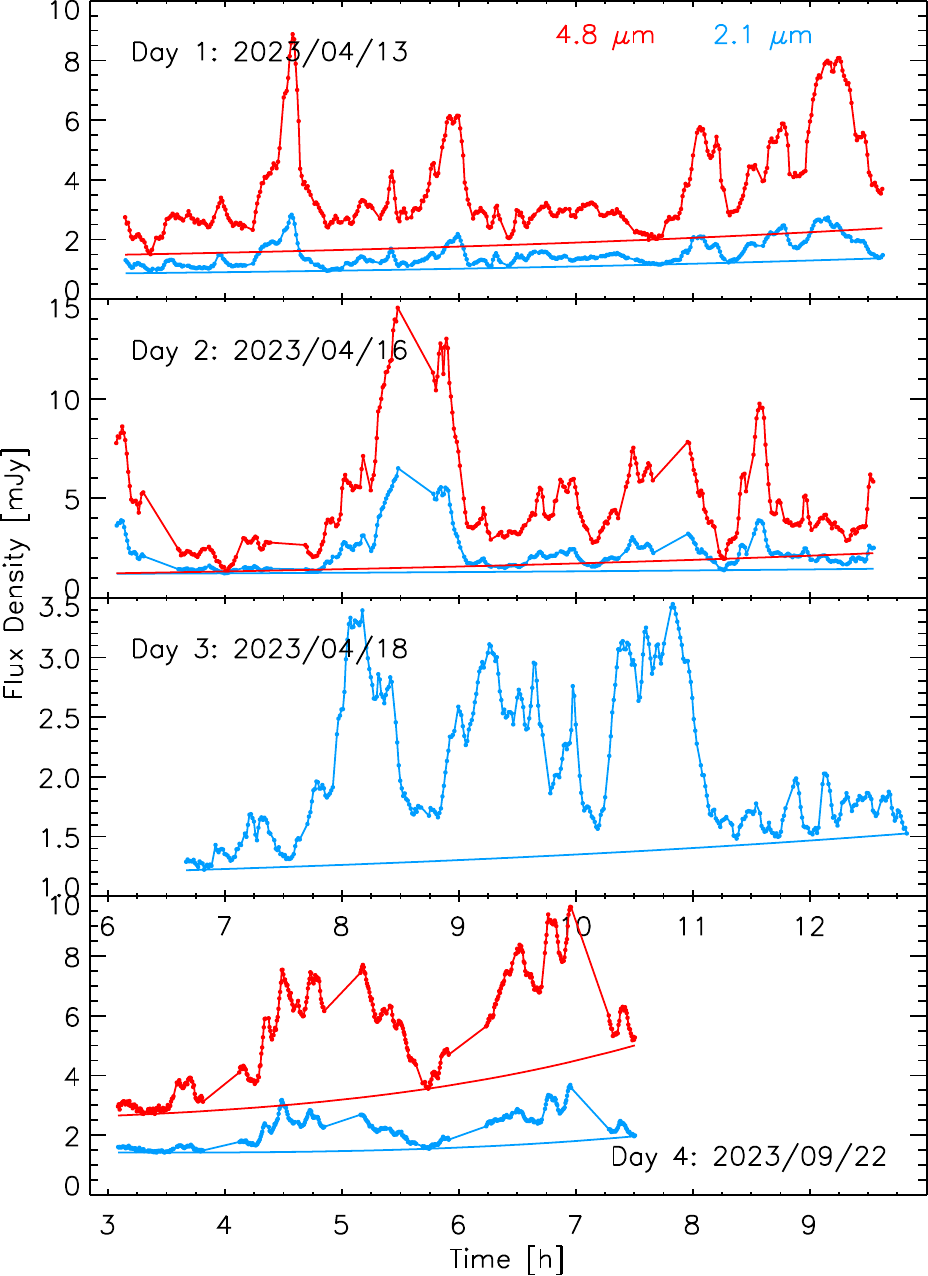}
  \includegraphics[width=3.4in]{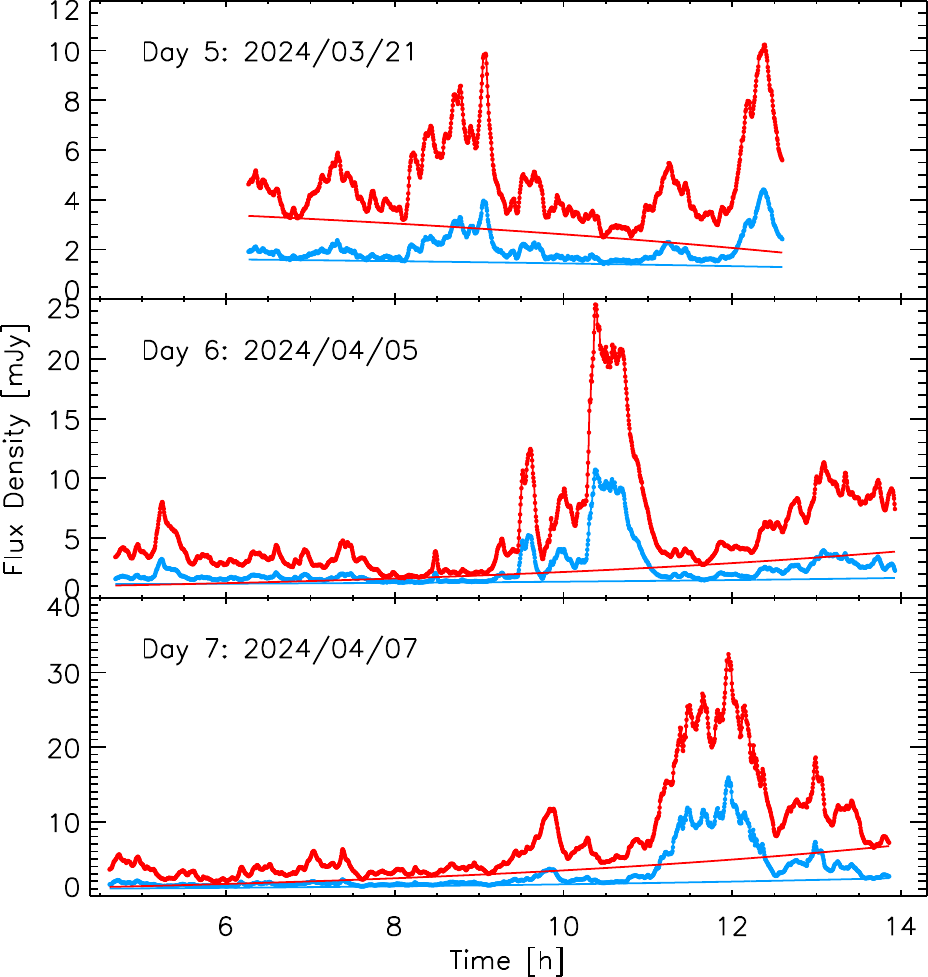}
   \caption{
{\it  Left} 
Four panels show light curves of Sgr A* by connecting the data points and  are  constructed for  Days  1-4, of 
observations in 2023.  
Blue and red color dots correspond to 2.1 and 4.8 $\mu$m, respectively. The sampling time is 46 and 25 sec in  Days 1-3 and Day 4, 
respectively. 
Correlated variability  is  detected. Extinction corrections have been applied.  
The extinction correction increases  the measured 2.1 and 4.8 $\mu$m flux densities by factors of 11.16 and 3.94, respectively. 
Linear fits to changing pedestal   have been made in all epochs  at 
2.1 (blue) and 4.8 (red) $\mu$m.  
To estimate the pedestal, the solid lines  are 4th degree polynomial fits to the data. 
The differing curves  indicate that the pedestal  varies in brightness and spectral 
index. 
There is a tendency to be relatively redder when flaring is brighter, but this could be an
artifact of the estimated background level that has been subtracted from all Sgr A* photometry.
 Note that the F480M filter was not available on Day 3. 
These figures show the evidence for long-term variability of the quiescent emission (pedestal) from Sgr~A*. 
{\it  Right} 
Similar to {\it Left}  except that  the light curves of Days 5-7 in 2024 are displayed.
The sampling time is 18.2 sec in Days 5 to 7  in 2024. 
Correlated variability in both  F210M and F480M bands are detected.  $1\sigma$ Error bars at 2.1 and 4.8 $\mu$m are 
0.0015 and 0.03 mJy, respectively,  which are too small to display. 
} 
\end{figure}

In  the 2024 observations (Days 5-7), no  dithering was done, allowing us to obtain continuous 
light curves with short sampling time  and no gaps. 
Days 5 and 6 each showed {\it three} flares,  
and Day 7 showed  {\it two}  flares. We also note long  flare  duration ($\ge$30min), multiple subflares,  and continuous weak 
fluctuations with short durations superimposed on flares in 2024.  
In particular, {\it six} subflares  noted the strongest intensity  $\sim$30 mJy at 4.8 $\mu$m 
has  the longest duration between $\sim$11h and $\sim$12.50h  UT on Day 7.

One striking result is  non-stop intensity fluctuations of the variable emission  from minute to hourly time scales,
probing   event-horizon-size scales, which will be discussed in 
$\S3.2$. 
Correlated variability between 2.1 and 4.8 $\mu$m emission 
is detected throughout all 7 epochs of observations. 
The  light curves suggest that the  subflares and low-level  fluctuations  
are superimposed on  strong flaring component  with longer duration. 
A trend that we note in multiple flares is that  subflares  become 
increasingly stronger on the rising and or the falling sides of flaring events \citep{fellenberg24}. 

%\bibitem[\protect\citeauthoryear{von Fellenberg et al.}{2024}]{fellenberg24} von Fellenberg S.~D., Witzel G., Bauboeck M., Chung H.-H., Marchili N., Martinez G., Sadun-Bordoni M., et al., 2024, A\&A, 688, L12. doi:10.1051/0004-6361/202451146

Some examples that are found are subflares  near  8.75 UT on Day 1,   10.50 UT on Day 3,   
 6.75 UT on Day 4,   8.50  on Day 5, 10.50 UT on Day 6  and 11.75 UT on Day 7.
The frequency of an  asymmetry in the   rise and fall of flares is most noticeable when subflares are detected on the rising side. 
The best example is on Day 5  where the flare starts rising near 8.0 UT with three  subflares followed by a peak near 9.00UT before the flare  finally 
falls steeply. 
If the subflares and flares are physically associated with each other, 
then one could consider that subflares are precursors of bright flares.
There are also events in which  subflares are noted on the falling side of bright flares. 
The best example is on Day 6 near 10.25h UT 
where  the  flare initially rises rapidly  to a peak of $\sim$24 mJy at 4.8 $\mu$m before  falling with multiple subflares near 10.75h UT. 
We can not quantify this morphological aspect of flaring events.

\subsubsection{Long-term flux variability}

The seven panels of Figure 3  shows that the pedestal level at which short-terms variability is detected is different on all days of observations. 
In addition, there is evidence that 
the pedestal flux level evolves as a function of time and shows a slope. 
The solid lines  are 4th degree polynomial fits to the data with the 
added constraint that the polynomials never exceed the data. Appendix A gives details of the chosen fits. 
 These polynomials are thus one way to estimate the slow variation of the pedestal  
emission of Sgr A* that underlies the flares. 
The differing curves at the two wavelengths on each day indicate that the pedestal  varies in color or spectral 
index as well as brightness.
The background can be blue or red in terms of spectral index. There is a
tendency to be relatively redder when flaring is brighter, but this could be an
artifact of chosen photometry backgrounds.
 The light curves 
also show pedestals  with amplitudes that vary between each epoch suggesting that there is long term variability of Sgr A* 
on daily, monthly and yearly time scales. The 
pedestal  is varying roughly by a factor of 2 over seven epochs of observations. These trends are best  illustrated in Figure 2
where all data are plotted on a single constant scale. 

Examples that show that Sgr A* pedestal emission 
is not constant can clearly  be viewed in the 
fits of all Days. 
A representation of this trend can be found in the top  and bottom panels  of Figure 3
where logarithm- and linear-scale light curves of  all seven epochs 
show non-stop variability of Sgr A*.  
The concatenated light curves in the top panel  where it seems more clearly evident that the 2.1 $\mu$m 
pedestal  emission on Day 1 is lower and on Day 7 is higher than on Days 2-6.
These suggest day-to-day, month-to-month  variations in the background intrinsic to Sgr A*.

Another trend that we note is the apparent correlation of strong flares with the pedestal emission. 
We note this trend is detected on multiple days where there are strong 
flaring events.  For example a strong flaring activity between 9 and 14h UT on Day 6   shows a stronger  pedestal 
at the end of the activity (right panel of Figure  3). 
In fact, in almost all days, we detect  an increase in the pedestal flux when there are groups of strong flares. 
It is difficult to quantify this correlation, given the frequency  of flaring events.  Nevertheless, this trend implies 
that strong flare emission arising from localized hot spots is correlated with global structure of the accretion flow, represented by varying 
pedestal flux. 
 
\begin{figure}[htbp]
   \centering
  \includegraphics[width=4in]{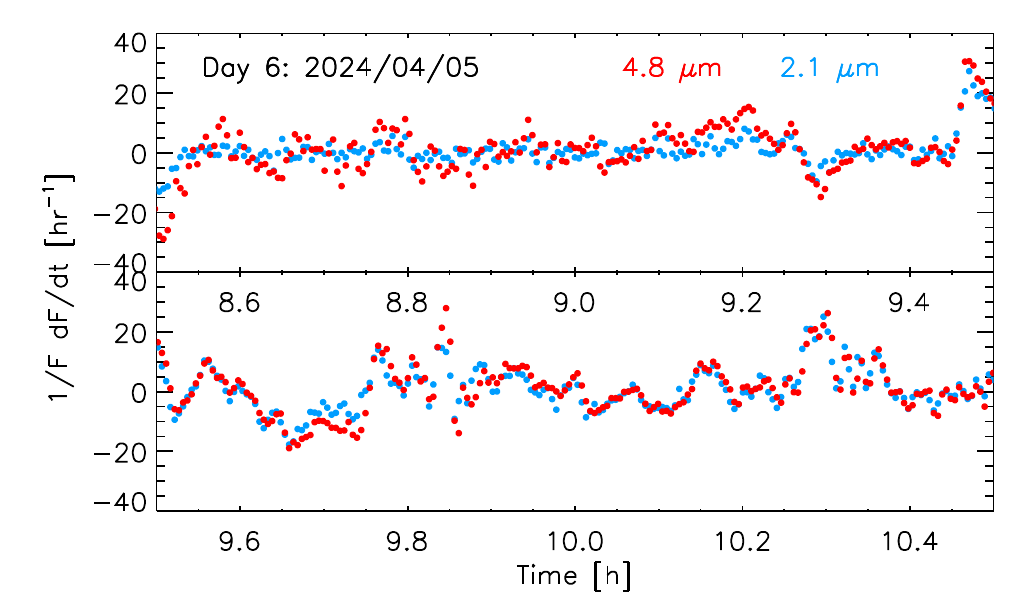}
   \caption{Values of the time derivative  of Day 6 data  normalized by the flux density as a function of time are presented in the top  and and bottom panels. 
Three  examples in which  values are highest   near 9.5, 9.85 and 10.3 UT,  showing  a change by a factor $\sim2$ in the flux density in $\sim2-3$ min. 
} 
\end{figure}

%Tab3
\begin{deluxetable}{lllccccc}   
\tablewidth{0pt}
\tablecaption{Flux density (mJy) log-normal (LN) and power law (PL)  fit parameters}
\tablehead{
\colhead{Wavelength [$\mu$m]} &  
\colhead{Model Fits} &  
\colhead{$\chi^2$} &  
\colhead{Peak} &  
\colhead{$\mu_{ln}$} &
\colhead{$\sigma_{ln}$} &
\colhead{$\beta$} &
\colhead{$x_{min}$}
}
\startdata
 2.1 & LN    &  163.16 &  460.26 & 0.64  &    0.21 &    &   \\
 2.1 & LN+PL &  61.48 &      504.76 & 0.64      &    0.19 &   -3.99    &   1.847 \\
 2.1 & 2LN  & 133.100 &      460.26 & 0.64       &    0.19 &                 &             \\
 2.1 &     &        &       11.37  & 2.30       &     0.065     &                 &              \\
\\
% 4.8 & LN    &  41.09 &  1178.96 & 1.18  & 0.48   &  &  \\
  4.8 & LN   &  113.35 & 1101.62 & 1.49  &    0.35 & &  \\
%  4.8 & LN+PL &  31.27  & 1163.10 & 1.15      & 0.46 &   -2.95    & 8.76 \\
   4.8 & LN+PL &  85.77  & 998.45  & 1.33      & 0.244 &   -2.96   & 4.02\\
% 4.8 & 2LN    &  19.18  &  1173.63 & 1.17  & 0.47   &  &  \\
 4.8 & 2LN     &  82.11  &  1097.98 & 1.48  &  0.35  &  &  \\
% 4.8 &      &        &   28.62   & 3.06       &   0.09    &                 &              \\
  4.8 &      &        &    33.44  &  3.11      &   0.08    &                 &              \\
\enddata
\end{deluxetable}

%distributions give failed solution to fit strong fluxes (see Table 3 and text). The histogram indicate two populations of the variable emission.  $\sigma \sqrt N$ 

%\addtocounter{figure}{-1}
%\stepcounter{subfig}
\begin{figure}[htbp]
   \centering
  \includegraphics[width=3in]{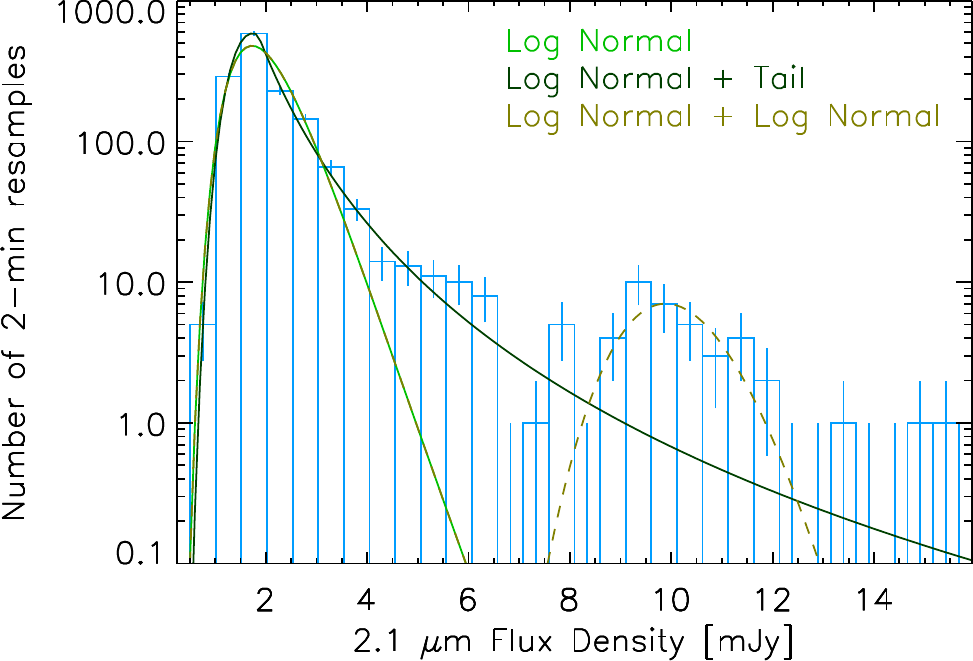}
  \includegraphics[width=3in]{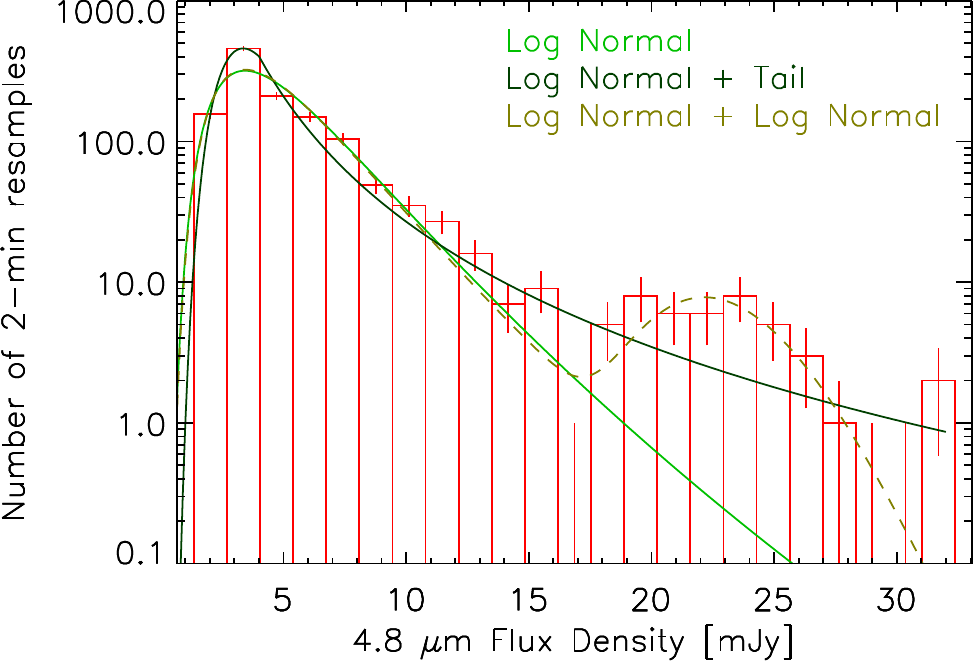}\\
  \includegraphics[width=3in]{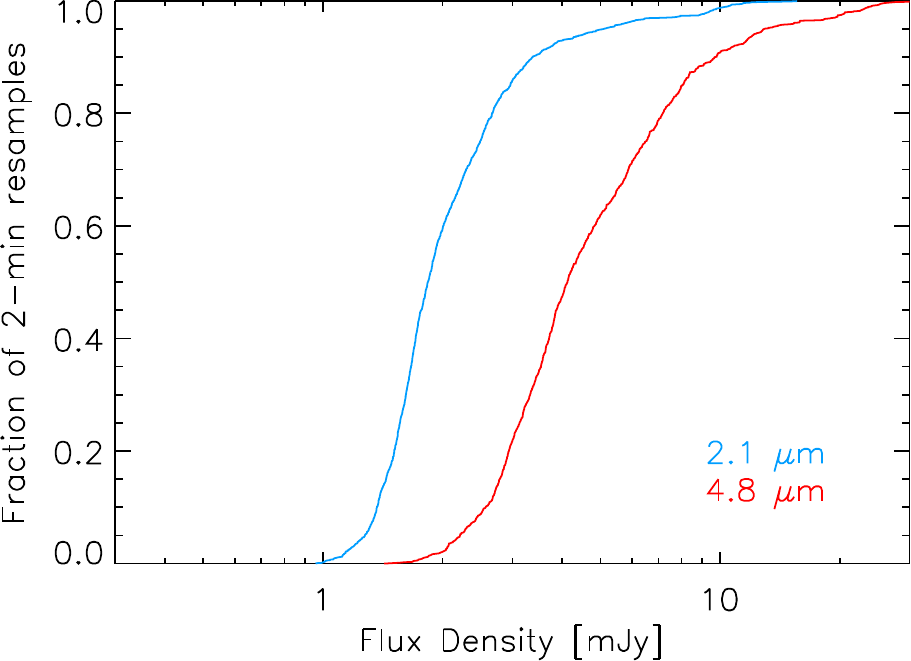}
   \caption{
Using all seven epochs of observations, 
histograms of the flux distribution of Sgr A* at 2.1 and 4.8 $\mu$m 
are displayed in top left and right panels, respectively.
Log-normal (light green line) and log-normal plus power-law (dark green line) and 2 log-normal (dashed green line) distributions fits to the histograms are shown.
The two log-normal, and   log-normal plus power-law fits are reasonable representations of weak 
fluctuations at 4.8 and 2.1 $\mu$m, respectively. Other
distributions give failed solutions to fit strong fluxes (see Table 3 and text). The histograms indicate two populations of the variable emission.  $\sigma = \sqrt N$ 
error bars are shown on the histogram bins.
The  bottom panel shows the fraction of time at which Sgr A* is observed below a given flux density. 
}
\end{figure}

\clearpage
\subsubsection{Flux density gradient of the variable emission}

Another feature that we notice in the light curves of Sgr A* is the sudden rise or drop in the flux density over a short time interval.
For all seven epochs, we determined the  flux-normalized 
derivative of the flux density as a 
function of time between any two adjacent data points normalized by the flux density.  
The two panels of Figure 4  show the intensity gradients of Day 6 where the highest  values  are seen.  
The mean gradients for all days 
roughly ranges between $\sim5\,  
\rm{hr}^{-1}$, during low-level intensity  fluctuations, and  $\sim 10-20\, \rm{hr}^{-1}$ for typical subflares.   
The highest values are noted near strong flares on  Day 6 near 9.5, 9.85  and 10.3 UT with a gradient  $\sim30\, \rm{hr}^{-1}$,   
which correspond to a factor of 2 in flux variation in  
1.39 min.  This time scale  is not limited by the  $\sim$18 sec sampling time measured on Day 6. 
This short time scale places a constraint on particle acceleration mechanism at work during an infrared flare at NIR wavelengths. 

%The low value of $\sim5\, \rm{hr}^{-1}$ corresponds to e-folding flux increase of decrease of $\sim 12$ min. $1/F dF/dt = 1/\tau$ where $\tau$ is 
%e-folding time to doubling time and the time to double the flux is $\tau_2 = ln(2)\times\tau$. However, the e-folding of the highest intensity gradient 
%is estimated to be $\sim2$ min

\subsubsection{Flux distribution of the variable emission}

As shown in Figures 2-4, Sgr A* shows a wide range of flux density values over 7 epochs of observations. 
The mean, median and 1-$\sigma$ error flux densities for each day are tabulated in Table 3.
The median flux densities at 2.1 and 4.8 $\mu$m ranges 1.39--2.03 and 2.05--4.70 mJy, respectively. 
The median flux density of GRAVITY measurements \citep{gravity20}
is 1.1$\pm0.3$  mJy at 2.1 $\mu$m.  The discrepancy in most likely due to flux variability measured in different epochs.

Figure 5 shows the histogram of the flux distribution of all 7-epoch data at 2.1 and 4.8 $\mu$m. All light curves were re-sampled at 2-min intervals for these 
histograms in order to avoid over-representing Day 5-7 observations, which employed higher sampling rates than the earlier days (see Table 1). 
The histogram 
suggests two populations of faint and bright emission 
with two peaks near 2 and 10 mJy at 2.1 $\mu$m. 
The bottom panel shows the cumulative fraction of time that 
the flux is below a certain value. In particular, we note that  there is shoulder between the fainter (subflare) emission and brighter (flare) emission at 
$\sim$3 and $\sim$9 mJy at 2.1 and 4.8 $\mu$m, respectively. 

The cumulative fraction of time of the variability of Sgr A* 
indicates that flare emission with flux densities  greater than $\sim$3  and $\sim$9 mJy at 2.1 
and 4.8 $\mu$m 
are  at a level of 10\% where bright flares occur. The remaining $\sim$90\% are 
detected at flux densities with a range between  1 and 3 mJy at 2.1 $\mu$m and 1.6 and 9 mJy at 4.8 $\mu$m.   
These limits suggest two different populations having different spectral indices.

%Tab4
\begin{deluxetable}{lcccccc}   
\tablewidth{0pt}
\tablecaption{Mean, median, sigma fluxes (mJy) for each day}
\tablehead{
\colhead{Epoch} &  
\colhead{Median}  &
\colhead{Mean} &
\colhead{Sigma} &
\colhead{Median} &
\colhead{Mean} &
\colhead{Sigma}\\
\colhead{}&
\colhead{2.1 $\mu$m}&  
\colhead{2.1 $\mu$m}&  
\colhead{2.1 $\mu$m}&  
\colhead{4.8$\mu$m}&  
\colhead{4.8$\mu$m}&  
\colhead{4.8$\mu$m}
}
\startdata
 Day 1  & 1.39&     1.52&     0.41&     3.05&     3.69&     1.53 \\
Day 2   & 2.03 &  2.28&     0.99&     3.97&     4.80&     2.57 \\
Day 3   & 1.84 &     2.07&     0.60&     0.00&     0.00&     0.00\\
Day 4  &2.15&     2.15 &    0.52&     5.70   &  5.49&     1.82\\
Day 5  &1.86 &    2.04  &   0.58 &    4.25  &   4.69 &    1.65\\
Day 6  &1.83  &   2.47   &  1.72  &   3.95 &    5.60  &   4.25\\
Day 7 &1.92    & 3.21     & 2.82   &  4.90&     7.77   &  6.49\\
\enddata
\end{deluxetable}

% changes in the proofs The parameters of the fits are listed in Table 3 which includes $\chi^2$, the amplitude of the peak, the natural log of the mean flux 

Following \cite{dodds-eden09}, 
we model the flux density  histograms with log normal (light green) and tailed log normal (dark green) distributions and 
the sum of two log normal distributions, as shown in  Figure 5. 
The parameters of the fits are listed in Table 3 which includes $\chi^2$, the amplitude of the peak, the natural log of the mean flux 
density ($\mu_{ln}$), the widths ($\sigma_{ln}$), the power law index ($\beta$) and lower-limit cut off ($x_{min}$)  values of three different 
distributions. 

%The log-normal (LN) distribution fits in Table 3 (line 1) followed by log-normal plus a power law (line 2) and then 2 log-normal distributions 
%(lines 3 and 4) are shown for both 2.1 and 4.8 $\mu$m in histograms in Figure 6.  

The tailed log normal (LN) distributions in Figure 5
represents the data better than the simple log normal distribution at 2.1 $\mu$m. At 2.1 $\mu$m, the mean ($\mu_{ln}$) of 
the LN distribution corresponds to $\sim$1.90 mJy. 
 The two-log 
normal distribution is  a better fit at 4.8 $\mu$m, but not at 2.1 $\mu$m. In this case, the mean flux densities for the two LN distributions are 
3.22 and 21.32 mJy. 
The result shows 
the presence of a bright component that is distinct from the fainter component, 
but with relatively little data at the brightest levels, it is impossible to precisely characterize the functional distribution of the bright emission.
However, the 4.8 $\mu$m data suggest 2-states described by  two LN distributions for faint and bright variability of Sgr A*, as found by   \cite{dodds-eden09}.

\begin{figure}[htbp]
\centering
   \includegraphics[width=2.6in]{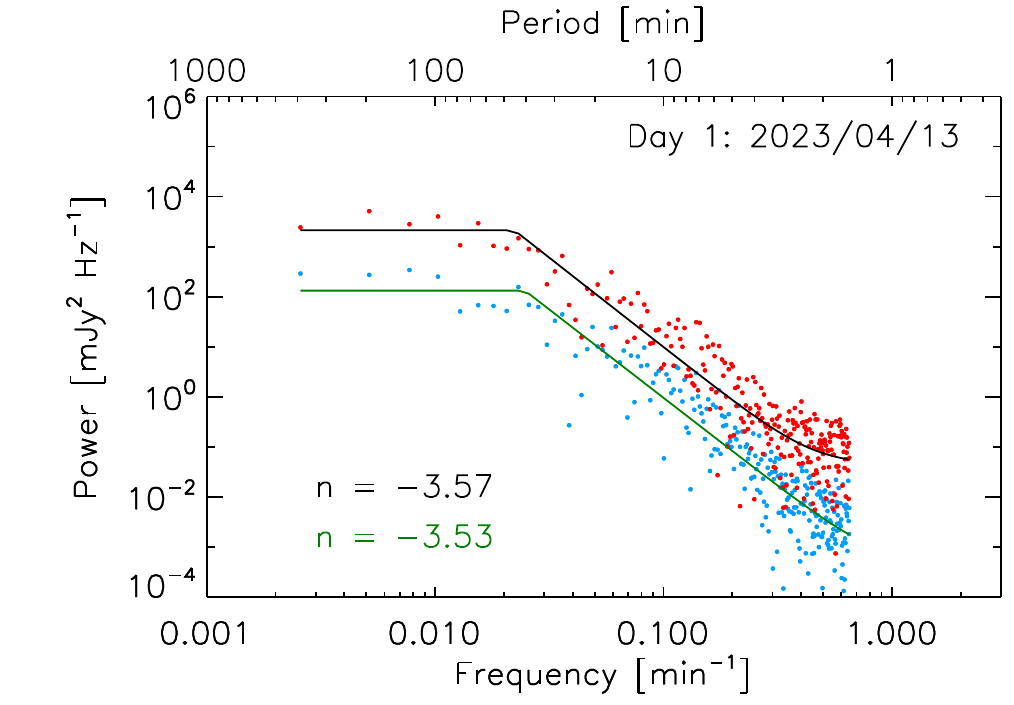}
   \includegraphics[width=2.6in]{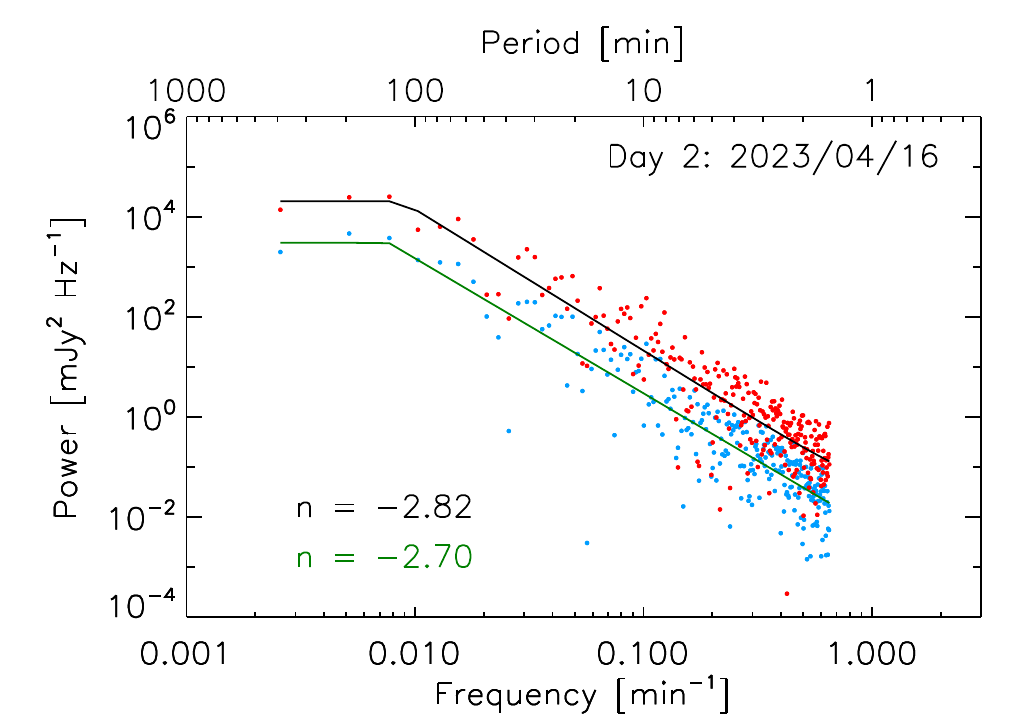}
   \includegraphics[width=2.6in]{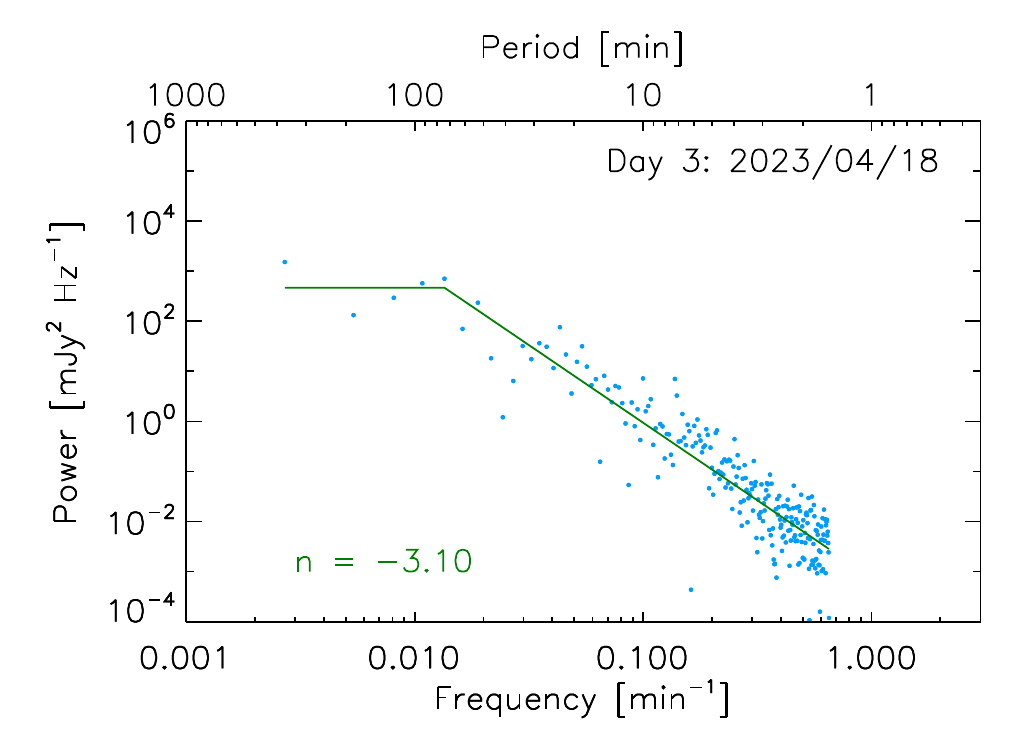}
   \includegraphics[width=2.6in]{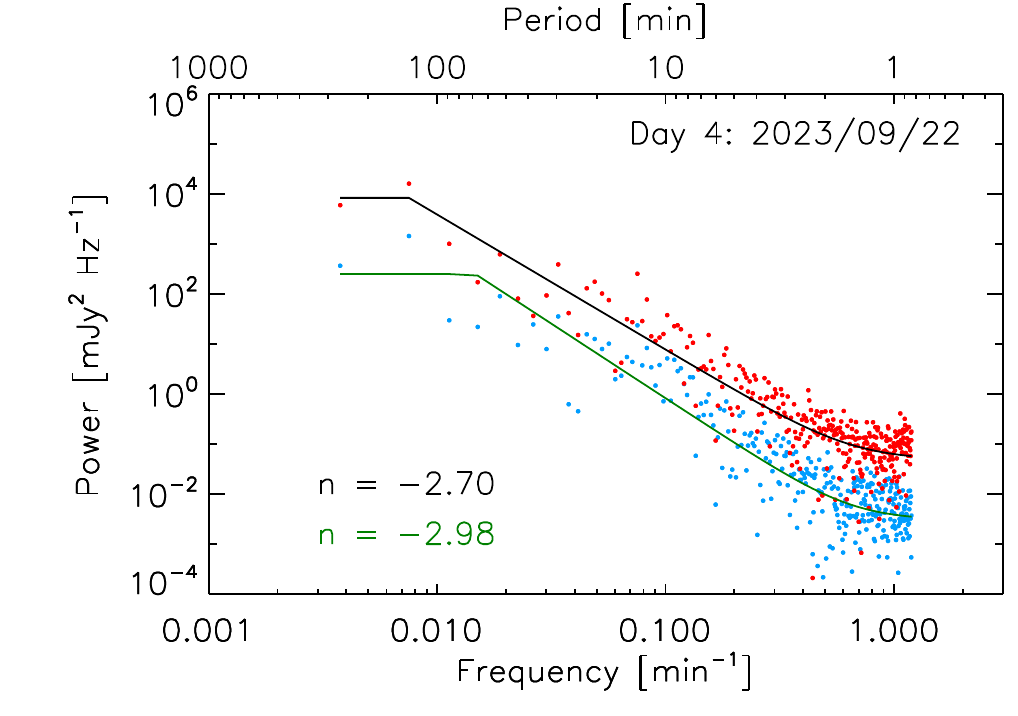}
   \includegraphics[width=2.6in]{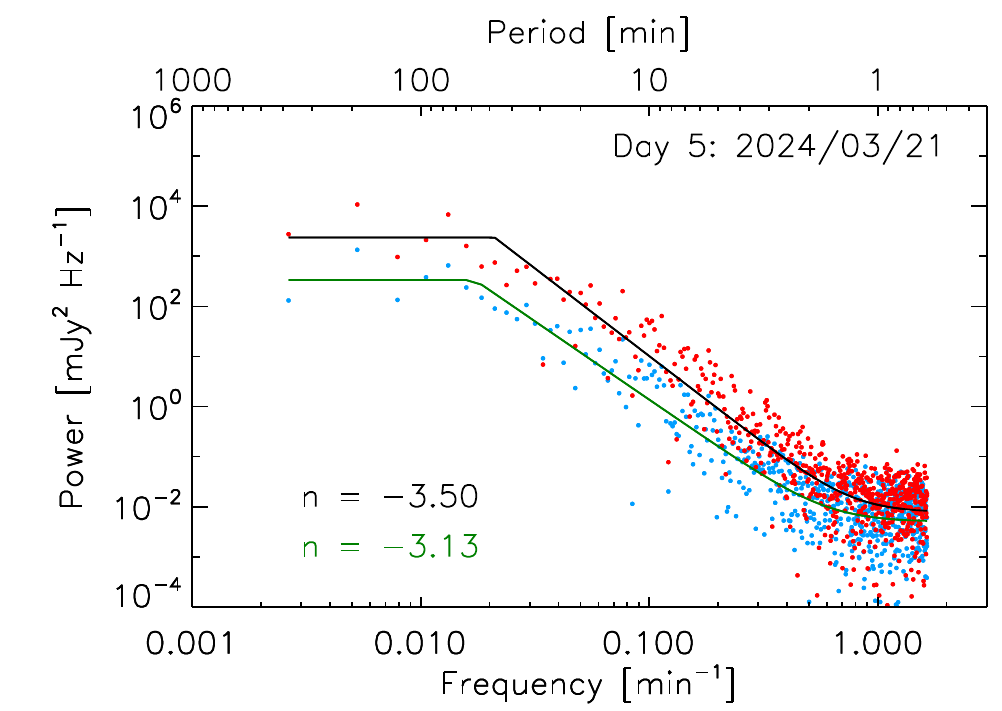}
   \includegraphics[width=2.6in]{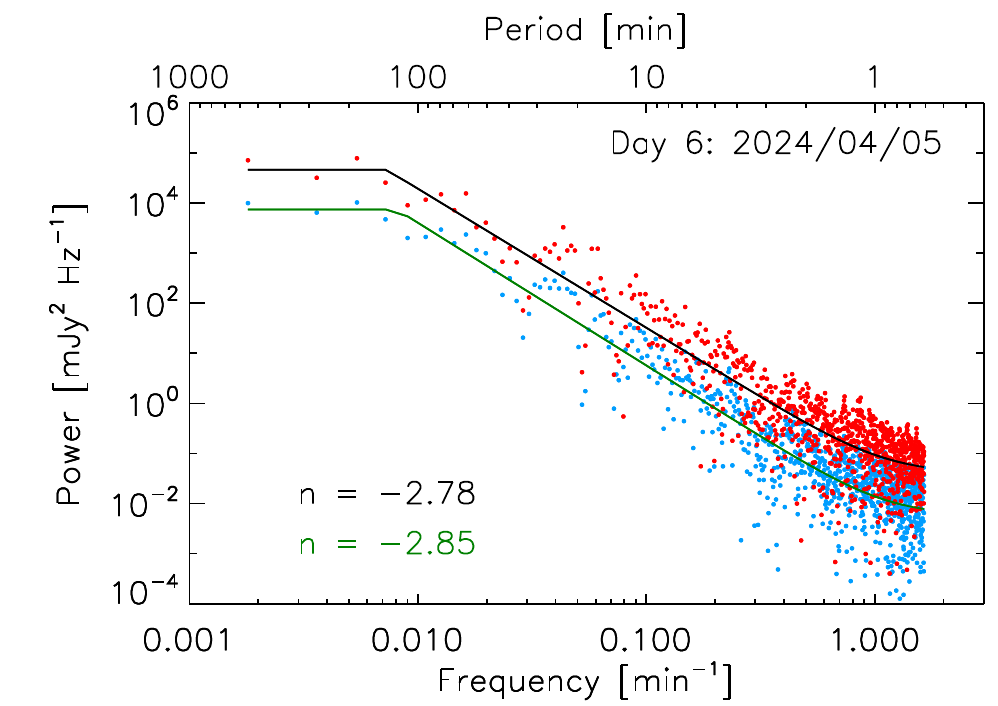}
   \includegraphics[width=2.6in]{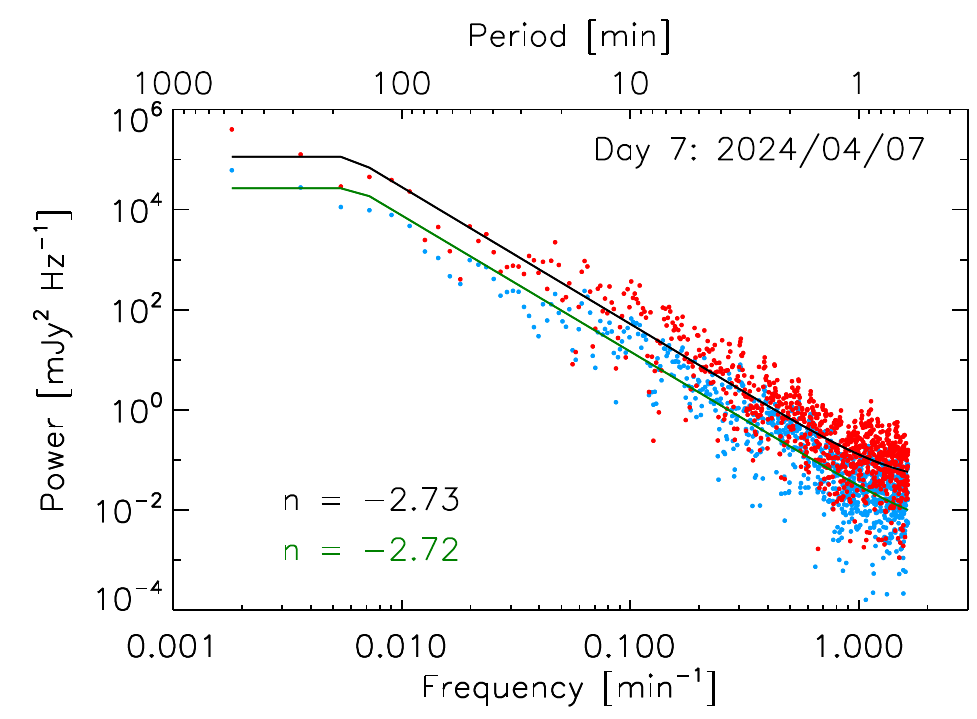}
 \caption{
Power spectra  of the temporal variations Sgr A* on a log frequency scale in the F210M 
(blue) and F480M (red) bands are presented for each epoch   of  observation.  
The  data for each epoch  are fitted with broken power-law trends over the 
   frequency ranges indicated by the solid lines with a transition 
frequency where the power-law distribution becomes flat at low $f_0$ and high frequencies $f_{1/2}$ (see Table 5). 
The power law index $n$ is indicated on the figure. 
These plots show evidence of Sgr A*'s variability,  displaying two different statistics on 
short and long-term time scales at frequencies $f\ge f_0$ and 
$f \lesssim f_0$, respectively. Note that the transition frequency 
$f_0$ changes in every epoch,  supporting log-term 
quiescent variability of  Sgr A*, though longer continuous observations are needed to confirm this result.    
}
\end{figure}

\subsection{Power spectrum of the flux variability}

The power spectra of the light curves for each day are shown in Figure 6. 
The dominant feature is a power-law slope. There is no 
indication of any strongly periodic behavior at any frequency. 
Most spectra do show flattening at high frequencies,  which is consistent 
with reaching the effective noise level of the observations. 
Most spectra also show flattening at the lowest frequencies, which is
an intrinsic feature of the light curves. 
The frequency turnover   noted at low frequencies
is indicative of  suppression of low frequency power that might have been expected 
(e.g. high pass filtering).
 However, the  turnover frequency from red spectrum to white spectrum changes in every epoch.

To quantify these trends, we fit each power spectrum with the 
sum of a broken power law and a white noise component 

\begin{eqnarray}
P(f) & = & P_0 \hspace{2in} f < f_0\\
     & = & (P_0-P_1) (f/f_0)^n + P_1 \hspace{0.8in} f \geq f_0.
\end{eqnarray}
where $P_0$ is the power at the break frequency, 
$f_0$, between the (assumed) flat spectrum and the power law,
$n$ is the power law index, and 
$P_1$ is the white noise power level.
An additional derived parameter, 
$f_{1/2}$, gives the frequency where the power law
and white noise components contribute equally to the total power   
\begin{equation}
f_{1/2} = f_0 [P_1/(P_0-P_1)]^{1/n}.
\end{equation}
These parameters  are listed in Table 5.
The quoted uncertainties are the standard deviations of 1000 Monte Carlo realizations of
power spectra calculated from data with random noise added (1-$\sigma$ is  the added random noise 
from the photometric uncertainties).
We note a strong anti-correlation between the derived $n$ and $f_0$
parameters on each day, thus 
higher  break frequencies (shorter periods) 
are  noted for steeper  power spectrum indices.

Unsurprisingly, the parameter $P_0$ largely tracks the overall brightness
of the flaring events on different days. The break frequency, $f_0$ does 
appear to vary from day to day, but its value is poorly constrained 
because is it relatively close to the lowest frequency of the observations.
Across the seven days, the power law indices have mean 
values of $n=-3.0\pm0.3$ at 2.1 $\mu$m,
and $n=-3.0\pm0.4$ at 4.8 $\mu$m. Shallower power law indices are generally
found on days when the flattening at high and or low frequencies is not
as clearly distinguished. 
There are strong covariances between the parameters
of these fits, so flatter power laws would be expected if $f_0$ and or $P_1$ 
are underestimated.

Table 5 shows the power spectrum break at low-frequency with 1-$\sigma$ error on each day  at 2.1 and 4.8 $\mu$m. 
The existence of this   break that varies from epoch to epoch 
suggests a physical mechanism that ties the low-frequency variability  to higher frequencies that follow a power-law spectrum. 
Longer duration of observations of Sgr A* are  required to confirm the break in the power spectrum at low frequencies. 
There is  variation from day-to-day timescale 
and the  time scales of  the break are much shorter than 8hr time scale estimated from  
submm data \citep{dexter14}. 

The physical 
mechanism that produces the power law  is altered or limited on longer 
time scales.  
If time scales correspond to light travel times, then this
would also suggest that the processes causing the variability do act
coherently on scales $\gtrsim 100$ light-minutes. Uncorrelated white noise  has also been reported 
at time scales greater than $\sim270^{+261}_{-94}$  min \citep{witzel18,witzel21} and $\sim128^{+329}_{-77}$ \citep{meyer09}.

%Tab5
\begin{deluxetable}{ccccccc}
\tablewidth{0pt}
\tablecaption{Power Spectra parameters}
\tablehead{
\colhead{Day} &
\colhead{Wavelength [$\mu$m]} &
\colhead{$P_0$ [mJy$^2$ Hz$^{-1}$]} &
\colhead{$n$} &
\colhead{$f_0$ [min$^{-1}$]} &
\colhead{$P_1$ [mJy$^2$ Hz$^{-1}$]} &
\colhead{$f_{1/2}$ [min$^{-1}$]}}
\startdata
 2023/04/13 &   2.1 &    132.4 $\pm$      2.5 &   -3.53 $\pm$    0.11 &    0.0247 $\pm$    0.0015 &     0.000 $\pm$     0.002 &    0.8567 \\
 2023/04/13 &   4.8 &   2127.2 $\pm$     25.3 &   -3.57 $\pm$    0.12 &    0.0222 $\pm$    0.0016 &     0.044 $\pm$     0.026 &    0.4556 \\
 2023/04/16 &   2.1 &   3008.4 $\pm$      9.4 &   -2.70 $\pm$    0.04 &    0.0077 $\pm$    0.0003 &     0.000 $\pm$     0.005 &  Infinity \\
 2023/04/16 &   4.8 &  20331.7 $\pm$     68.6 &   -2.82 $\pm$    0.07 &    0.0088 $\pm$    0.0004 &     0.023 $\pm$     0.058 &    1.1305 \\
 2023/04/18 &   2.1 &    468.8 $\pm$      2.7 &   -3.10 $\pm$    0.07 &    0.0135 $\pm$    0.0005 &     0.000 $\pm$     0.002 &  Infinity \\
 2023/04/18 &   4.8 &      \nodata &     \nodata &     \nodata &      \nodata &       \nodata \\
 2023/09/22 &   2.1 &    248.8 $\pm$     12.1 &   -2.98 $\pm$    0.08 &    0.0147 $\pm$    0.0003 &     0.003 $\pm$     0.002 &    0.6536 \\
 2023/09/22 &   4.8 &   8277.3 $\pm$    597.2 &   -2.70 $\pm$    0.06 &    0.0075 $\pm$    0.0000 &     0.048 $\pm$     0.016 &    0.6560 \\
 2024/03/21 &   2.1 &    336.0 $\pm$      2.0 &   -3.13 $\pm$    0.08 &    0.0172 $\pm$    0.0008 &     0.005 $\pm$     0.001 &    0.6007 \\
 2024/03/21 &   4.8 &   2377.6 $\pm$     11.2 &   -3.50 $\pm$    0.07 &    0.0211 $\pm$    0.0006 &     0.008 $\pm$     0.004 &    0.7792 \\
 2024/04/05 &   2.1 &   7527.6 $\pm$     10.3 &   -2.85 $\pm$    0.05 &    0.0081 $\pm$    0.0006 &     0.006 $\pm$     0.002 &    1.1222 \\
 2024/04/05 &   4.8 &  46598.2 $\pm$     87.3 &   -2.78 $\pm$    0.04 &    0.0073 $\pm$    0.0006 &     0.040 $\pm$     0.008 &    1.1146 \\
 2024/04/07 &   2.1 &  26604.8 $\pm$     26.6 &   -2.72 $\pm$    0.03 &    0.0063 $\pm$    0.0002 &     0.003 $\pm$     0.002 &    2.3447 \\
 2024/04/07 &   4.8 & 113080.7 $\pm$    142.7 &   -2.73 $\pm$    0.03 &    0.0060 $\pm$    0.0002 &     0.033 $\pm$     0.009 &    1.4829 \\
\enddata
\end{deluxetable}

\subsection{The spectral index vs.  the  flux density}

Given the unique capability of NIRCam to  observe two NIR wavelengths simultaneously,
the spectral index $\alpha$
of bright and faint flares can be determined as a function of time. 
The top  left panel of Figure 7 shows the variation of 4.8 
$\mu$m against 2.1 $\mu$m flux density for all seven days. 
A correspondingly colored version of the light curves for all data is shown in the top right panel Figure 7. 
The colors denote  six days of observations at 2.1 and 4.8 $\mu$m, as indicated in the light curves.
The strongest flaring activity displayed in red on Day 7 tends to 
show a shallower slope. 
A striking result is the anti-correlation and correlation reflecting  a variation of the 
slope that becomes  shallower as the flux density increases.  This is consistent with two different populations of energetic particles producing  
faint  and bright variable emission.  
 The variation of the spectral index 
as a function of F210M flux density is displayed  in the middle  left panel. 
The mean  spectral index tends to  become steeper with increasing brightness up to 
$\sim 3$ mJy at 2.1 $\mu$m, without any further steepening at higher brightnesses. 
The weak  anti-correlation of the spectral index with brightness (up 
to a limit), changes and shows a shallower spectral index with increasing brightness. The spectral index eventually saturates with 
increasing brightness with values close to $\alpha\sim-1$. The middle  right panel of Figure 7 shows the spectral index variation as a function of time for all epochs. 
The four panels of all data presented  in  Figure 7 are shown for individual days in 
Appendix B, Figures 10-15.

One of the most challenging aspects of this  unusual behavior of  positive spectral index 
 is the background emission  due to  contamination by stars in 
the vicinity of Sgr A*. 
In order to investigate the effect of varying the estimated background  emission, 
scatter plots of the 2.1 and 4.8 $\mu$m flux densities for all days, 
is shown  in the upper left panel of Figure 7. 
Overlaid  on this Figure   are 
 lines indicating  loci of constant spectral index. 
The lines in color are linear fits to the photometry on each day, which will be discussed in \S3.3.1. 
 We note that the fainter data points have $\alpha \sim -1.58$ and the spectral index becomes shallower  to $\alpha \sim -0.85$ for 
the brighter data points.

 We also examined  the impact of changing the background  flux level at 4.8 $\mu$m of Day 7 data only (red color in Fig. 7c), as
shown in the  bottom   panel of Figure 7e.
It is clear that the overall trend remains unchanged but
the positive  spectral index of the weak  flux densities
disappear when the background  emission  is increases  by 1 and 2 mJy, which has the effect of increasing the brightness of Sgr A*.
There is clearly an uncertainty in the spectral index of
the weakest  emitting fluctuations.
However, the key point is that in spite of this uncertainty in the 4.8 $\mu$m  background flux density  by 1 or 2 mJy,
the slope in the flux vs flux plots are independent of the background level.
The slopes are independent of the background level, thus supporting bimodality of flare emission from Sgr A*. 
How those slopes translate into spectral indices does depend on the choice of background level.

%f4a
\setcounter{subfig}{0}
\begin{figure}[htbp]
  \centering
   \includegraphics[width=3.5in]{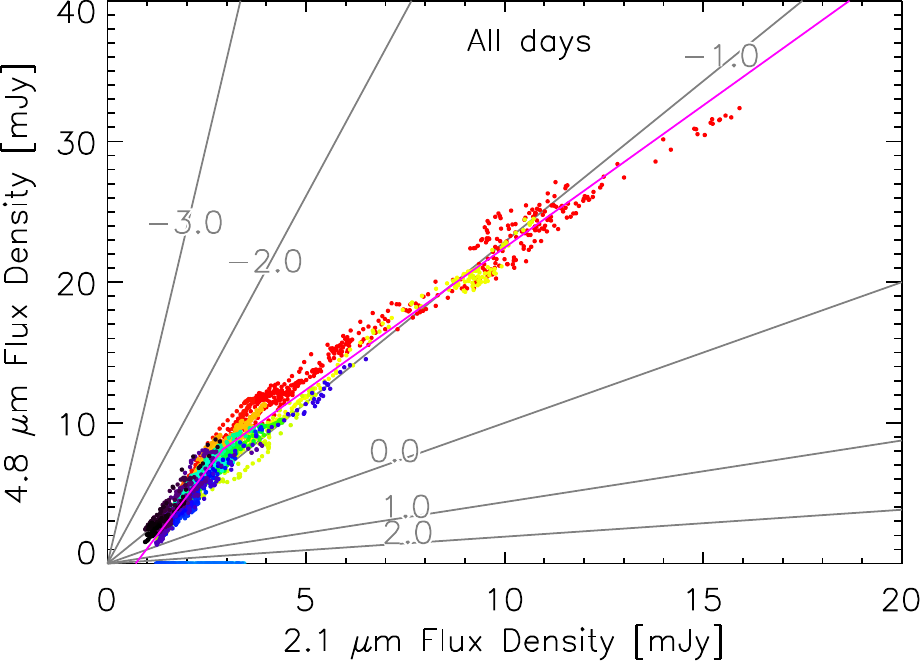}
   \includegraphics[width=3.5in]{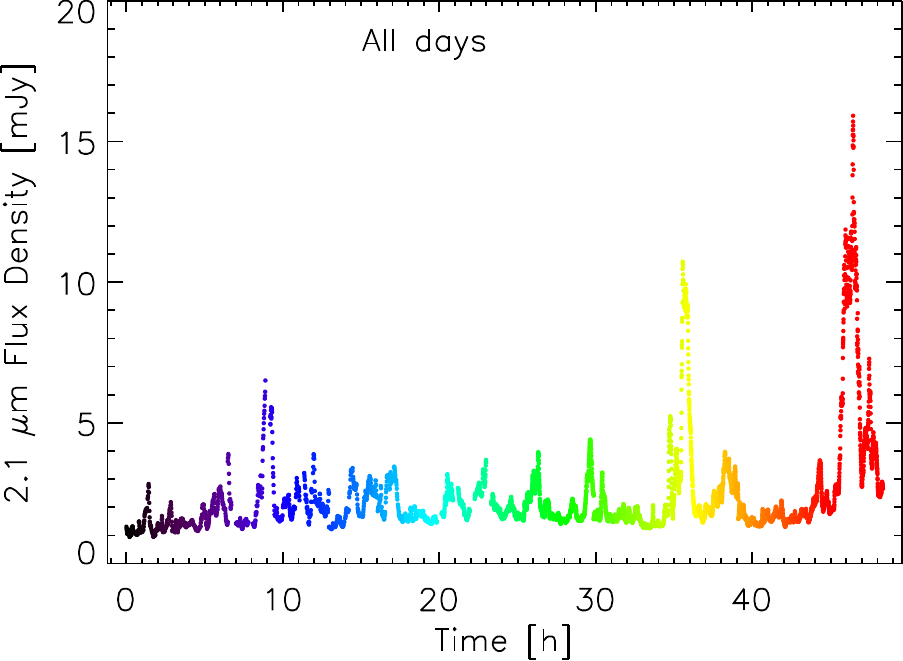}\\
   \includegraphics[width=3.5in]{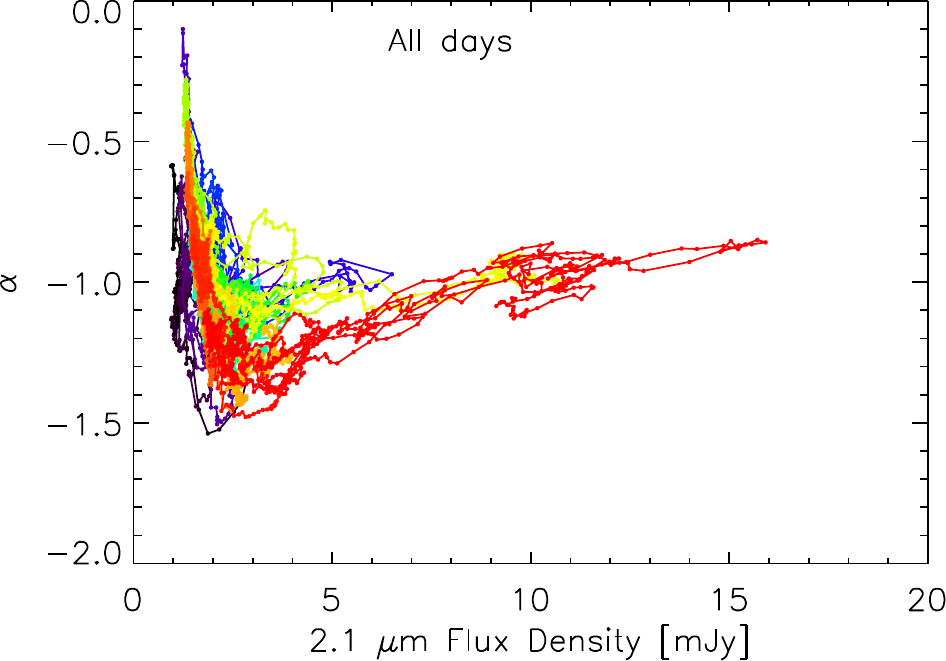}
   \includegraphics[width=3.5in]{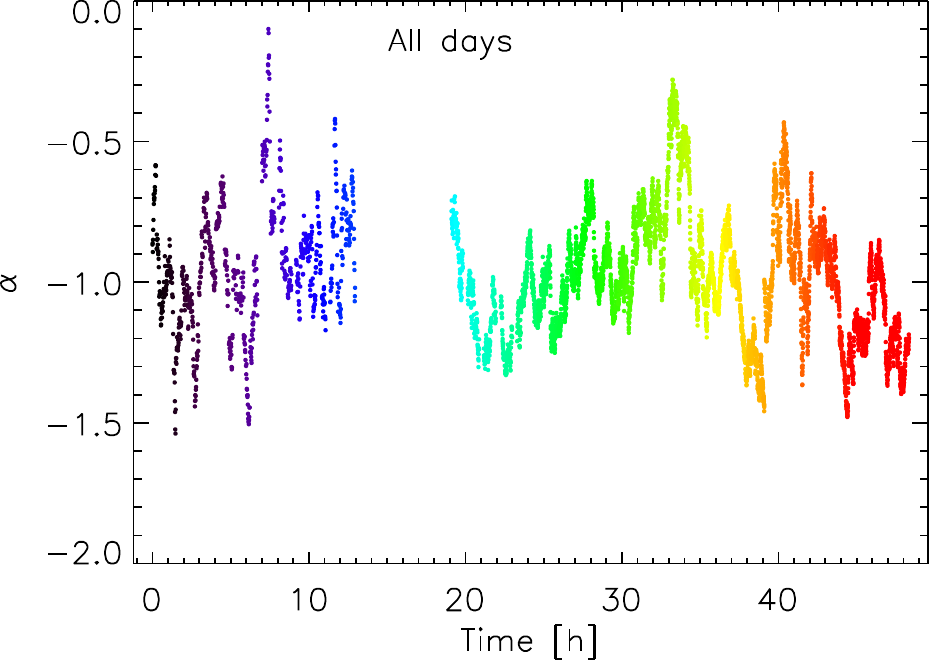}
    \includegraphics[width=3.5in]{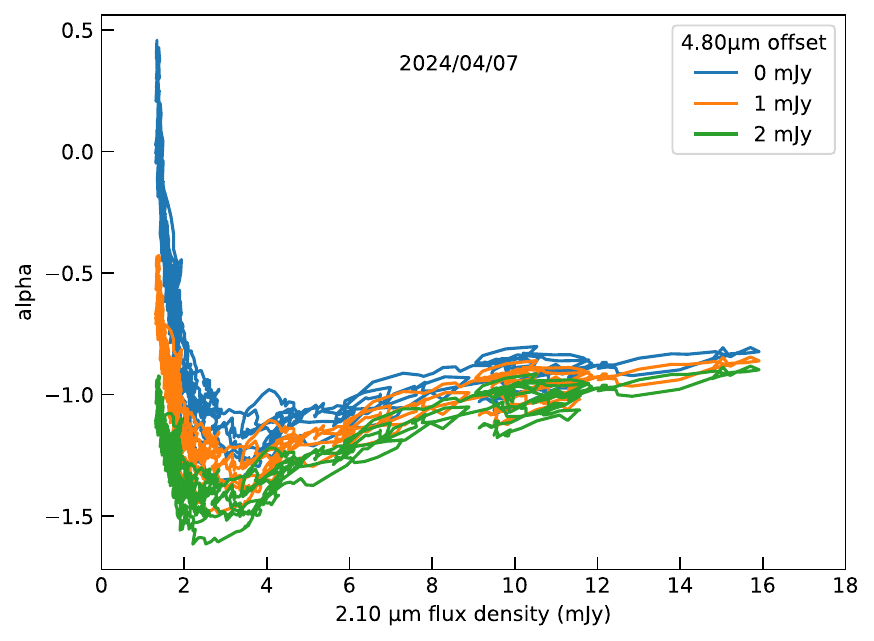}
\caption{ 
{\it Top Left}
A scatter plot of the F210M and F480M flux densities on all days 
 overlaid with  loci of constant  spectral index.
The magenta line is a broken linear fit to the data, with parameters listed in Table 6. 
 The shape of the fainter data points correspond to a  spectral index $\alpha 
\sim -1.58$ whereas the slope for the brighter data points flattens to  $\alpha \sim -0.85$. 
The change in the slope 
 suggests two different populations with different spectral indices.
{\it Top  Right}
Light curves of Sgr A* at 2.1 $\mu$m for all epochs. 
{\it Middle Left}
The  spectral index of Sgr A* as a function 2.1 $\mu$m flux density 
for all six epochs of  observations. 
{\it Middle Right}
The spectral index evolution for  all six epochs. 
The colors correspond to the times indicated for each epoch. 
{\it Bottom}
The spectral index vs.  2.1 $\mu$m flux density on Day 7 if the 4.8 $\mu$m subtracted background is reduced  by 1 and 2 mJy.
Given that the background 4.8 $\mu$m flux is
more likely to be contaminated by nearby stars than the 2.2$\mu$m flux of Sgr A*,
the positive spectral index of faint
variable emission (shown in blue color)
disappears  when the 4.8 $\mu$m flux is  increased by 1  mJy.
However, the bright flare's spectral indices and the trend are  not affected by the background change.
}
\end{figure}  

  \subsubsection{Linear fits to the slope of  2.1 vs.  4.8 $\mu$m flux density}

We have fitted broken linear fits to the flux density of 4.8 $\mu$m vs. 2.1 $\mu$m for all days, as shown in 
Figures 7 (top left panel).  
Fits to the flux density for individual days are shown in Figures 10-15  in Appendix B. 
The fits are given by the following equation 

\begin{equation} 
S(4.8) = S_0(4.8) + a_{\pm} \times (S(2.1)-S_0(2.1)) 
\end{equation} 
where  S(2.1) and  S(4.8) are flux densities  at 2.1 and 4.8 $\mu$m, respectively,  S$_0$ is the 
flux density where the slope changes,  
a$_{+}$ applies for S(2.1) $\ge$  S$_0$(2.1),  
a$_{-}$ applies for S(2.1) $\le$  S$_0$(2.1). 
These fits use  4 parameters,  as given in Table 6. 
The values $\alpha_{-}$ and $\alpha_{+}$ are the slopes, a, translated into constant power law indices.
The mean spectral index over all  days are
-1.58$\pm0.150$ and  -0.87$\pm0.16$ for the faint and strong variable emission, as listed in the last column.
The  values would change if the offsets in either band were changed, so 
their values could differ. The a$_{\pm}$ values are important in that they are  independent 
of whatever offset or background adjustment is  made.  

 The fact that the slope values  for all days and individual days, as displayed in Appendix B,   seem to cluster around the 
$``$all days'' values of 3.7 and 2.0,  
suggests  a consistent picture that there are two  populations of 
flares with a  spectral index that  changes  from steep to shallow, as flares become brighter. 
The  dependence of the spectral index on the flux density resembles  another characteristic that distinguishes 
weak and bright flares, namely,  different  statistics applied to flux distribution of faint and bright flares, as described in $\S3.1.4$.
2.1 $\mu\rm{m} \sim 3$ mJy seen here is another motivation for the distinction between flares and subflares as defined previously.

%Tab6
\begin{deluxetable}{lllcccc}   
\tablewidth{0pt}
\tablecaption{Parameters of linear fits to 2.1 $\mu$m vs  4.8~$\mu$m flux density}
\tablehead{
\colhead{Date} &  
\colhead{S$_0 (2.1 \mu\rm{m})$ } &  
\colhead{S$_0 (4.8 \mu\rm{m})$} &  
\colhead{a$_{-}$} &
\colhead{a$_{+}$} &
\colhead{$\alpha_{-}$} & 
\colhead{$\alpha_{+}$} 
}
\startdata
All   &      2.90  &     8.10   &    3.70  &     2.02 & -1.58 & -0.85 \\
\\
Day 1 & -- &     -- &      3.33 &   -- & -1.46          &     -- \\
Day 2 & 2.93  &     7.03   &    3.13     &  2.10 & -1.37          &     -0.89   \\
Day 3 &  0         &      0          &     0            &   0 & 0 & 0      \\
Day 4 & 2.61  &     7.27  &     3.71 &      2.28 & -1.59 &              -0.99    \\
Day 5 &  2.36 &      6.08 &      3.58 &       2.31 & -1.54     &          -1.01   \\
Day 6 & 2.74   &    8.05   &   4.41  &     2.24 & -1.80       &        -0.98   \\
Day 7 & 3.35   &    10.39    &   4.04  &     1.82 & -1.69       &        -0.73\\
\enddata
\end{deluxetable}

\subsection{Time Delays}

Given the remarkable capability of NIRCam's simultaneous observations at 2 wavelengths, a comparison of the 2.1 and 4.8 
$\mu$m light curves were made to determine if there is a time delay.  
The cross correlation of the light curves at these two 
wavelengths indicate for the first time that there is evidence for a delay in the 4.8 $\mu$m variable emission with respect to 
2.1 $\mu$m in all  observations. 
Figure 8  shows the cross correlation of the light 
curves on six  separate days showing time delays ($\Delta\, t$) labeled on each plot.  The time delays range between 3 and 40 seconds. 

Another characteristic of the observed flares is the presence of loops in the flux density vs. spectral index plots. The loop feature is clearly 
noted in individual spectral index plots (bottom left panels of Figs 7-9  and Figs. 10-15 in Appendix C.),
The loop structures and time delays are modeled in the discussion \S4.

%f5
%\setcounter{subfig}{0}
\begin{figure}[htbp] 
   \centering
   \includegraphics[width=2in]{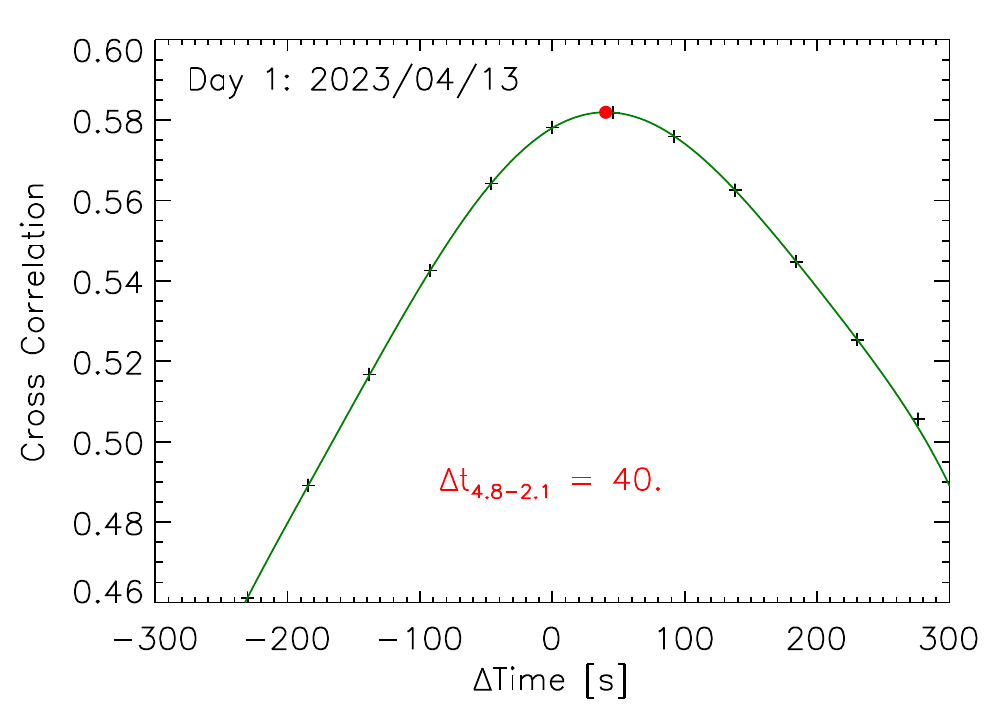}
   \includegraphics[width=2in]{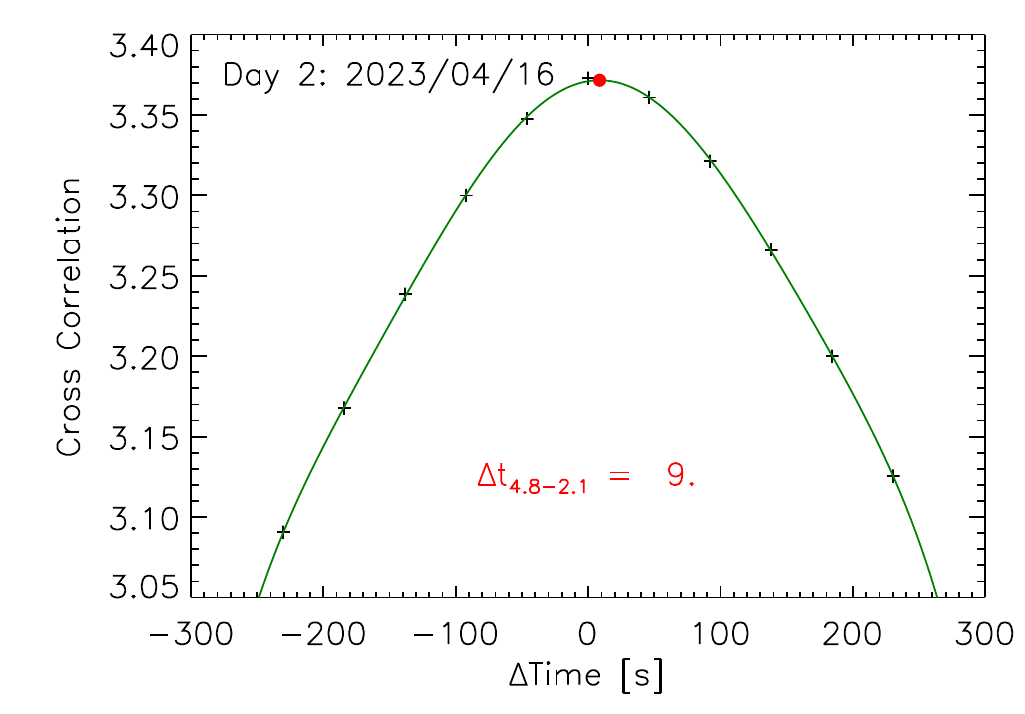}
   \includegraphics[width=2in]{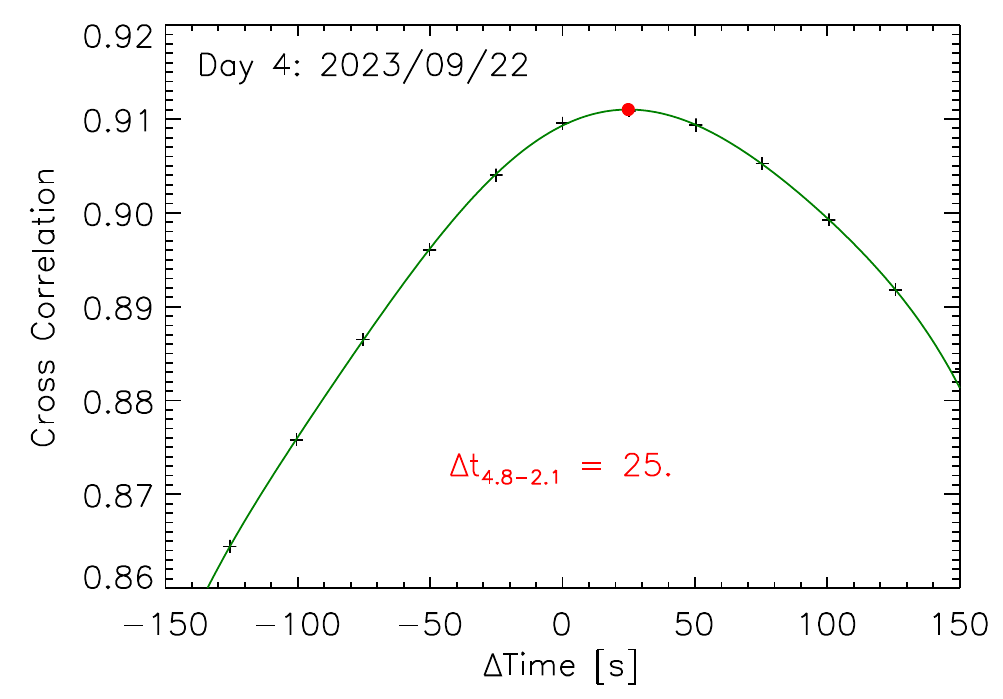}
   \includegraphics[width=2in]{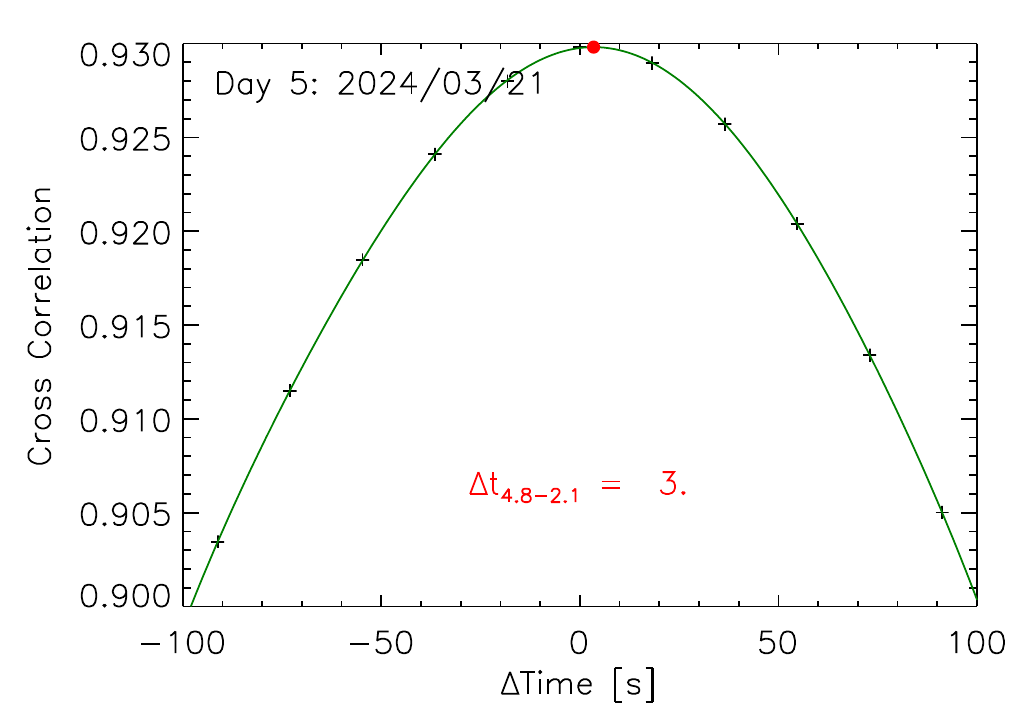} 
   \includegraphics[width=2in]{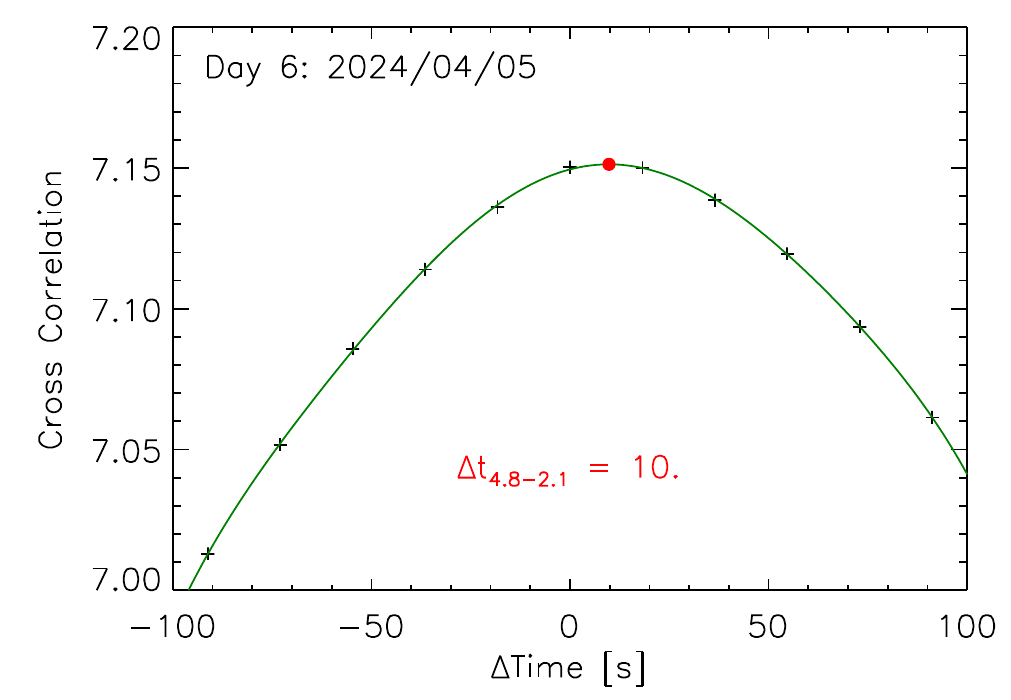} 
   \includegraphics[width=2in]{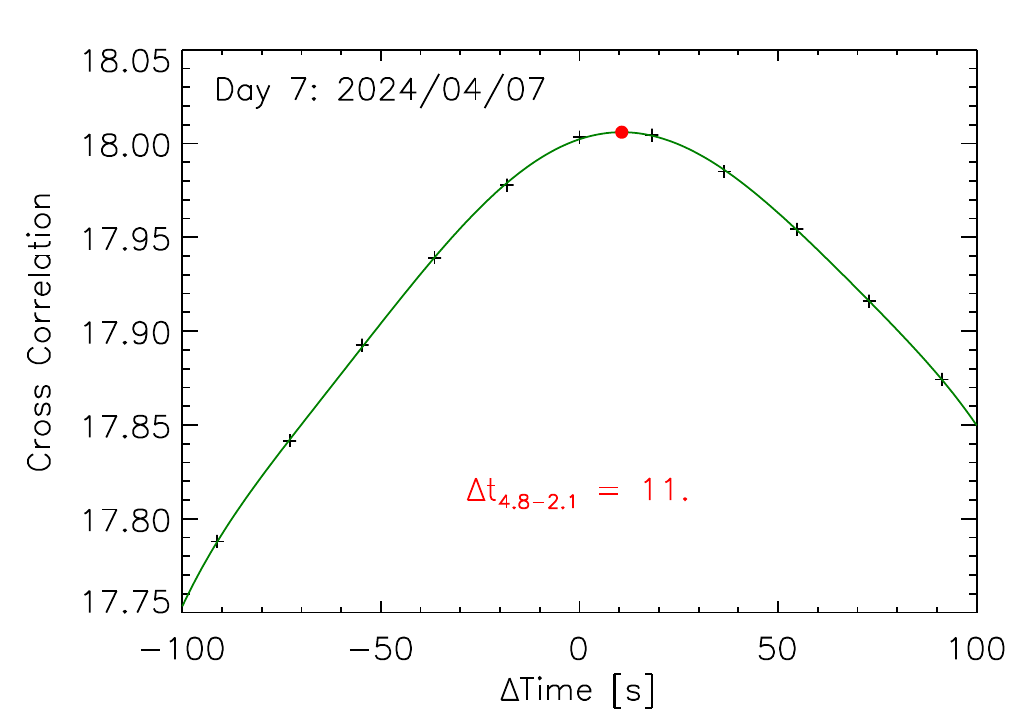} 
   \caption{
Cross correlation of the Sgr A* light curves in the F210M and F480M bands as a function of time are shown for each of the
six  epochs of observations.  Epochs 1, 2, and 4 are displayed in the top 3 panels whereas 
epochs 5, 6 and 7 are displayed in the bottom panel.  The cross correlations indicate  time delayed in  the range between 3 and 40 seconds.
Positive $\Delta t$, as shown in red on each figure,  corresponds to F210M variations preceding  those at F480M.
}
\end{figure}

\subsection{Loop Structures}

Figure 9 shows  the variation of the spectral index as a function of F480M flux density for five subflares (left columns).  
Loop structures in these plots reveal  the spectral index increase and decrease as a 
flare emission rises and decays.  
Additional loops are displayed in Figure 16,  Appendix C. 
In all these cases, the variations of 
the spectral index trace counterclockwise loops. 
The  right panels show the light curves of 
individual 
flares whereas 
the left panels 
display 4.8 $\mu$m flux density of these events  as a function of the spectral index. Each fluctuation shows a counter-clockwise loop pattern as 
the flux density rises and decays. The pattern shows steepening of the spectral index as 4.8 $\mu$m flux density rises followed by flattening of 
the spectral index as the flux density decreases.

%f6
%\setcounter{subfig}{0}
\begin{figure}[htbp] 
%\addtocounter{figure}{-1}
%\stepcounter{subfig}
%\begin{figure}[htbp] 
  \centering
   \includegraphics[width=2.3in]{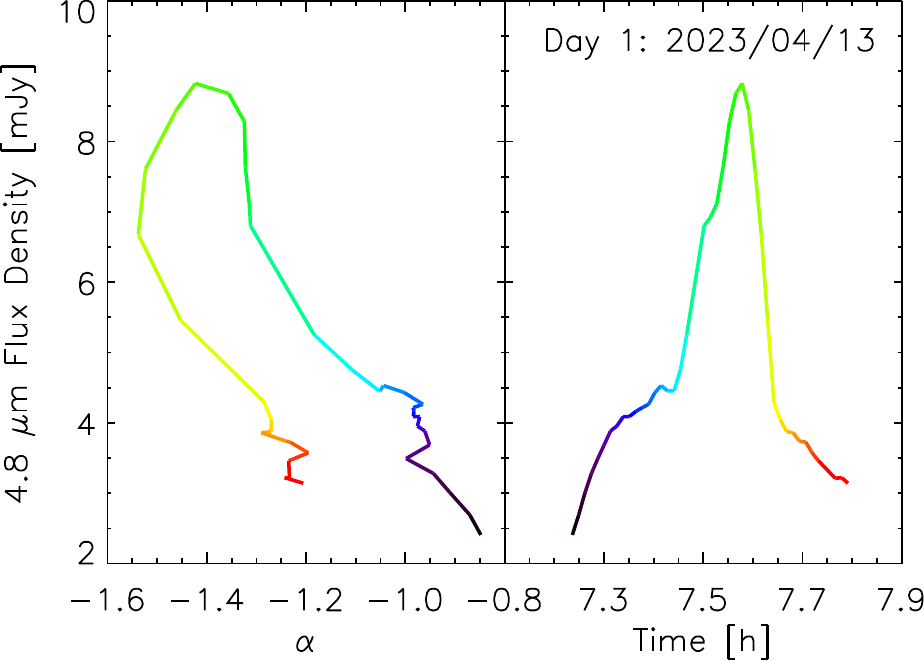}
\hspace{0.5in}
   \includegraphics[width=3in]{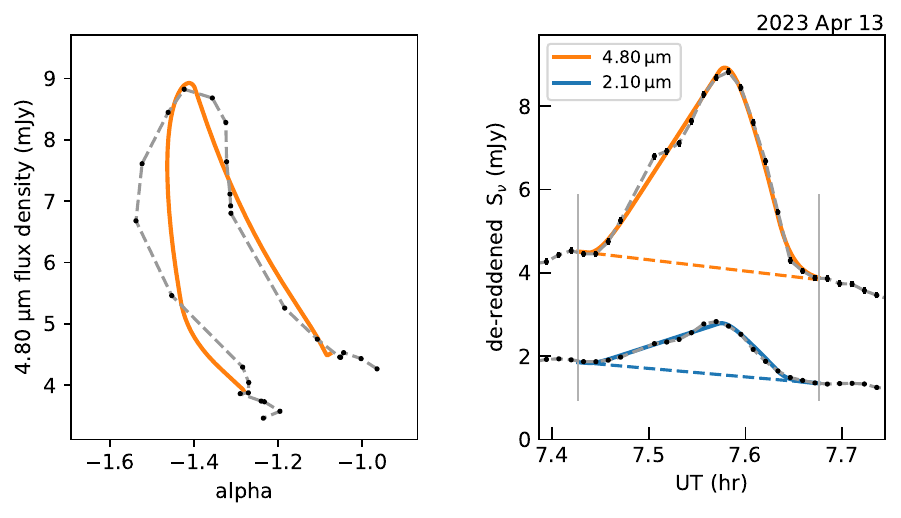}\\
   \includegraphics[width=2.3in]{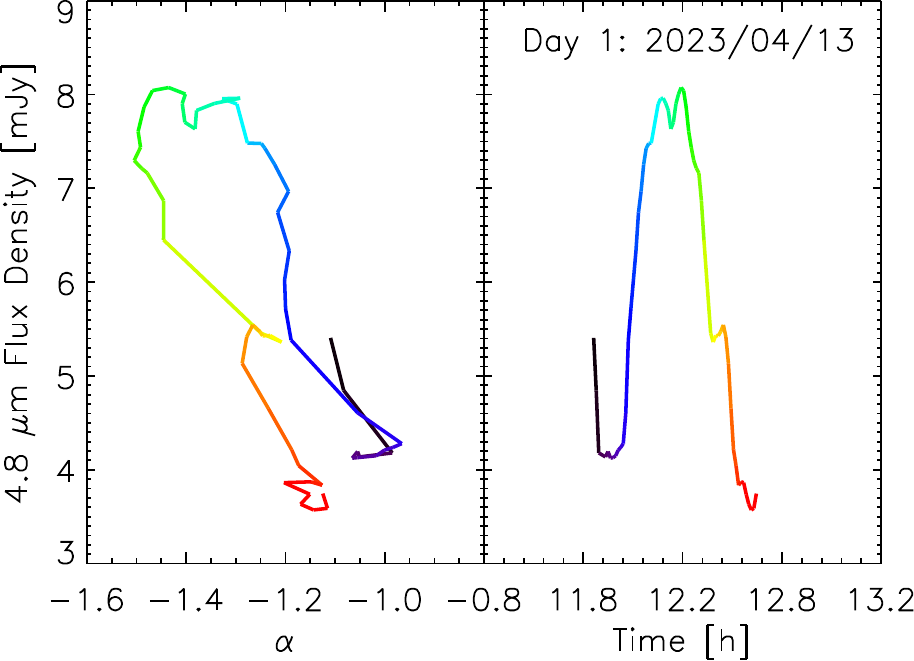}
\hspace{0.5in}
   \includegraphics[width=3in]{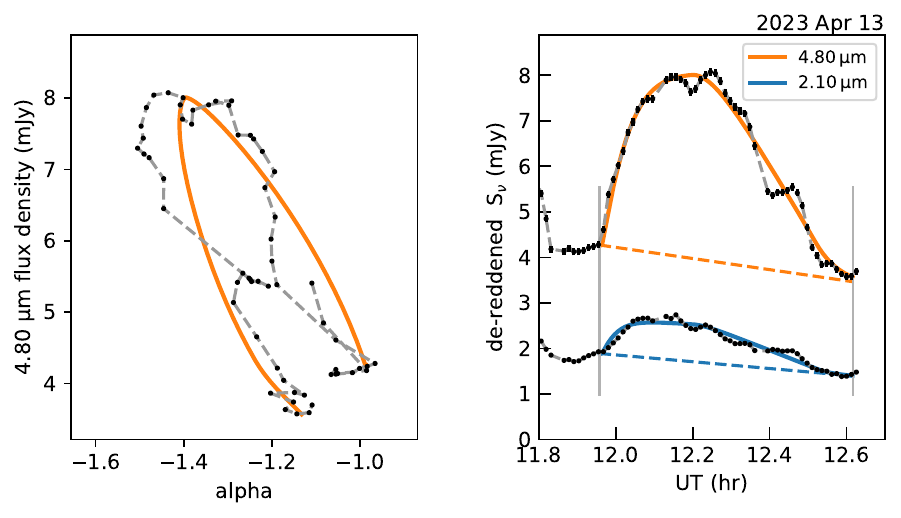}\\
   \includegraphics[width=2.3in]{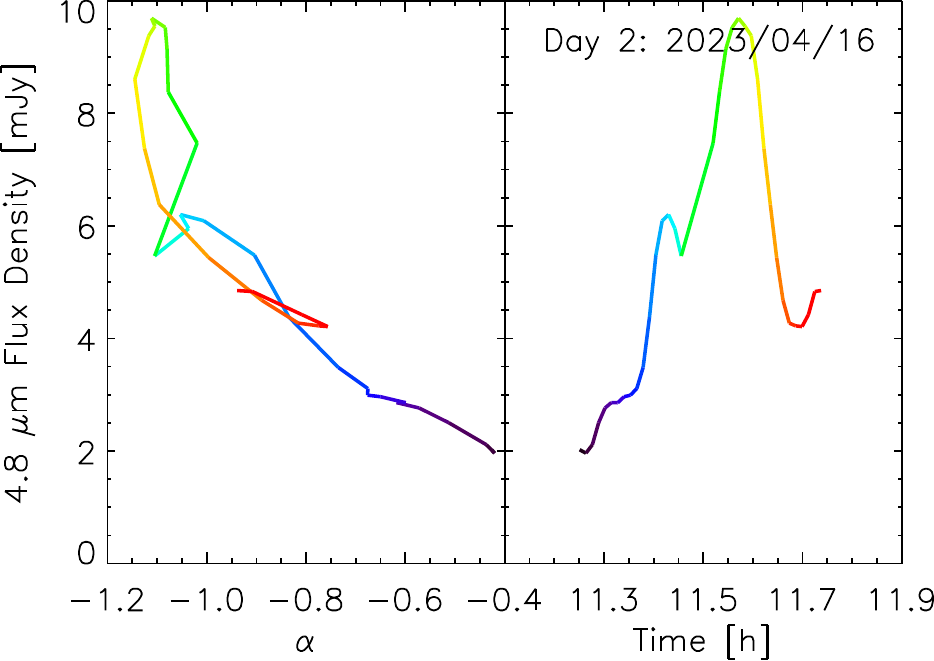}
\hspace{0.5in}
  \includegraphics[width=3in]{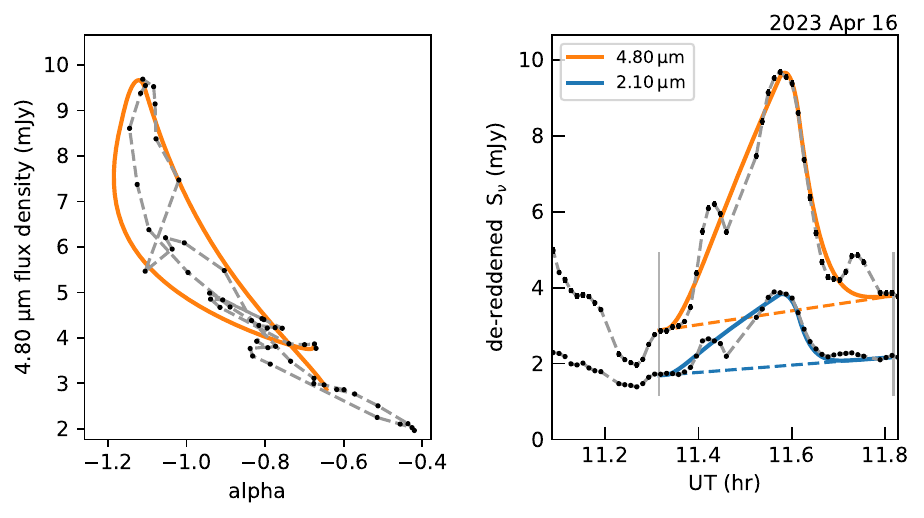}\\ 
\includegraphics[width=2.3in]{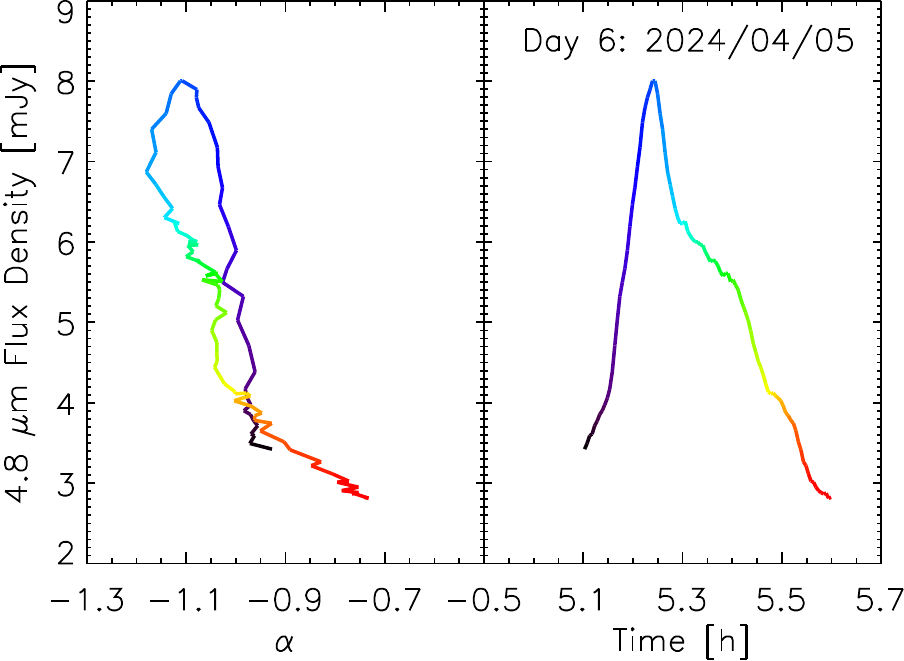}
\hspace{0.5in}
  \includegraphics[width=3in]{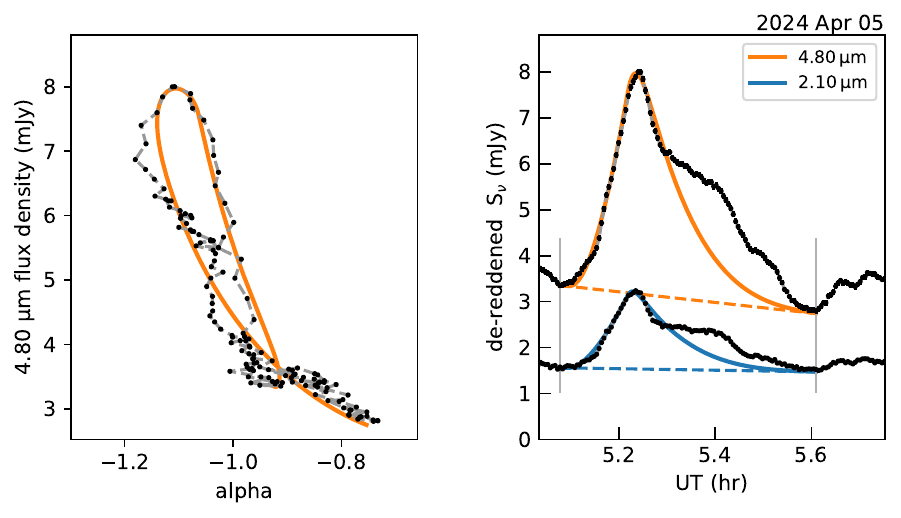}\\
   \includegraphics[width=2.3in]{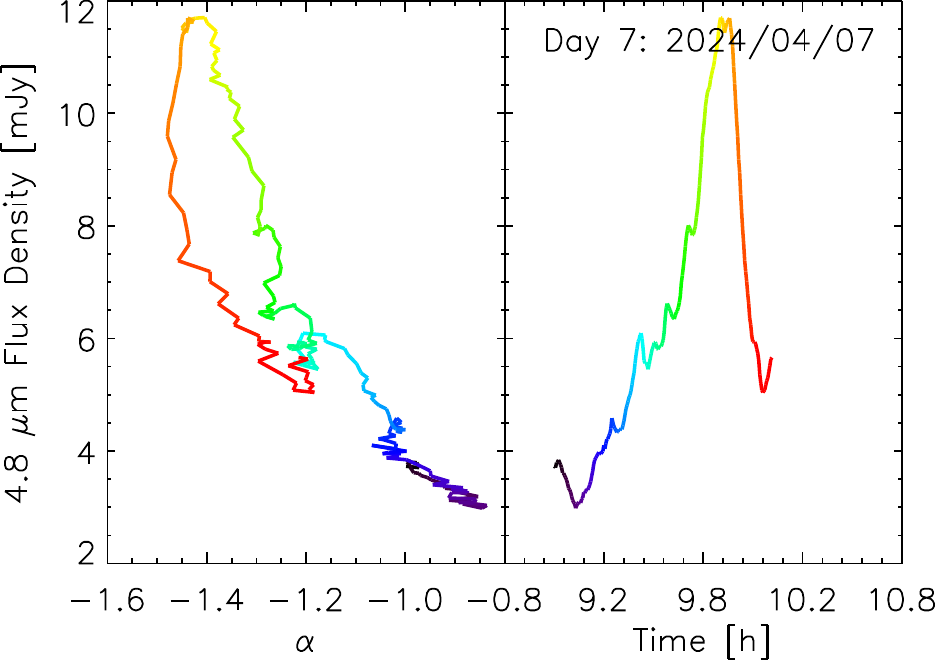}
\hspace{0.5in}
  \includegraphics[width=3in]{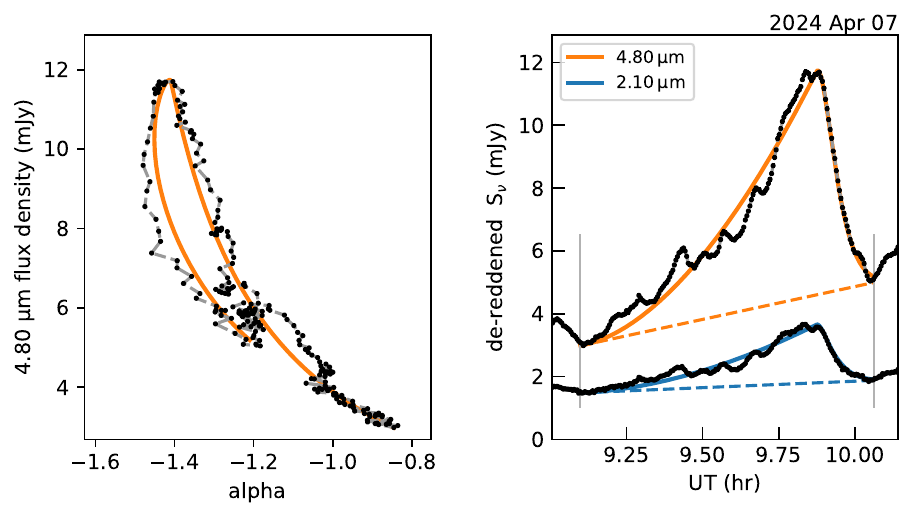}
%\vspace{-0.8in}
   \caption{  
   {\it Left two columns } 
   The variation of the spectral index as a function of F480M flux density   is
   shown  for  5 flares. The colors denote time.  The variations of the spectral index trace counterclockwise loops.  
 {\it Right two columns} 
The  displayed modeled time delays and loop diagrams are  interpreted to be due to synchrotron cooling,
as the 4.8$\mu$m emission lags  in time behind the 2.1 $\mu$m emission.
These fits provide physical characteristics of individual  flares (see Table 7). 
}
\end{figure}

\section{Discussion}

We have described  some of the salient features such as sub-minute and long time-scale 
quiescent variability, 2-minute rise of the variable emission from Sgr A*,  
the time delays between 2.1  and 4.8 $\mu$m emission, and  the dependence of  the spectral index on the  flux density of flare emission. 
In addition, we presented in seven epochs that  Sgr A* is constantly fluctuating at NIR wavelengths with 
no evidence for quiescent state, consistent with earlier ground-based 
and NICMOS observations \citep{fyz09,weldon23}. We had limited data to fit the flux distribution of Sgr A* but our analysis  indicates
that two different populations of particles are  needed to explain the 
flux distribution of Sgr A*. Two  stochastic processes with 
log-normal distributions at 4.8 $\mu$m suggest that there are characteristic energy scales of the accelerating events for 
bright and faint fluxes. 
There are a number of  new characteristics of flaring activity of Sgr A* as  
revealed in  seven epochs of  monitoring observations, as described in more detail below. 

\subsection{The origin of loop diagrams and time delays}

The characteristic loop structure that individual flares exhibit in the spectral index vs 2.1\,\micron\ flux plots in Figure 9 (see also Fig. 10-15) 
 occurs because at a given flux 
level, the spectral index of the falling part of the light curve is steeper than during the rise.  This steepening is suggestive of synchrotron cooling.  
The loop 
structures presented in Figure 9  are simpler to model because there are not many low-level fluctuations, thus easier to identify. 
To explore 
this, we adopt a  model in which electrons are injected into the source region with an $E^{-2}$ spectrum and upper cutoff energy $E_c$; the injected electrons 
subsequently suffer synchrotron losses in a constant magnetic field $B$.  Injection commences, peaks, and ends at times $t_0$, 
$t_1$, and $t_2$, respectively, with the injection rate rising as $(t-t_0)^{\beta_1}$ 
for $t_0<t<t_1$, and declining as $(t_2-t)^{\beta_2}$ for $t_1<t<t_2$ (described 
in detail in Appendix D). The evolution of the electron population and the resultant synchrotron emission at 2.1 and 4.8 $\mu$m are then computed.  Note that as the 
synchrotron emission is optically thin the results depend only on the total number of electrons and their energy distribution, not on the size of the region so this 
does not enter as a parameter.

We apply this  model to five typical flares displayed in in Figure 9 (left two panels). 
 A linear baseline is subtracted from each flare and the model light curves are fit to the residual 
light curves in the two IR bands.
The observed light curves and best-fit model are compared in the left and right pairs of panels of Figure 9.  
The synchrotron emission at 4.80 and 2.10 
$\mu$m from the evolving, age-stratified electron population reproduces the observed light curves quite well. Table 7 lists the parameters of the 
best-fit model in each case.  The fits require a variety of injection rate shapes:  the rise may be convex or concave and the ratio of the time 
intervals for the rise and fall varies between 4 and 0.5. The inferred magnetic field strengths are in the range 38--92\,G, consistent with the field 
strength in GRMHD simulations of the inner accretion flow \citep{ressler20a,ressler20b}.
The source size cannot be directly determined because the emission is optically thin, but if magnetic reconnection is responsible for powering the 
flare, then one expects that the source volume encloses magnetic energy of order a few times the energy dumped into the particles and subsequently 
radiated away, ie $E\approx 10^{37.5}$ erg.  This implies a radius a few times $\sim (8\pi E/B^2)^{1/3} \approx 0.5\,r_g$.

\begin{table}
\caption{Flare models}
\begin{center}
\begin{tabular}{rrrrrrrrrrrrr}
    Date    & UT$_0$\,(hr) &   $S_{4.8}$\,(mJy) & $\Delta_1$\,(s) &  $\beta_1$ & $\Delta_2$\,(s) & $\beta_2$  & $E_c$\,(MeV) &$B$\,(G) & $r_\mathrm{min}/r_g$ & $n_{e,\mathrm{max}}$\,(cm$^{-3}$)\\
2023 April 13 &    7.44 &     4.83 &    489 &   0.98 &      220 &   0.92 &   424 &   92.7 &  0.32 & $ 5.21\times10^5$ \\
2023 April 13 &   11.96 &     4.04 &    841 &   0.05 &     1154 &   0.95 &   537 &   38.3 &  0.83 & $ 0.57\times10^4$ \\
2023 April 16 &   11.33 &     6.31 &    871 &   0.83 &      147 &   0.51 &   719 &   48.0 &  0.74 & $ 1.01\times10^4$ \\
2024 April 05 &    5.09 &     4.80 &    455 &   1.53 &     165 &   4.00 &   736 &  88.5 &  0.33 & $ 4.64\times10^5$ \\
2024 April 07 &    9.10 &     7.13 &   2798 &   1.83 &     1036 &   4.56 &   489 &   72.5 &  0.48 & $ 2.82\times10^4$
\end{tabular}
\end{center}
\label{table:flaremodels}
\end{table}

\subsection{The origin of two components to the NIR flux}

The piecewise-linear form of the flux-flux plots and the dual-component flux histograms indicate that there are two different processes contributing to 
the NIR variability.  The first is a faint, continual flickering with $S(2.1\micron)\la 3$\,mJy, and a steep spectral index $\alpha\sim -1.6$.  The 
second, bright component consists of isolated flares with $\alpha \sim -0.85$ and $S(2.1\micron)\ga 3 $\,mJy.

A notable feature of our observations is that the qualitative appearance of the NIR light curves varies from day to day (see, e.g., the upper and lower 
panels of Fig.~ 2).  The emission underlying the bright flares secularly increases over the course of the observations on the first and final days.  
This is reflected in the piecewise linear fits to the flux-flux plots on individual days (see Table 7), with the transition flux $S_0$ at 2.1\,\micron\ 
varying between about 2.4 and 4.3\,mJy.  Notably, the spectral index of the faint component only varies from $-1.4$ to $-1.8$ between the different 
epochs, consistent with electron power-laws in the range $E^{-3.8}$ -- $E^{-4.6}$.  The flaring component also shows day-to-day variations: the spectral 
index varies between $-0.63$ and $-1$ (see Table 7).  In addition, the flares on the final two days have notably larger amplitude than the earlier days.

A plausible scenario is that the faint component is that synchrotron emission from a high-energy power-law tail (typically $\sim E^{-4.2}$) of the 
electron population in the inner accretion flow, and that this is moderated by the turbulent fluctuations in density and magnetic field strength. 
(e.g.~Gregorian \& Dexter 2024). This implies that the accretion flow evolves significantly on $\sim10$\,hr time scales, as is apparent in many GRMHD 
simulations of Sgr A*. Our data suggests that the bright flare does not occur independently of the fainter variations suggesting that non-thermal 
electrons are accelerated by occasional reconnection events within the bulk of the accretion flow (e.g., \cite{scepi22,dimitro24,grigorian24} or, 
alternatively, from reconnection-driven ejection of plasmoids out of the disk plane (e.g., \citep{aimar23}).

\section{Summary}

Simultaneous observation at two NIR wavelengths in seven epochs have accurately characterized the IR flare's spectral evolution. This tracks the evolution 
of the accelerated electron spectrum, have yielded information about the particle acceleration process, and the subsequent cooling of the highest energy 
particles. The additional information of detecting a flaring event at two wavelengths has provided estimates of the magnetic field strength, and electron 
density of hotspots in the accretion flow. Our findings include:\\

$\bullet$ 
Many measurements  have found  log-normal or power-law or a  combinations of LN and PL in the past. 
Our data are  not consistent with single log-normal or tailed log-normal.  
The result shows the presence of a bright component that is distinct from the fainter component, 
suggesting  2-states described by  two LN distributions for faint and bright variability of Sgr A* \citep{dodds-eden09}.
\vspace{0.2in}

$\bullet$ 
The power spectrum of the NIR variability follows a red-noise process with an index that varies in each 
epoch ranging between -3.57 and -2.70. 
Across the seven days, the power law indices have mean  values of $-3.0\pm0.3$ at 2.1 $\mu$m,
and $-3.0\pm0.4$ at 4.8 $\mu$m. 
The power spectra   break, where the red-noise turns to  white noise,   
at frequencies ranging  between 0.0060 and 0.0247 min$^{-1}$  corresponding to 166 and 40 min, respectively. 
After the removal of the white noise at high frequencies, we note 
the highest  frequencies are estimated 
between  0.45 and  2.34 min$^{-1}$ corresponding to time scales of  $\sim$2.2 and 0.43 min, respectively.  
\vspace{0.2in}

$\bullet$ 
Reaching the white noise level at high frequencies implies  that
a faster sampling rate cannot reveal the power law at high frequencies
because shorter exposures have insufficient sensitivity for precise
flux density measurements. Investigation of the intrinsic variability at 
higher frequencies might be possible by using wide band filters rather than 
medium band filters. Longer wavelength observations (where Sgr A* is brighter)
could also help, but confusion may be  more of an issue with the lower
angular resolution.   
\vspace{0.2in}

$\bullet$ 
There is evidence of intensity gradient showing 
flux density variation by a factor two in $\le$2 min. 
This suggests that the variability is arising from small region of the accretion flow on a scale of few gravitational radii or less. 
\vspace{0.2in}

$\bullet$ 
A trend that we note in multiple flares is that  subflares  become increasingly stronger on the rising and or the falling sides of flaring events. 
If the subflares and flares are physically associated with each other, 
then subflares are precursors of bright flares.
\vspace{0.2in}

$\bullet$ 
The light curves also show pedestal emission  with amplitude that vary between each epoch suggesting that there is long term variability of Sgr A* 
on daily, monthly and yearly time scales. The 
pedestal is varying roughly by a factor of 2 over seven epochs of observations. 
\vspace{0.2in}

$\bullet$ 
The flux density histogram 
suggests two populations of faint and bright emission with two peaks near 2-3 and 10 mJy at 2.1  and 4.8 $\mu$m, respectively.  
\vspace{0.2in}

$\bullet$ 
We note an  anti-correlation of the spectral index with brightness (up 
to a limit), before it changes and shows a shallower spectral index with increasing brightness. 
The spectral index eventually saturates with 
increasing brightness with values close to $\alpha\sim-1$.
\vspace{0.2in}

$\bullet$ 
The cross correlation of the light curves at 2.1 and 4.8 $\mu$m  
show a systematic few-second  delay in the 4.8 $\mu$m variable emission with respect to that 
of 2.1 $\mu$m in all  observations. 
Another characteristics of the observed flares is the presence of loops in the flux density vs. spectral index plots. The loop feature is 
noted in individual spectral index plots and is a loop pattern revealing the spectral index increase and decrease as a 
flare emission rises and decays. 
\vspace{0.2in}

$\bullet$ 
The power spectrum shows no evidence of periodicity but there is evidence of turnover at low frequencies in all our observations.  
If the power spectrum didn't turn over at low frequencies, the slow variations in the pedestal would be larger than the flares. 
\vspace{0.2in}

$\bullet$ 
We suggested that the frequent low-amplitude variability arises from a high-energy power-law tail of the electron population in the accretion 
flow, with the fluctuations arising due to bright spots where turbulent motions or convection has temporarily compressed the plasma.  The less common 
bright flares may be due to occasional magnetic reconnection events in the flow, as has been suggested by several authors (see the discussion \S 4.2). 
\vspace{0.2in}

$\bullet$ 
Our determination of the spectral index of each component, and the power spectra represent observational signatures that models of these event need to 
reproduce.  In addition,  models of the bright flares must be able to explain the short rise time of bright flare events (i.e. $\sim 2$\,min), and the 
duration and spectral “loop” evolution that is consistently found in our measurements. \vspace{0.2in}

$\bullet$ 
A trend that we note is that subflares are precursors to strong flares. 
We noted some apparent qualitative behavior that we were unable to quantify: smaller subflares appear to be precursors to strong flares, and the 
underlying pedestal emission is often increased immediately after a flare.  These may be signatures of correlated activity: for example the emission from 
bright flares may feed back on the accretion flow to enhance its NIR emission.

\begin{acknowledgments}
{Some of the data presented in this article were obtained from
the Mikulski Archive for Space Telescopes (MAST) at the Space Telescope Science Institute.
All the {\it JWST}  data used in this paper can be found in MAST: doi:\dataset[10.17909/0z7e-gf46]{https://dx.doi.org/10.17909/0z7e-gf46}.
This work is based on observations made with the NASA/
ESA/CSA James Webb Space Telescope. The data were
obtained from the Mikulski Archive for Space Telescopes at
the Space Telescope Science Institute, which is operated by the
Association of Universities for Research in Astronomy, Inc.,
under NASA contract NAS 5-03127 for JWST. These
observations are associated with  NASA's JWST-2235 and JWST-3559 programs.
This work is also partially supported by the grants AST-2305857 from the NSF.
JMM is supported by an NSF Astronomy and Astrophysics Postdoctoral Fellowship under award AST-2401752.
The National Radio Astronomy Observatory is a facility of the National Science
 Foundation operated under cooperative agreement by Associated Universities, Inc.
Work by R.G.A. was supported by NASA under award number 80GSFC24M0006.
Lastly, we thank  S. Markoff, G. Bower, D. Haggard and K. Hada for useful  comments.}  
\end{acknowledgments}

\appendix

\section{Polynomial Fits to Pedestal Emission}
%\vspace{0.2in}

Fourth order was chosen purely subjectively, in order to allow more freedom 
for variation (changes in slope) during the interval while avoiding too much freedom that 
would allow the pedestal  to start tracing the behavior of the larger flaring events.
On each day of observations a lower envelope $f'(t)$ was fit to
the light curves $f(t)$ using a fourth-degree polynomial.
The choice of degree was subjective, with the intent
to allow more freedom for variation than a simple linear fit,
while avoiding too much freedom that would allow the
fit to start tracing the behavior of the larger flaring events.
The exact procedure used is that the polynomial fit is specified by
\begin{equation}
f'(t) = a_0 + \sum_{i=1}^4 a_i t^i
\end{equation}
with the constraint that
\begin{equation} 
a_0 = \min\left(f(t)-\sum_{i=1}^4 a_i t^i\right)
\end{equation}
such that
\begin{equation}
f(t)-f'(t) \ge 0.
\end{equation}
The IDL procedure MPFITFUN \citep{2009ASPC..411..251M}
was used to optimize the parameters [$a_1$, $a_2$, $a_3$, $a_4$]
to minimize $\chi^2$.

\section{The Spectral Index vs. the  Flux Density for Individual Days}

Figures 10-15  are very similar Figure 7 showing four  panels describing the correlation and anti-correlation of the spectral index vs flux density for Day 1 to Day 7 
(excluding Day 3). These figures show that the slope of the faint flux is steeper than that of the bright flux,
suggesting two different populations with different spectral indices.
For each  day of observation, the four panels 
display the flux density at 2.1 $\mu$m vs 4.8 $\mu$m, the colored light curves, 
the spectral index as a function of 2.1 $\mu$m, and the colored spectral index as a function of time. 
Similar trends noted for combined data can also be seen for individual data sets.

\section{Complex Loop Structures}
%\vspace{0.2in}

Loop structures reveal  the spectral index increase and decrease as a 
flare emission rises and decays.  
Figure 9 shows  
the variation of the spectral index as a function of F480M flux density for six complex subflares.  
In all these cases, the variations of 
the spectral index trace counterclockwise loops. 
The  right panels show the light curves of 
individual colored flares and subflares  on Days 4-7. 
We note low-level fluctuations or subflares are superimposed on these flares, thus harder to model. 
 The left panels 
display 4.8 $\mu$m flux density of these events  as a function of the spectral index. Each fluctuation shows a counter-clockwise loop pattern as 
the flux density rises and decays. The pattern shows steepening of the spectral index as 4.8 $\mu$m flux density rises followed by flattening of 
the spectral index as the flux density decreases. There are counter-clockwise loops corresponding to each subflare.

\section{Electron Population in an Evolving Injection Rate and Magnetic Field}

Here we derive expressions for the electron population in a region given an evolving injection rate $Q(E,t)$ (electrons cm$^{-3}$\,erg$^{-1}$\,s$^{-1}$) 
and subject to synchrotron losses in a magnetic field $B(t)$.  We start by considering the energy lost by a single electron, from this we derive how a 
population evolves due to synchrotron losses.

An ultra-relativistic electron with energy $E$ suffers synchrotron losses at rate 
\begin{equation}
    \frac{dE}{dt} = -b B^2 E^2\,,
    \label{eqn:Edot}
\end{equation}
where 
\begin{equation}
    b = \frac{4e^3}{9m_e^4c^5}\,.
    \label{eqn:b}
\end{equation}

For simplicity we have neglected the dispersion in loss rates due to different pitch angles, and adopted the loss rate averaged over an isotropic 
pitch-angle distribution.  Thus, at time $t$, an electron of age $\tau$ with energy $E$ was injected at time $t-\tau$ with initial energy $E_0$, where
\begin{equation}
    \frac{1}{E_0} = \frac{1}{E} - b \int_0^\tau [B(t-\tau')]^2\,d\tau'\,.
    \label{eqn:E0vsE}
\end{equation}

Now consider a cohort of electrons, $N(E,t)$ (electrons\,erg$^{-1}$), that were injected at time $t-\tau$, and now have age $\tau$.  The electrons that 
were injected with initial energy in an interval $[E_0,E_0+dE_0]$ are now in the interval $[E,E+dE]$ , where $E$ and $E_0$ are given by Eq. 
(\ref{eqn:E0vsE}).  Thus the initial and present populations are related via $N_0(E_0,t-\tau)\,dE_0 = N(E,t)\,dE$, where $dE_0/E_0^2 = dE/E^2$ and so the 
initial and current populations are related via $N(E,t) = E_0^2\,N(E_0,t-\tau)/E^2$.

For injection rate $Q(E,t)$, the initial population of electrons with ages in the interval $[\tau,\tau+d\tau]$ is $N(E_0)=Q(E_0,t-\tau)\,d\tau$ and the 
contribution to the population at time $t$ is $E_0^2\,Q(E_0,t-\tau)\,d\tau/E^2$.  The net population at time $t$ is obtained by integrating over all ages, 
i.e.
\begin{equation}
    N(E,t) = \frac{1}{E^2}\int_0^\infty \!\! E_0^2\,Q(E_0,t-\tau) \,d\tau\,.
    \label{eqn:NEt}
\end{equation}
Note that $E_0$ depends on $\tau$ via eq \ref{eqn:E0vsE}.

The synchrotron luminosity of the population is, assuming an isotropic electron pitch-angle distribution and averaging over an isotopic distribution of 
field orientations,
\begin{equation}
   L_\nu  = \frac{\sqrt{3}e^3 B}{m_e c^2}\int_0^\infty\!\!N(E,t)\,G(x)\,dE
    \label{eqn:Lnu}
\end{equation}
where 
\begin{equation}
    x = \frac{4\pi m_e c \,\nu}{3eB} \left(\frac{E}{m_e c^2}\right)^{\!\!-2}
    \label{eqn:x}
\end{equation}
and 
$G(x)$ can be approximated to within 0.2\% by
\begin{equation}
    G(x) \approx
    \frac{1.808\,x^{1/3}}{\sqrt{1+3.4\,x^{2/3}}}\;
    \frac{1+\,2.21\,x^{2/3} +0.347\,x^{4/3}}
                                      {1+1.353\,x^{2/3}+0.217\,x^{4/3}} \;\exp(-x)
    \label{eqn:Gapprox}
\end{equation}.

\begin{equation}
    R(t) = \begin{cases}
    \left(\frac{t-t_0}{t_1-t_0}\right)^{\beta_1} & \text{if}\ t_0<t<t_1\,,\\
    \left(\frac{t_2-t}{t_2-t_1}\right)^{\beta_2} & \text{if}\ t_1<t<t_2\,, \text{or}\\
    \quad0 & \text{otherwise\,.}
    \end{cases}
    \label{eqn:R}
\end{equation}

For simplicity, we assume that $B$ is constant, and adopt an injection rate
\begin{equation}
    Q(E,t) \;=\; Q_0 \, R(t)\left(\!\frac{E}{E_1}\right)^{-p} \text{ for } E_1<E<E_c\,,
    \label{eqn:Q}
\end{equation}
and 0\,  otherwise.  We set $E_1 = 2$\,MeV and $p=2$, and treat the cutoff energy $E_c$ as a free parameter.  The injection envelope,
rises monotonically from 0 to 1 between $t=t_0$ and $t_1$, then declines monotonically to 0 at $t=t_2$; $\beta_1$ and $\beta_2$ control the shape of the 
rise and fall.

Once the 6 free parameters -- $B$, $Q_0$, $E_c$, $t_0$, $t_1$, and $t_2$ -- are specified, we compute the light curves $S_\nu(t) = L_\nu / (4\pi d^2)$ 
using Eq. \ref{eqn:Lnu} at 143 and 62.5\,THz, i.e. the frequencies equivalent to 2.1 and 4.8\,\micron, respectively.

%f7a
%\addtocounter{figure}{-1}
%\stepcounter{subfig}
%\begin{figure}[htbp]

%\setcounter{subfig}{1}
\begin{figure}[htbp] 
   \centering
   \includegraphics[width=3in]{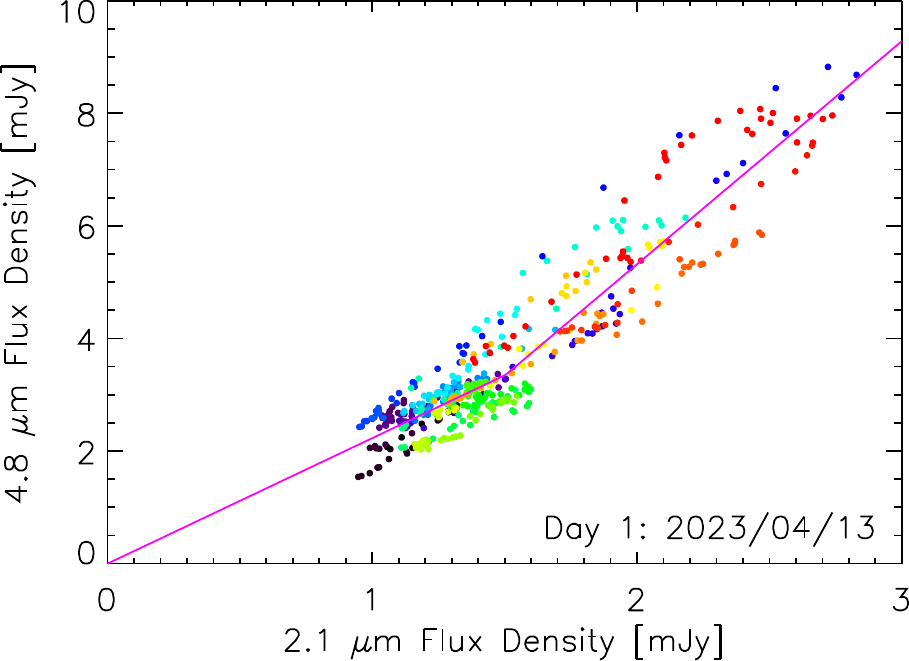}
  \includegraphics[width=3in]{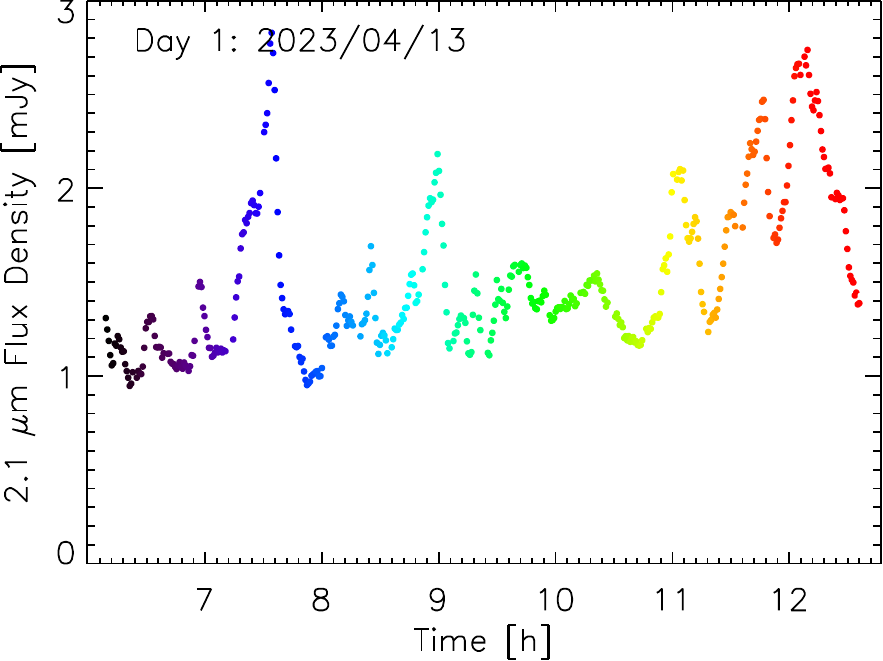}
   \includegraphics[width=3in]{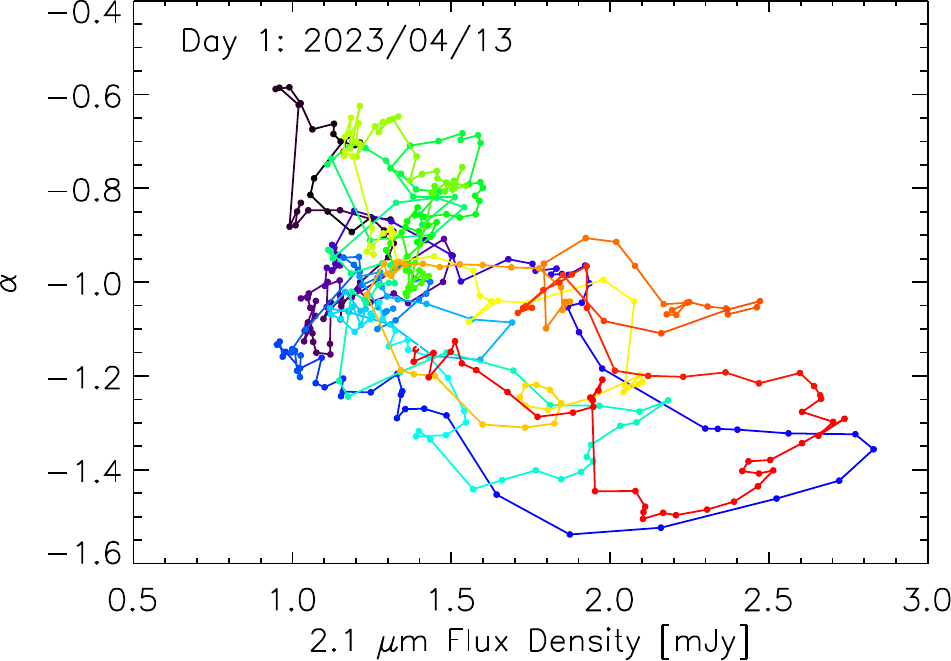}
   \includegraphics[width=3in]{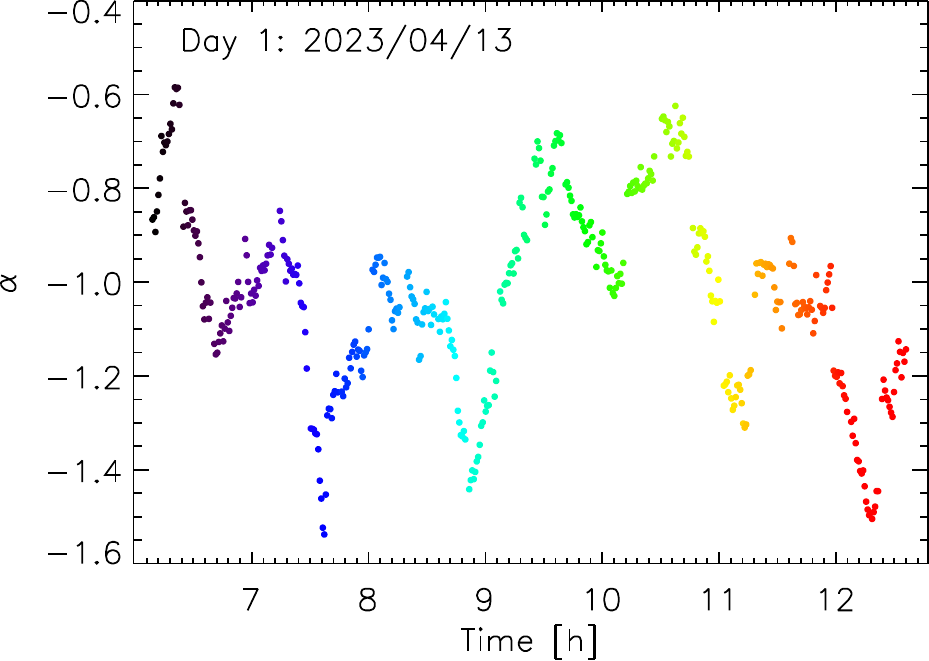}
   \caption{
Similar to Figure 7  where data for all days are shown. 
The top left panel shows the colored variation of the 4.8 $\mu$m flux against the 2.1 $\mu$m flux density. 
The top right  panel shows light curves of Sgr A* in color at 2.1 $\mu$m. 
The bottom left   panel shows 
the  spectral index of Sgr A* in color as a function 2.1 $\mu$m flux density. 
The bottom right panel shows the spectral index as a function of each time in color. 
} 
\end{figure}

%f7b
%\addtocounter{figure}{-1}
%\stepcounter{subfig}
\begin{figure}[htbp]
   \centering
   \includegraphics[width=3in]{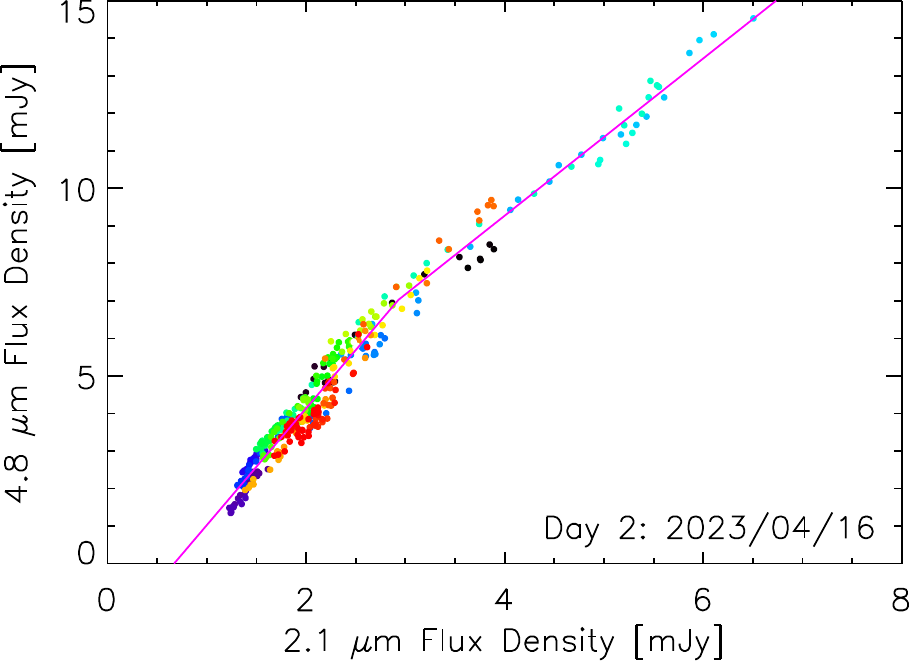}
   \includegraphics[width=3in]{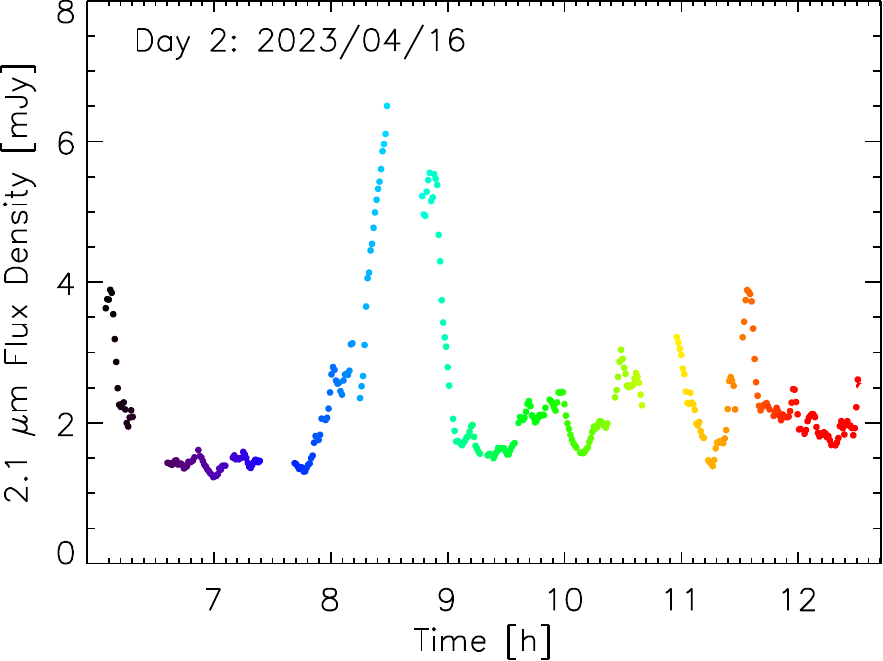}
   \includegraphics[width=3in]{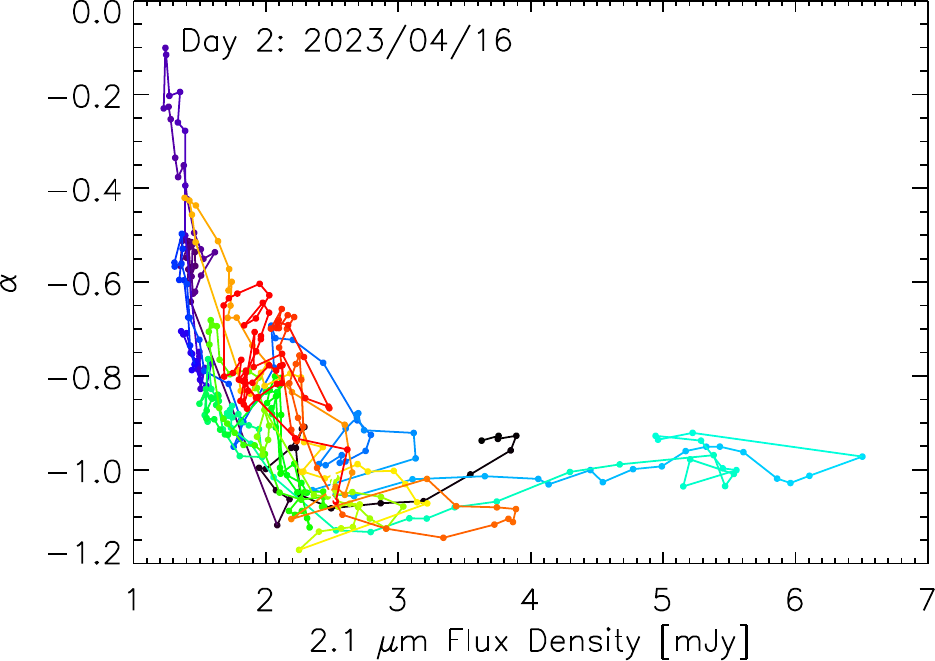}
   \includegraphics[width=3in]{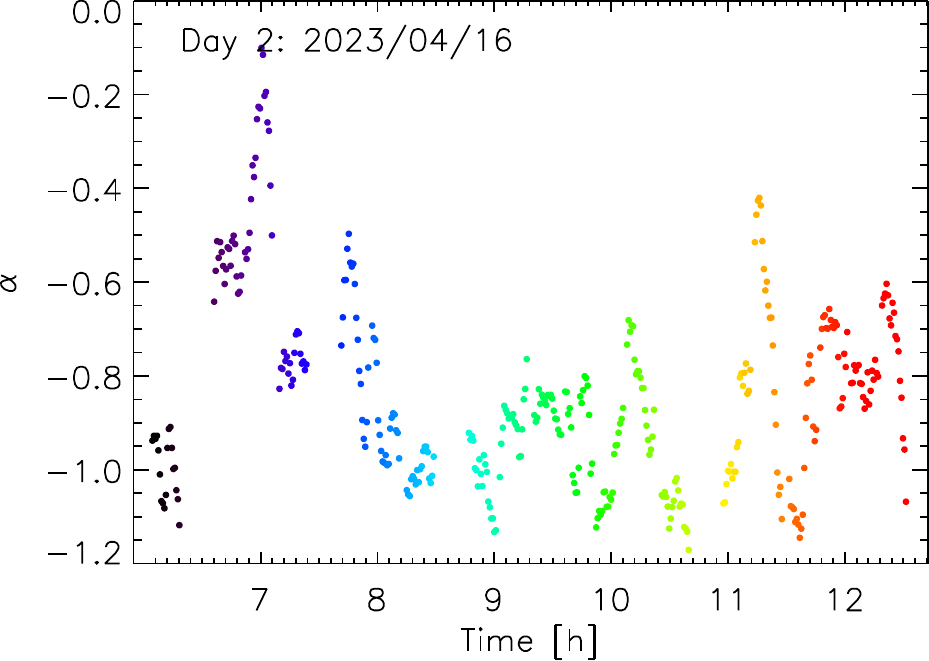}
   \caption{
Same as  Figure 7 but showing Day 2 data.
}
\end{figure}

%f7c
%\addtocounter{figure}{-1}
%\stepcounter{subfig}
\begin{figure}[htbp]
   \centering
   \includegraphics[width=3in]{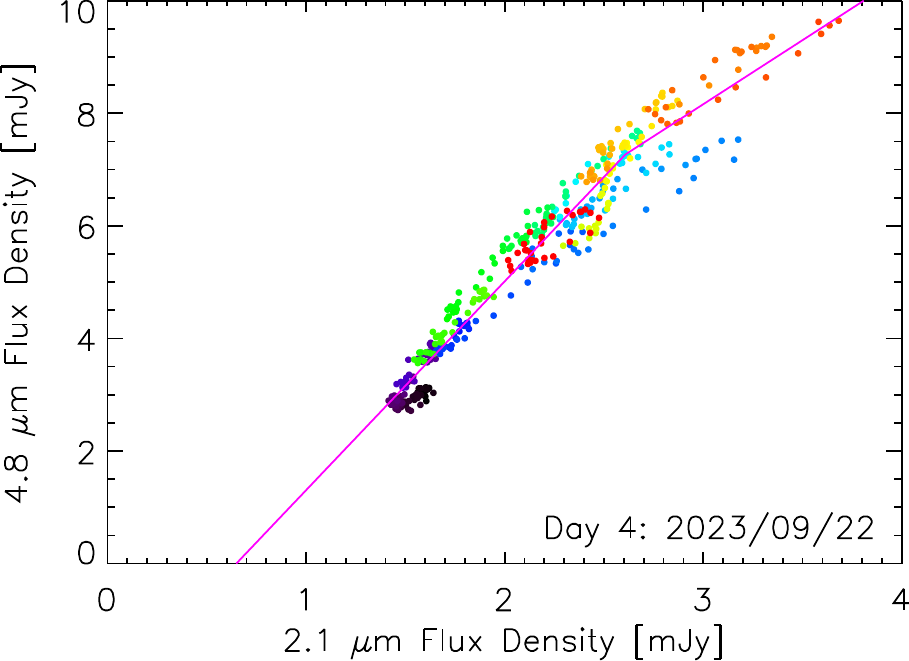}
   \includegraphics[width=3in]{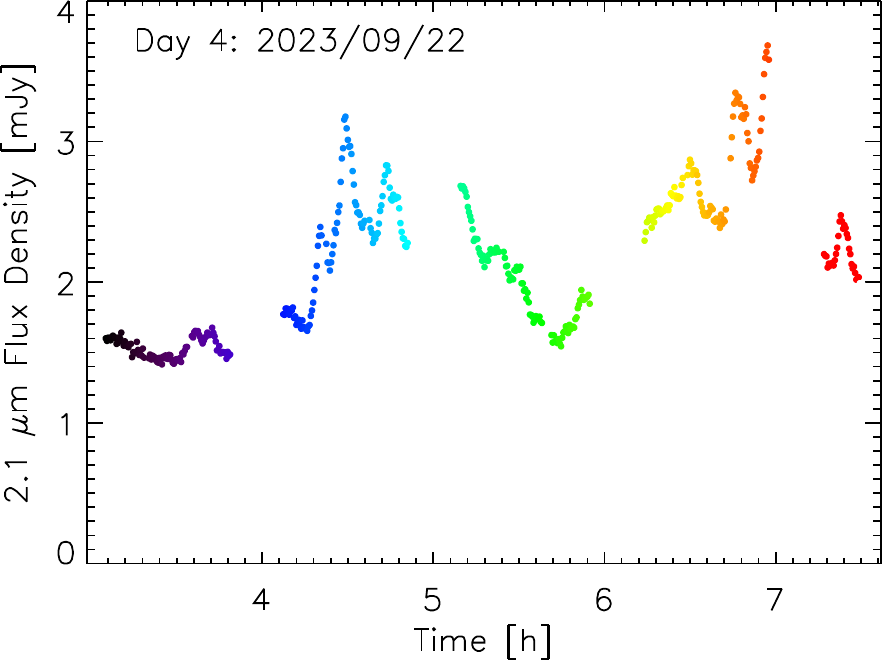}\\
   \includegraphics[width=3in]{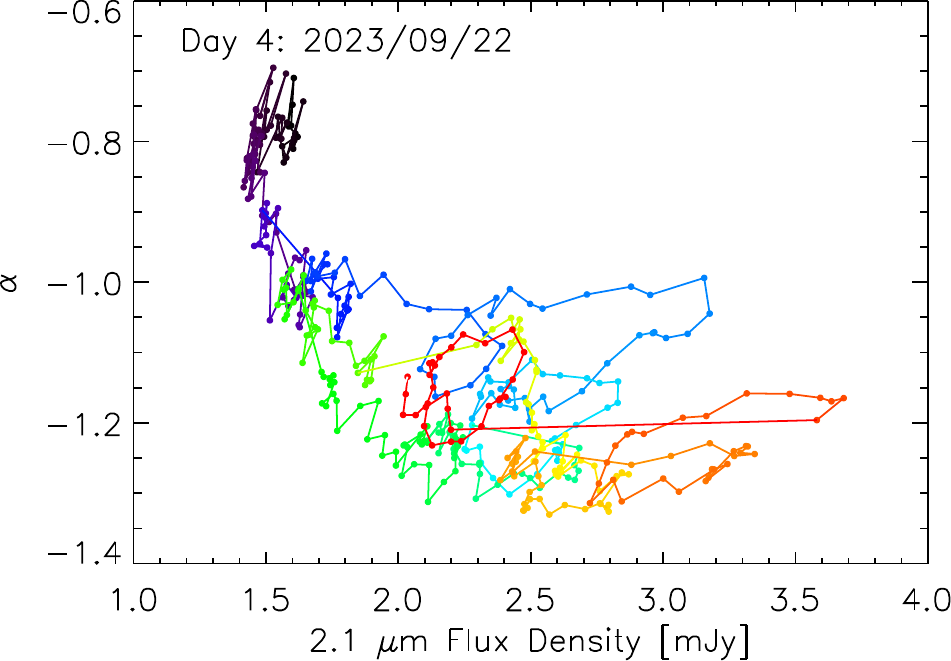}
   \includegraphics[width=3in]{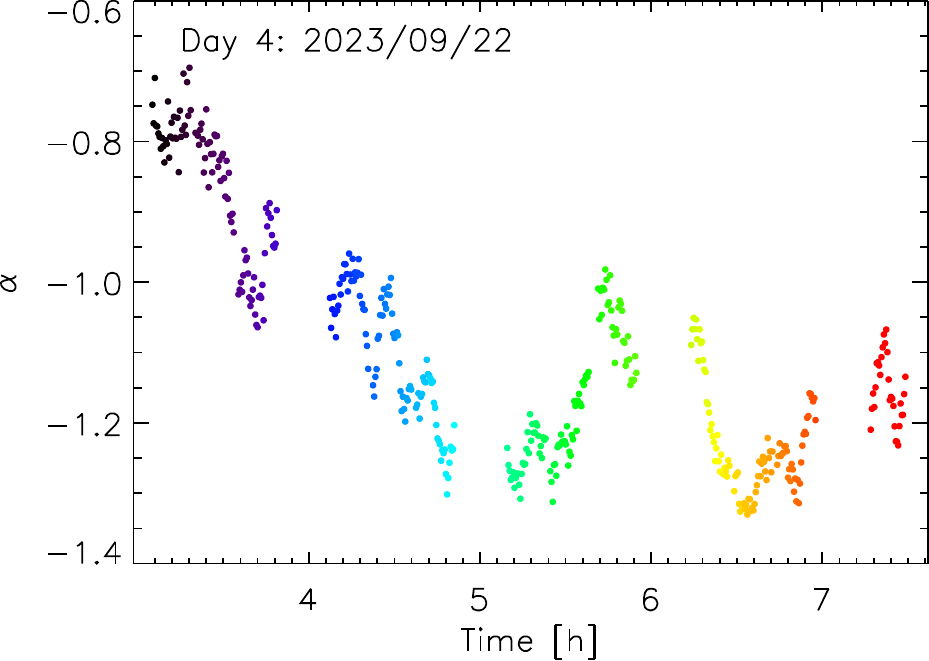}
   \caption{
Same as  Figure 7 but showing Day 4 data.
}
\end{figure}

%f7d
%\addtocounter{figure}{-1}
%\stepcounter{subfig}
\begin{figure}[htbp]
   \centering
   \includegraphics[width=3in]{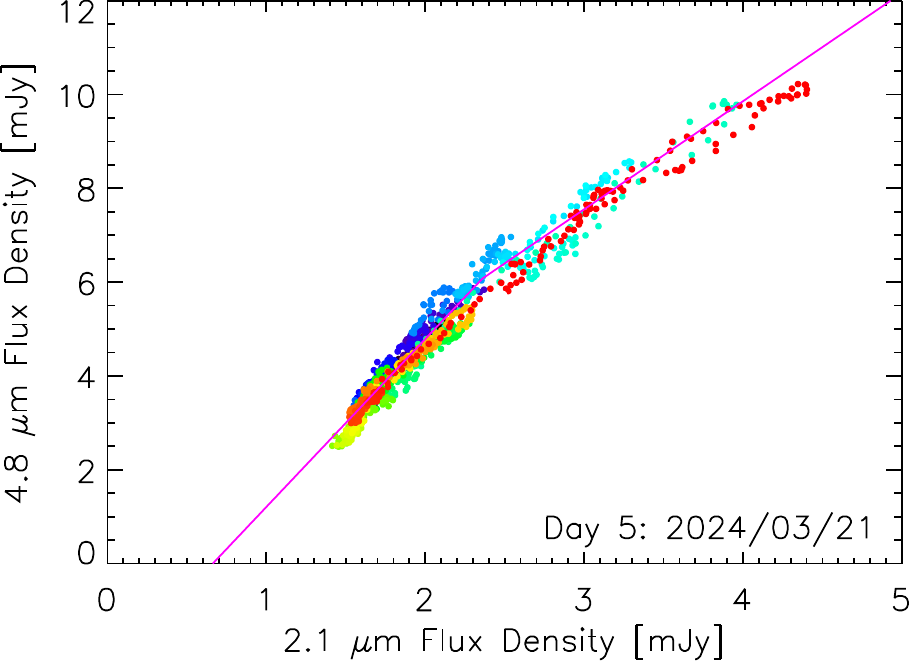}
   \includegraphics[width=3in]{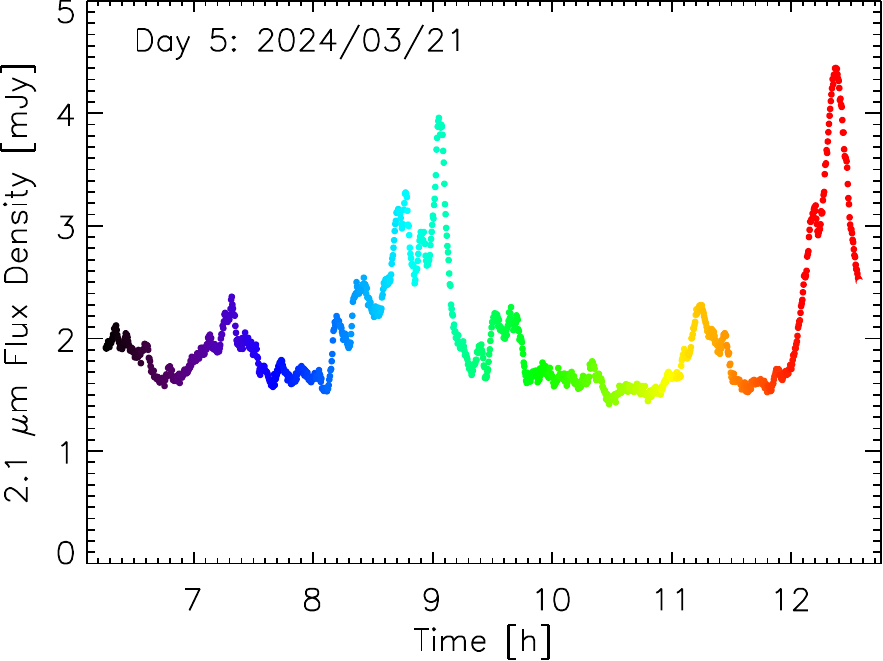}
   \includegraphics[width=3in]{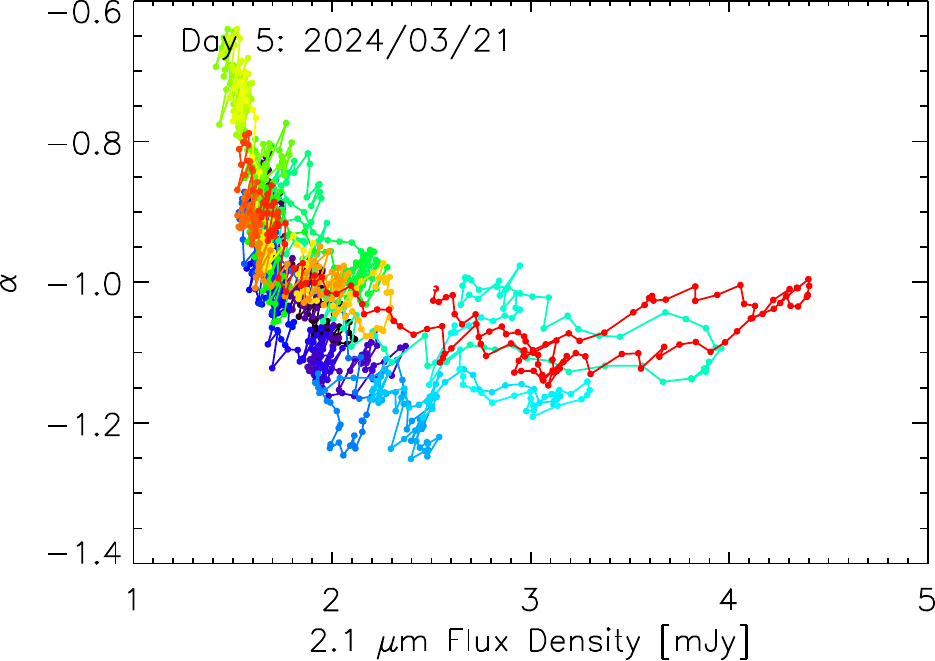}
   \includegraphics[width=3in]{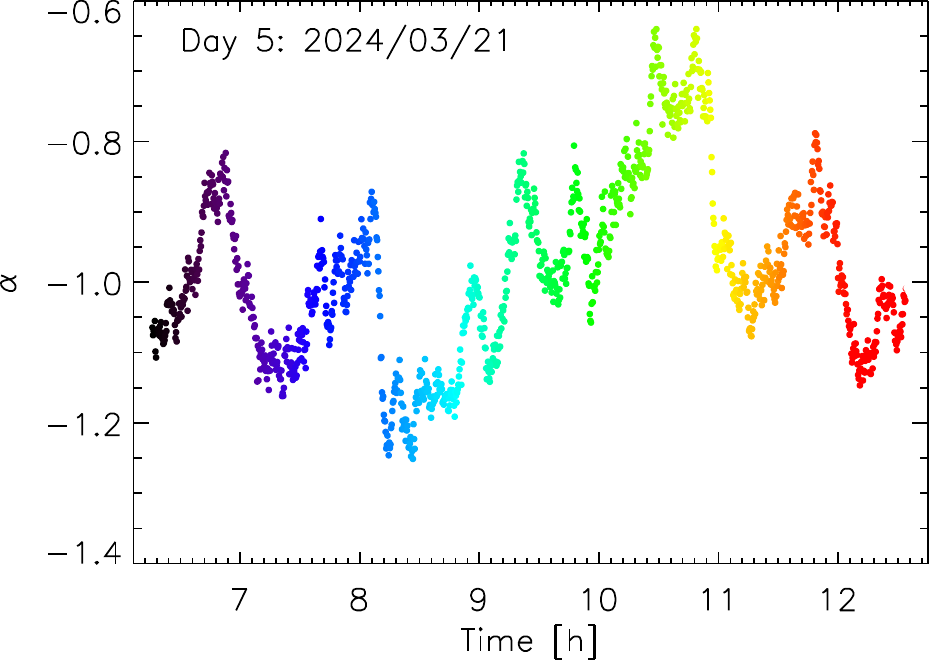}
   \caption{
Same as  Figure 7  but showing Day 5 data.}
\end{figure}

%f7e
%\addtocounter{figure}{-1}
%\stepcounter{subfig}
\begin{figure}[htbp]
   \centering
   \includegraphics[width=3in]{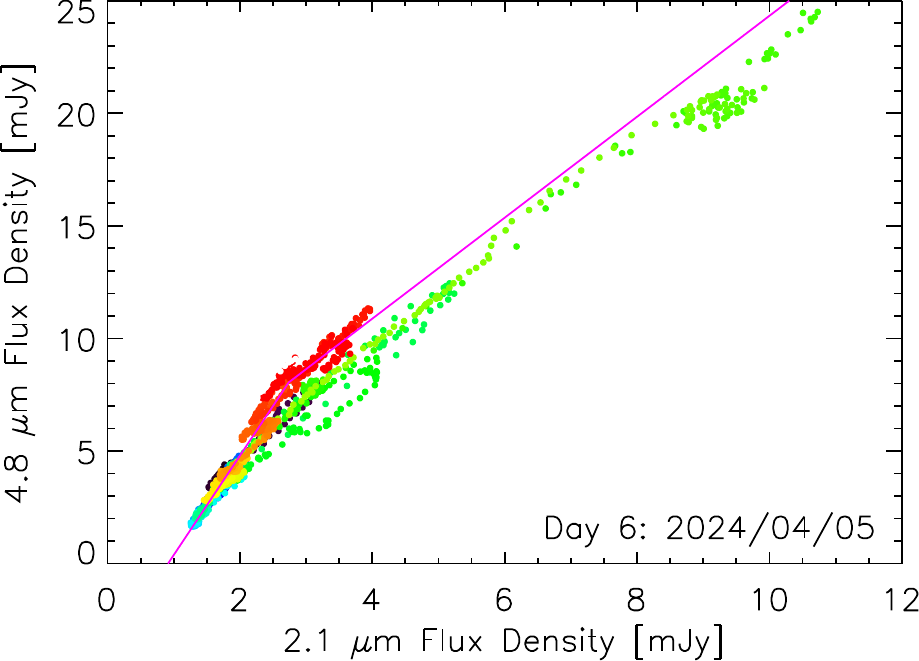}
   \includegraphics[width=3in]{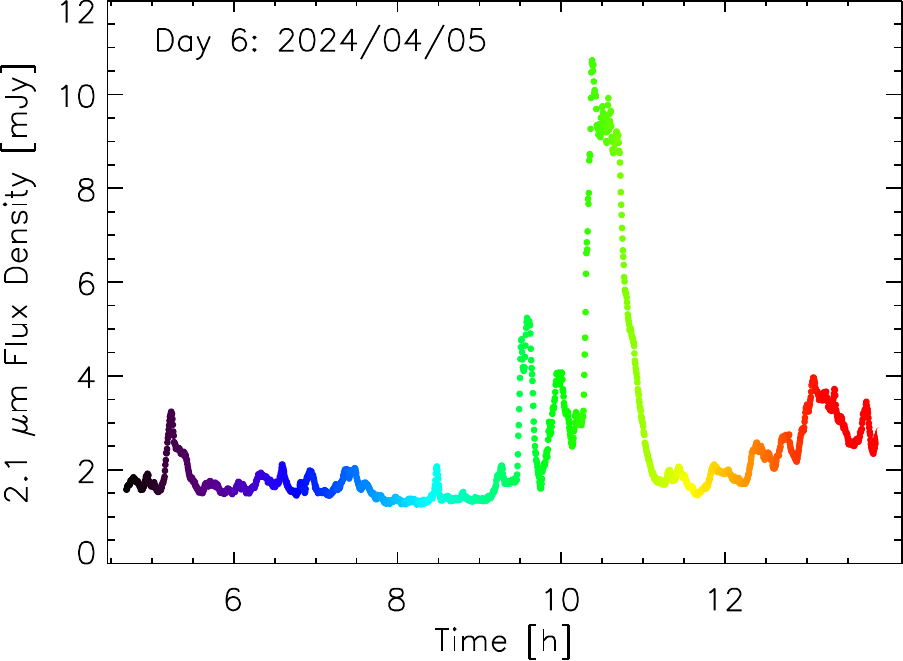}
   \includegraphics[width=3in]{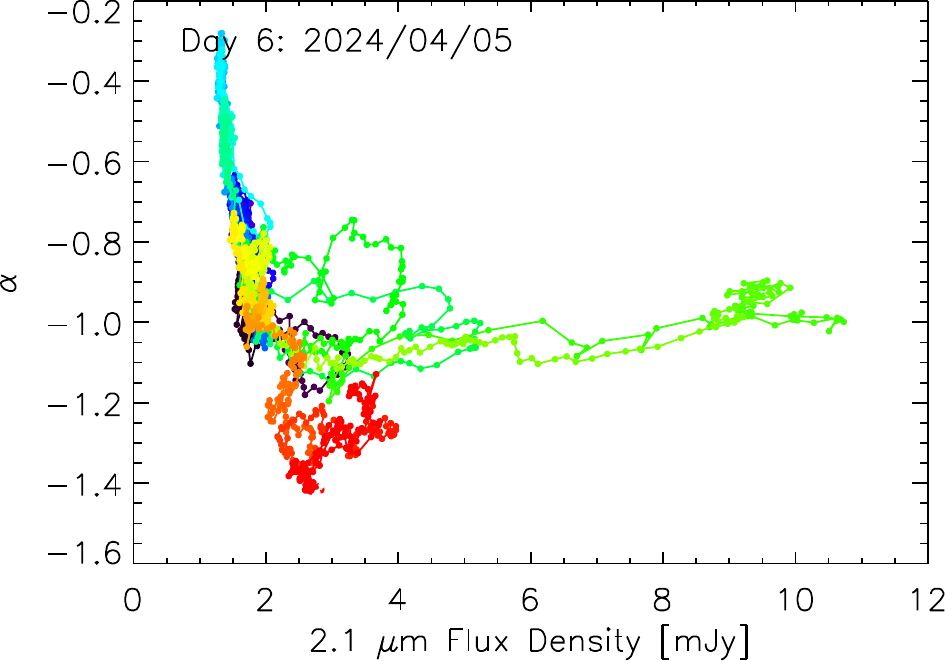}
   \includegraphics[width=3in]{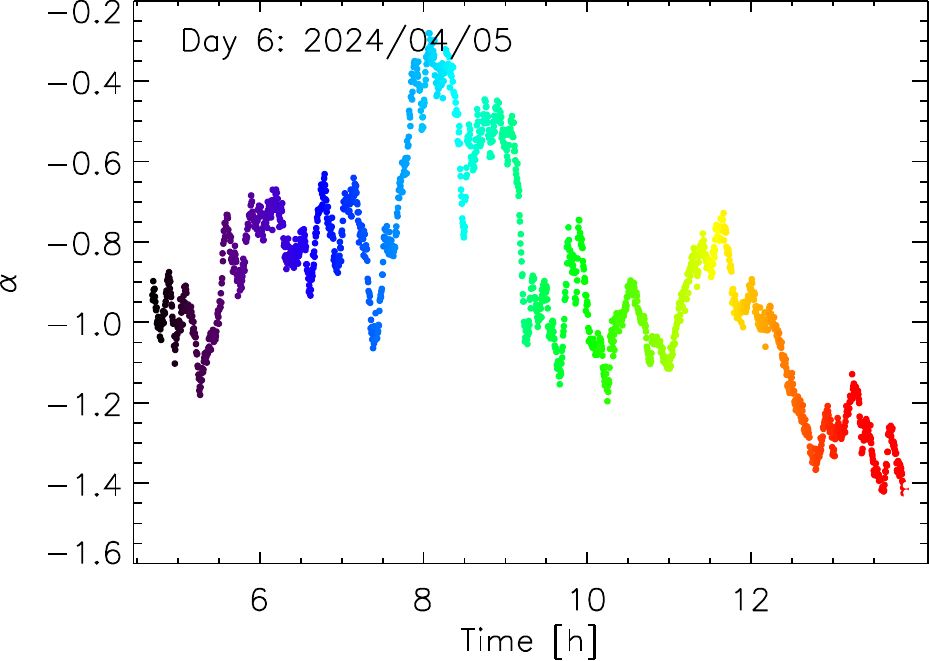}
   \caption{
Same as  Figure 7  but showing Day 6 data.
}
\end{figure}

%f7f
%\addtocounter{figure}{-1}
%\stepcounter{subfig}
\begin{figure}[htbp]
   \centering
 %    \vspace{-1.0in}
   \includegraphics[width=3in]{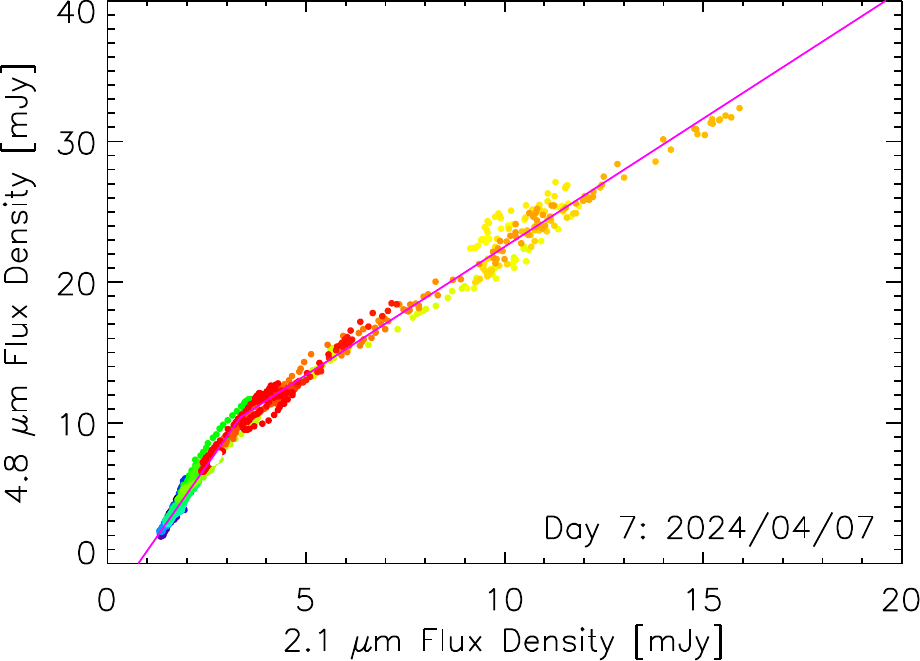}
   \includegraphics[width=3in]{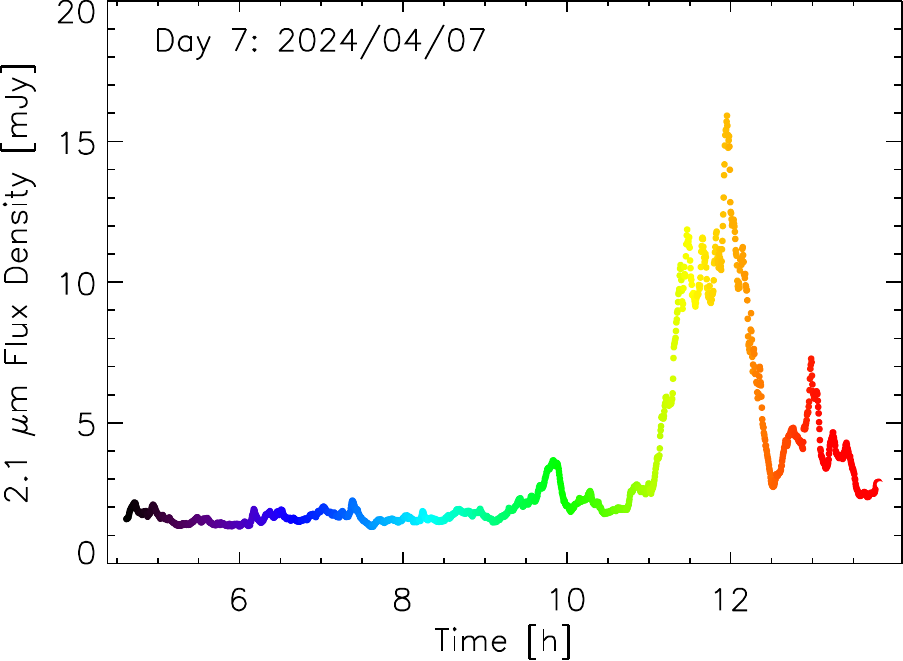}
   \includegraphics[width=3in]{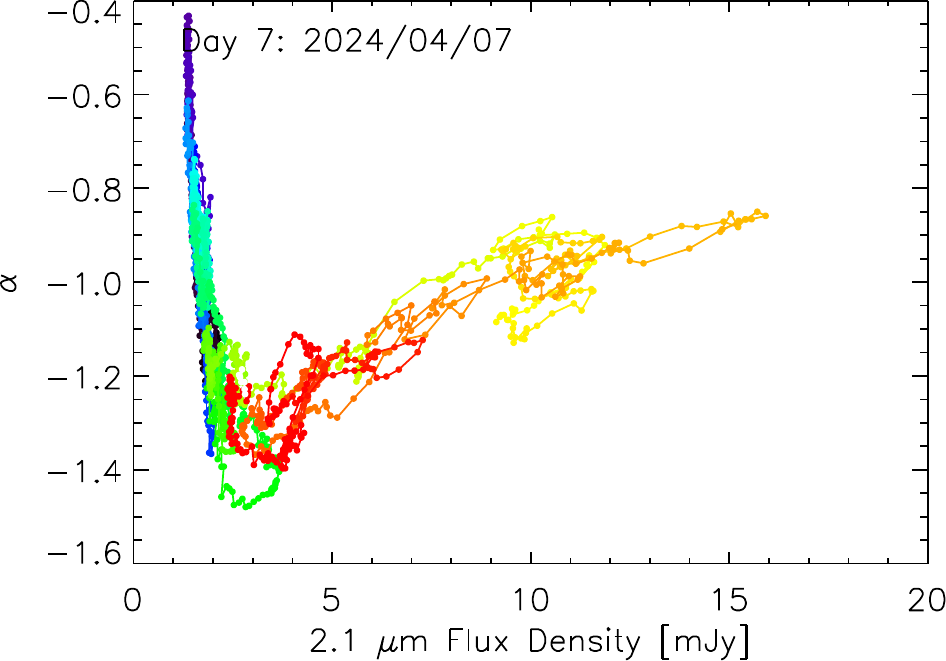}
   \includegraphics[width=3in]{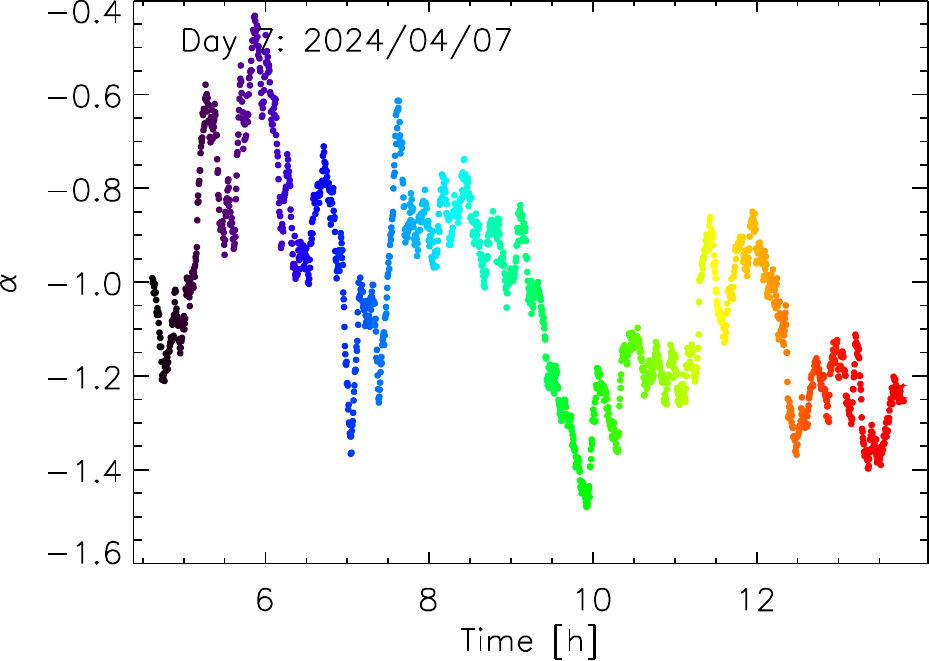}
   \caption{
Same as  Figure 7  but showing Day 7 data.
}
\end{figure}

%f8
%\setcounter{subfig}{0}
\begin{figure}[htbp] 
   \includegraphics[width=3in]{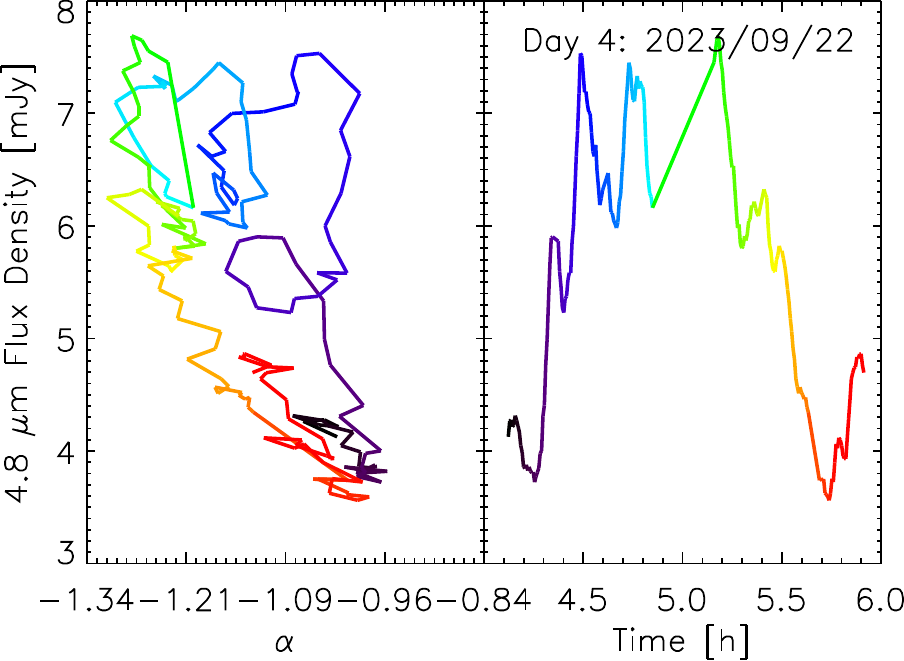}
   \includegraphics[width=3in]{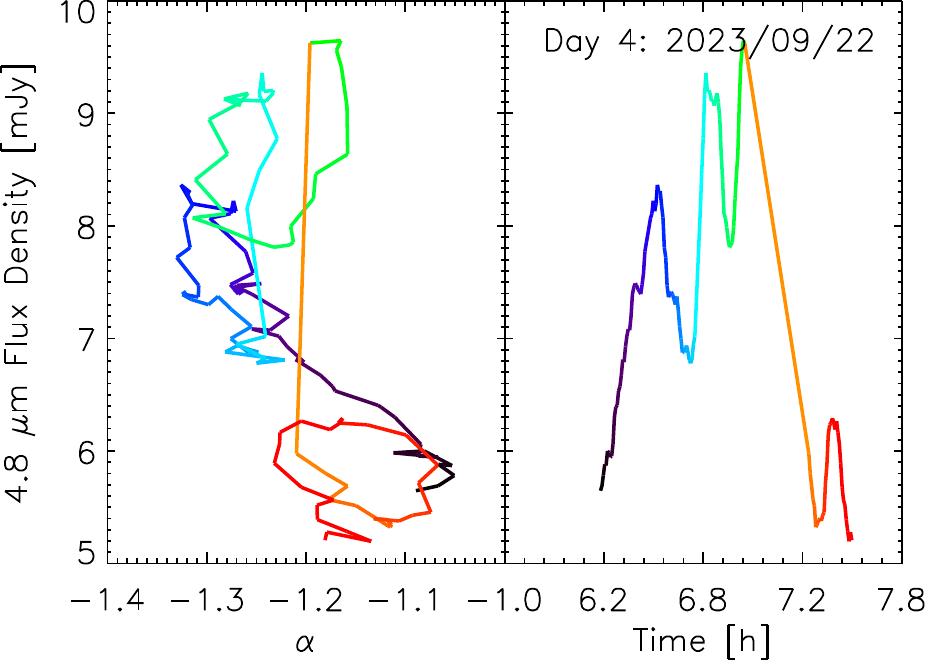}\\
   \includegraphics[width=3in]{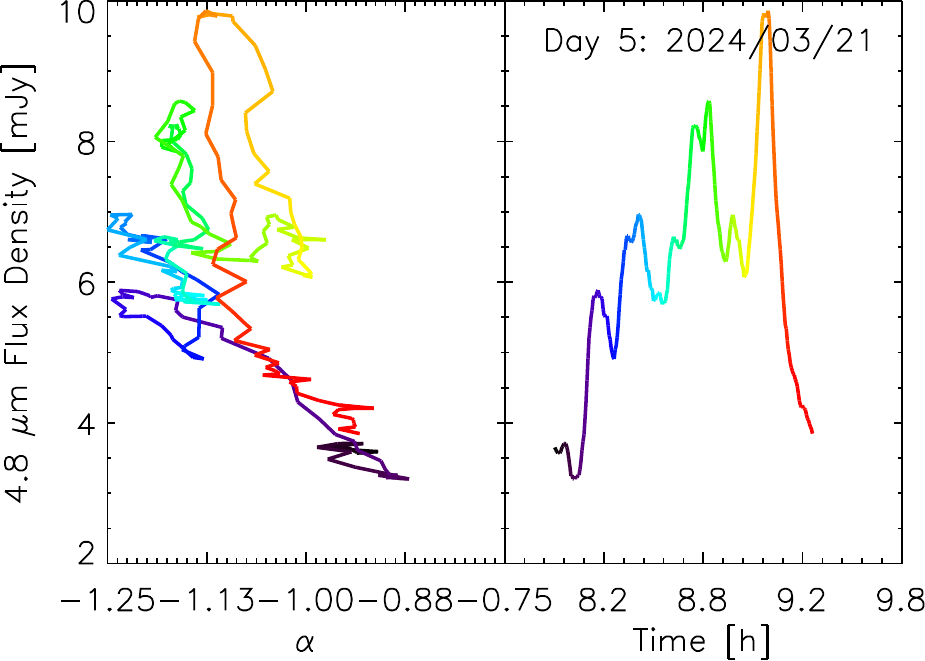}
   \includegraphics[width=3in]{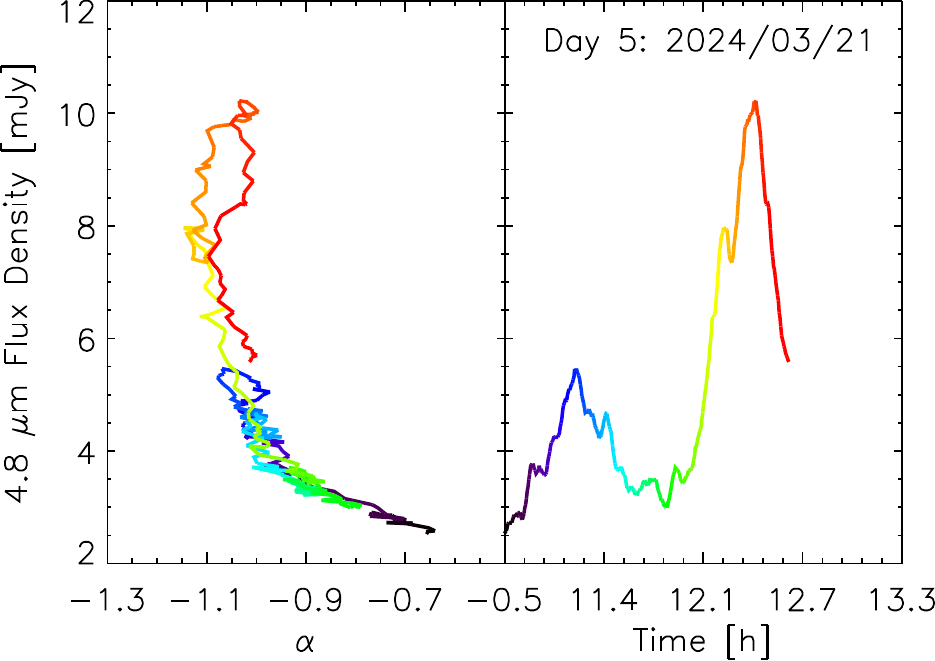}\\
   \includegraphics[width=3in]{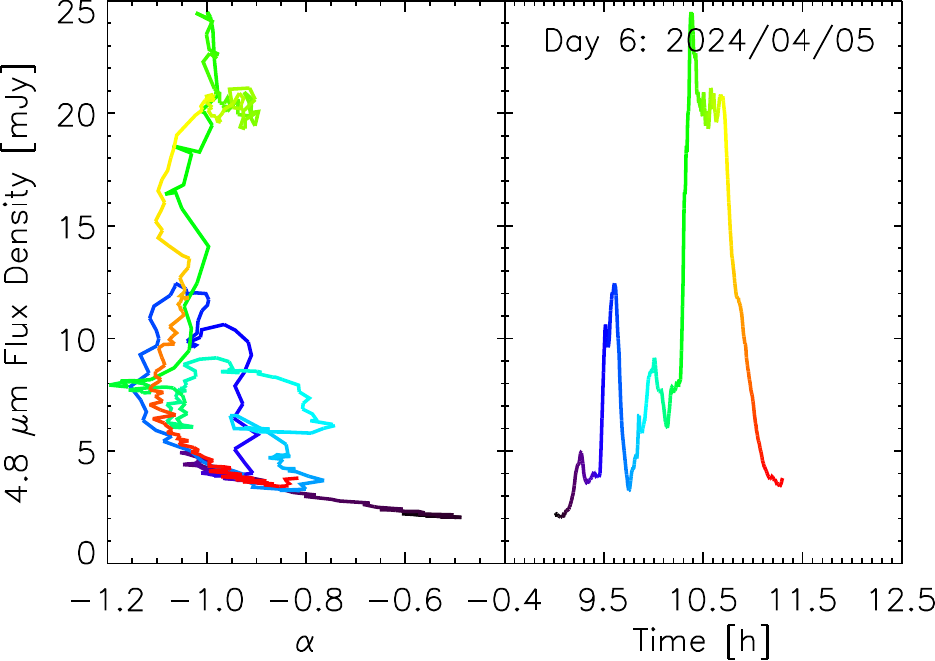}
   \includegraphics[width=3in]{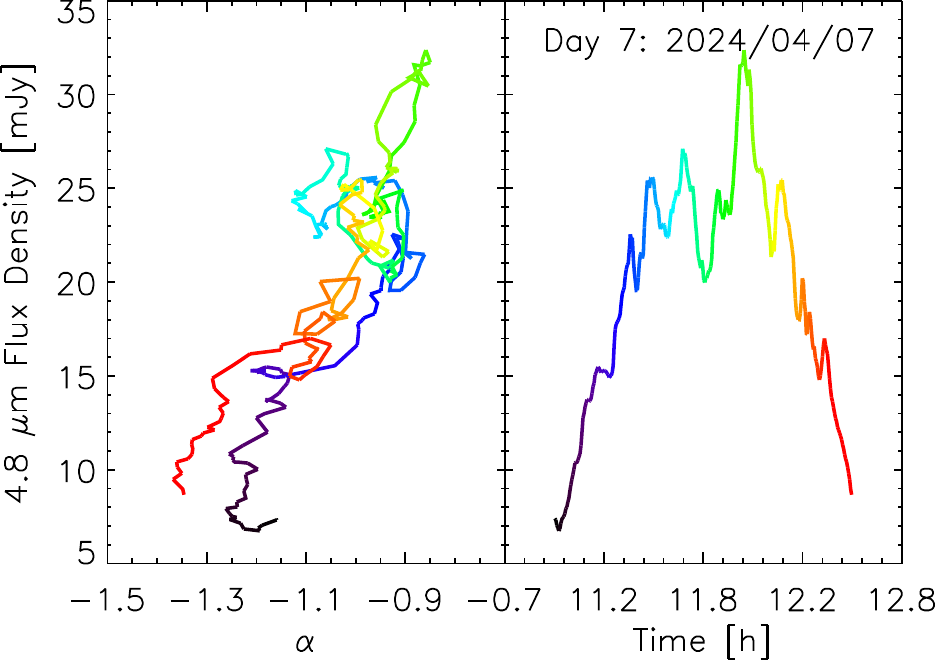}
   \caption{
Similar to Figure 9 (left two columns)  except that more complex  flares with multiple subflares 
are selected. 
}
\end{figure}

\clearpage
%\vfill\eject

%[ et al. (xx)]
%%%%%%%%%%%%%%%%%%%%%%%%%%%%%%%%%%%%%%%%%%%%%%%%%%%%%%%%%%%%%%%%%%%%%%%%%%%
\bibliographystyle{aasjournal}
 %***

\clearpage

\end{document}